\documentclass{siamart1116}
\usepackage{etoolbox}
\apptocmd{\sloppy}{\hbadness 10000\relax}{}{}

\usepackage{amsfonts,amsmath,amssymb}
\usepackage{mathtools}
\usepackage{graphicx}
\usepackage{epstopdf}
\usepackage{algorithm}
\usepackage[noend]{algpseudocode}
\usepackage{subcaption}
\usepackage{float}
\usepackage{multirow}

\ifpdf
  \DeclareGraphicsExtensions{.eps,.pdf,.png,.jpg}
\else
  \DeclareGraphicsExtensions{.eps}
\fi

\numberwithin{theorem}{section}

\newcommand{\TheTitle}{Avoiding communication in primal and dual block coordinate descent methods}
\newcommand{\TheAuthors}{Aditya Devarakonda, Kimon Fountoulakis, James Demmel, and Michael W. Mahoney}

\headers{Avoiding communication in machine learning}{A. Devarakonda, K. Fountoulakis, J. Demmel, and M. W. Mahoney}

\title{{\TheTitle}}

\author{
  Aditya Devarakonda\thanks{EECS Department, University of California, Berkeley, Berkeley, CA 94709
    (\email{aditya@eecs.berkeley.edu}).}
  \and
  Kimon Fountoulakis\thanks{ICSI and Statistics Department, University of California, Berkeley, Berkeley, CA 94709 (\email{kfount@berkeley.edu},
    \email{mmahoney@stat.berkeley.edu}).}
  \and
  James Demmel\thanks{Mathematics and EECS Department, Univeristy of California, Berkeley, Berkeley, CA 94709 (\email{demmel@berkeley.edu})}
  \and
  Michael W. Mahoney\footnotemark[2]
}

\usepackage{amsopn}

\DeclareMathOperator*{\argmin}{arg\,min}

\ifpdf
\hypersetup{
  pdftitle={\TheTitle},
  pdfauthor={\TheAuthors}
}
\fi

\newcommand*{\data}{./fig/num}

\begin{document}
\maketitle
\begin{abstract}
  Primal and dual block coordinate descent methods are iterative methods for solving regularized and unregularized optimization problems. Distributed-memory parallel implementations of these methods have become popular in analyzing large machine learning datasets. However, existing implementations communicate at every iteration which, on modern data center and supercomputing architectures, often dominates the cost of floating-point computation. Recent results on communication-avoiding Krylov subspace methods suggest that large speedups are possible by re-organizing iterative algorithms to avoid communication. We show how applying similar algorithmic transformations can lead to primal and dual block coordinate descent methods that only communicate every $s$ iterations--where $s$ is a tuning parameter--instead of every iteration for the \textit{regularized least-squares problem}. We show that the communication-avoiding variants reduce the number of synchronizations by a factor of $s$ on distributed-memory parallel machines without altering the convergence rate and attains strong scaling speedups of up to $6.1\times$ on a Cray XC30 supercomputer.
\end{abstract}



\section{Introduction}
%
The running time of an algorithm depends on computation, the number of arithmetic operations ($F$), and communication, the cost of data movement. The communication cost includes the ``bandwidth cost", i.e. the number, W, of words sent either between levels of a memory hierarchy or between processors over a network, and the ``latency cost", i.e. the number, L, of messages sent, where a message either consists of a group of contiguous words being sent, or is used for interprocess synchronization. On modern computer architectures, communicating data often takes much longer than performing a floating-point operation and this gap is continuing to increase. Therefore, it is especially important to design algorithms that minimize communication in order to attain high performance on modern computer architectures. Communication-avoiding algorithms are a new class of algorithms that exhibit large speedups on modern, distributed-memory parallel architectures through careful algorithmic transformations \cite{Ballard14}. Much of direct and iterative linear algebra have been re-organized to avoid communication and has led to significant performance improvements over existing state-of-the-art libraries \cite{Ballard14, Ballard13, Carson15, Hoemmen10, Solomonik14, williams14}. The results from communication-avoiding Krylov subspace methods \cite{Carson15, demmel07, Hoemmen10} are particularly relevant to our work.

The origins of communication-avoiding Krylov subspace methods lie in the $s$-step Krylov methods work. Van Rosendale's $s$-step conjugate gradients method \cite{rosendale83}, Chronopoulos and Gear's $s$-step methods for preconditioned and unpreconditioned symmetric linear systems \cite{chronopoulos89a,chronopoulos89b}, Chronopoulos and Swanson's $s$-step methods for unsymmetric linear systems \cite{chronopoulos96} and Kim and Chronopoulos's $s$-step non-symmetric Lanczos method \cite{kim92} were designed to extract more parallelism than their standard counterparts. $s$-step Krylov methods compute $s$ Krylov basis vectors and perform residual and solution vector updates by using Gram matrix computations and replacing modified Gram-Schmidt orthogonalization with Householder QR \cite{walker88}. These optimizations enable $s$-step Krylov methods to use BLAS-3 matrix-matrix operations which attain higher peak hardware performance and have more parallelism than the BLAS-1 vector-vector and BLAS-2 matrix-vector operations used in standard Krylov methods. However, these methods do not avoid communication in the $s$ Krylov basis vector computations. Demmel, Hoemmen, Mohiyuddin, and others \cite{demmel07, Hoemmen10, mohiyuddin12, mohiyuddin09} introduced the matrix powers kernel optimization which reduces the communication cost of the $s$ Krylov basis vector computations by a factor $O(s)$ for well-partitioned matrices. The combination of the matrix powers kernel along with extensive algorithmic modifications to existing $s$-step methods and derivation of new $s$-step methods resulted in what Carson, Demmel, Hoemmen and others call communication-avoiding Krylov subspace methods \cite{Carson15, demmel07, Hoemmen10}.
\begin{table}[t!]
\begin{center}
\small
\begin{tabular}{c| c | c | c}
\hline
\multicolumn{4}{c}{Summary of Ops and Memory costs}\\ \hline \hline
Algorithm & Data layout & Ops cost (F) & Memory cost (M)\\ \hline
{BCD} & \multirow{2}{*}{1D-column} & {$O\left(\frac{Hb^2fn}{P} + Hb^3\right)$} & {$O\left(\frac{fdn + n}{P} + b^2 + d\right)$}\\ \cline{1-1} \cline{3-4}
{CA-BCD} &  & {$O\left(\frac{Hb^2sfn}{P} + Hb^3\right)$} & $O\left(\frac{fdn + n}{P} + b^2s^2 + d\right)$\\ \hline
{BDCD} & \multirow{2}{*}{1D-row} & {$O\left(\frac{H'b'^2fd}{P} + H'b'^3\right)$} & {$O\left(\frac{fdn + d}{P} + b'^2 + n\right)$}\\ \cline{1-1} \cline{3-4}
{CA-BDCD} &  & {$O\left(\frac{H'b'^2sfd}{P} + H'b'^3\right)$} & $O\left(\frac{fdn + d}{P} + b'^2s^2 + n\right)$\\ \hline
\end{tabular}
\quad
\begin{tabular}{c| c | c | c}
\hline
\multicolumn{4}{c}{Summary of Communication costs}\\ \hline \hline
Algorithm & Data layout  & Latency cost (L) & Bandwidth cost (W)\\ \hline
{BCD} & \multirow{2}{*}{1D-column} &  $O\left(H\log P\right)$ & $O\left(Hb^2\log P\right)$\\ \cline{1-1} \cline{3-4}
{CA-BCD} &  & $O\left(\frac{H}{s}\log P\right)$ & $O\left(Hb^2s\log P\right)$ \\ \hline
{BDCD} & \multirow{2}{*}{1D-row} & $O\left(H'\log P\right)$ & $O\left(H'b'^2\log P\right)$ \\ \cline{1-1} \cline{3-4}
{CA-BDCD} & & $O\left(\frac{H'}{s}\log P\right)$ & $O\left(H'b'^2s\log P\right)$\\ \hline
\end{tabular}
\end{center}
\caption{Ops (F), Latency (L), Bandwidth (W) and Memory per processor (M) costs comparison along the critical path of classical BCD (Thm. \ref{thm:bcd1dcol}), BDCD (Thm. \ref{thm:bdcd1drow}) and communication-avoiding BCD (Thm. \ref{thm:cabcd1dcol}) and BDCD (Thm. \ref{thm:cabdcd1drow}) algorithms for 1D-block column and 1D-block row data partitioning, respectively. $H$ and $H'$ are the number of iterations and $b$ and $b'$ are the block sizes for BCD, and BDCD. We assume that $X \in \mathbb{R}^{d \times n}$ is sparse with $fdn$ non-zeros that are uniformly distributed, $0 < f \leq 1$ is the density of $X$ (i.e. $f = \frac{nnz(X)}{dn}$), $P$ is the number of processors and $s$ is the recurrence unrolling parameter. $fbn$ is the non-zeros of the $b \times n$ matrix with $b$ sampled rows from $X$ at each iteration, and $fb'd$ is the non-zeros of the $d \times b'$ matrix with $b'$ sampled columns from $X$ at each iteration. We assume that the $b \times b$ and $b' \times b'$ Gram matrices computed at each iteration for BCD and BDCD, respectively, are dense.}
\label{tbl:sumres}
\end{table}

We build on existing work by extending those results to machine learning where scalable algorithms are especially important given the enormous amount of data. Block coordinate descent methods are routinely used in machine learning to solve optimization problems \cite{nesterov12, richtarik14, Wright15}. Given a sparse dataset $X \in \mathbb{R}^{d \times n}$ where the rows are features of the data and the columns are data points, the block coordinate descent method can compute the regularized or unregularized least squares solution by iteratively solving a subproblem using a block of $b$ rows of $X$ \cite{nesterov12,richtarik14,Wright15}. This process is repeated until the solution converges to a desired accuracy or until the number of iterations has reached a user-defined limit. If $X$ is distributed (in 1D-row or 1D-column layout) across $P$ processors then the algorithm communicates at each iteration in order to solve the subproblem. As a result, the running time for such methods is often dominated by communication cost which increases with $P$. 

\begin{table}[t]
\small
\begin{center}
\begin{tabular}{c|c|c}
\hline
\multicolumn{3}{c}{Ops and Memory Costs Comparison}\\ \hline \hline
Algorithm & Ops cost (F) & Memory cost (M)\\ \hline
BCD (section \ref{sec:anal} Thm. \ref{thm:bcd1dcol})& $O\left(\frac{Hb^2fn}{P} + Hb^3\right)$ & $O\left(\frac{fdn + n}{P} + b^2 + d\right)$\\ \hline
BDCD (section \ref{sec:anal} Thm. \ref{thm:bdcd1drow})& $O\left(\frac{H'b'^2fd}{P} + H'b'^3\right)$ & $O\left(\frac{fdn + d}{P} + b'^2 + n\right)$\\ \hline
Krylov methods \cite{Ballard14}& $O\left(\frac{kfdn}{P}\right)$ & $O\left(\frac{fdn}{P} + \min({d, n}) + \frac{\max({d, n})}{P}\right)$ \\ \hline
\end{tabular}
\quad
\begin{tabular}{c|c|c}
\hline
\multicolumn{3}{c}{Communication Costs Comparison}\\ \hline \hline
Algorithm & Latency cost (L) & Bandwidth cost (W)\\ \hline
BCD (section \ref{sec:anal} Thm. \ref{thm:bcd1dcol}) & $O(H\log P)$ & $O(Hb^2\log P)$\\ \hline
BDCD (section \ref{sec:anal} Thm. \ref{thm:bdcd1drow})& $O\left({H'}\log P\right)$ & $O\left({H'}b'^2\log P\right)$\\ \hline
Krylov methods \cite{Ballard14}& $O(k\log P)$ & $O\left(k\min(d,n)\log P\right)$  \\ \hline
\end{tabular}
\end{center}
\caption{Computation and communication costs along the critical path of BCD, BDCD, Krylov and TSQR methods. $ H, H',~\text{and}~ k$ are the total number of iterations required for BCD, BDCD and Krylov methods, respectively, to converge to a desired accuracy. $b$ and $b'$ are the block sizes for BCD and BDCD, respectively. For Krylov methods we assume a 1D-block row layout if $n < d$ (1D-block column if $n > d$) and replicate the $\min(d,n)$-dimensional vectors and partition the $\max(d,n)$-dimensional vectors.}
\label{tbl:bcdflops}
\end{table}
There are some frameworks and algorithms that attempt to reduce the communication bottleneck. For example, the CoCoA framework \cite{cocoa} reduces communication by performing coordinate descent on locally stored data points on each processor and intermittently communicating by summing or averaging the local solutions. CoCoA communicates fewer times than coordinate descent -- although not provably so -- but changes the convergence behavior. HOGWILD! \cite{recht11} is a lock-free approach to stochastic gradient descent (SGD) where each processor selects a data point, computes a gradient using its data point and updates the solution without synchronization. Due to the lack of synchronization (or locks) processors are allowed to overwrite the solution vector. The main results in HOGWILD! show that if the solution updates are sparse (i.e. each processor only modifies a part of the solution) then running without locks does not affect the final solution with high probability.

In contrast, our results reduce the latency cost in the primal and dual block coordinate descent methods by a factor of $s$ on distributed-memory architectures, for dense and sparse updates without changing the convergence behavior, in exact arithmetic. Hereafter we refer to the primal method as block coordinate descent (BCD) and the dual method as block dual coordinate descent (BDCD). The proofs in this paper assume that $X$ is sparse with $fdn$ non-zeros that are uniformly distributed where $0 < f \leq 1$ is the density of $X$ (i.e. $f = \frac{nnz(X)}{dn}$). Each iteration of BCD samples\footnote{uniformly, without replacement.} $b$ rows of $X$ (resp. $b'$ columns of $X$ for BDCD). The resulting $b \times n$ (resp. $d \times b'$ for BDCD) sampled matrix contains $fbn$ (resp. $fb'd$ for BDCD) non-zeros. These assumptions simplify our analysis and provide insight into scaling behavior for ideal sparse inputs. We leave extensions of our proofs to general sparse matrices for future work.

The principle behind our communication-avoiding approach is to unroll the BCD and BDCD vector update recurrences by a factor of $s$, compute Gram-like matrices for the next $s$ iterations, and use linear combinations of the $s$ gradients to update the solution vector. Table \ref{tbl:sumres} summarizes our results for BCD with $X$ stored in a 1D-block column layout and 1D-block row layout for BDCD. Our communication-avoiding variants reduce the latency cost, which is the dominant cost, by a factor of $s$ but increase the bandwidth and flops cost by a factor of $s$. The algorithms we derive also avoid communication for other data layout schemes, however, we limit our discussion in this paper to the 1D-block column and 1D-block row layouts.
\\
\subsection{Contributions}
We briefly summarize our contributions:
\begin{itemize}
\item We present communication-avoiding algorithms for block coordinate descent and block dual coordinate descent that {\it provably} reduce the latency cost by a factor of $s$.
\item We analyze the operational, communication and storage costs of the classical and our new communication-avoiding algorithms under two data partitioning schemes and describe their performance tradeoffs.
\item We perform numerical experiments to illustrate that the communication-avoiding algorithms are numerically stable for all choices of $s$ tested.
\item We show performance results to illustrate that the communication-avoiding algorithms can be up to $6.1\times$ faster than the standard algorithms on up to 1024 nodes of a Cray XC30 supercomputer using MPI.
\end{itemize}

\subsection{Organization}
The rest of the paper is organized as follows: Section \ref{sec:rwork} summarizes existing methods for solving the regularized least squares problem and the communication cost model used to analyze our algorithms. Section \ref{sec:deriv} presents the communication-avoiding derivations of the BCD and BDCD algorithms. Section \ref{sec:anal} analyzes the operational, communication and storage costs of the classical and communication-avoiding algorithms under the 1D-block column and 1D-block row data layouts. Section \ref{sec:eval} provides numerical and performance experiments which show that the communication-avoiding algorithms are numerically stable and attain speedups over the standard algorithms. Finally, we conclude in Section \ref{sec:conclusion} and describe directions for future work.

\section{Background}\label{sec:rwork}
%

We begin by describing the cost model used to analyze the running time of the standard and new, communication-avoiding algorithms. Then we survey existing methods for solving regularized least squares problems. We compare the algorithm costs of these methods and describe their tradeoffs to motivate the need for communication-avoiding block coordinate descent methods.
\subsection{Modeling Communication}
Algorithms have traditionally been analyzed by counting arithmetic (the number of floating-point operations). However, data movement (communication) is another important cost that often dominates arithmetic cost \cite{fuller11,graham05}. By combining the arithmetic and communication costs we obtain the following running time model
\begin{equation}\label{eq:comm}
\text{T}_{\text{algorithm}} = \underbracket{\gamma F}_{\text{Computation Cost}} + \underbracket{\alpha L + \beta W}_{\text{Communication Cost}}
\end{equation}
where $\gamma, \alpha,~\text{and}~\beta$ are machine-specific parameters that correspond to the time per operation, overhead time per message, and time per word moved, respectively. $F$, $L$, and, $W$ are algorithm-specific parameters that represent the total number of floating-point operations computed, the number of messages sent and the number of words moved, respectively. Communication models have been well-studied in literature from the LogP \cite{culler93} and LogGP \cite{alexandrov97} models to the $\alpha$-$\beta$ model (eq. \ref{eq:comm}). The LogP and LogGP models are refinements of the $\alpha$-$\beta$ model, therefore, we use the latter for simplicity. The $\alpha$-$\beta$ model applies to both sequential and parallel computations but we focus on the latter in this paper.
\begin{figure}[t!]
\centering
\includegraphics[trim = 0.in 2in 0.1in 2in, clip,width=.8\textwidth]{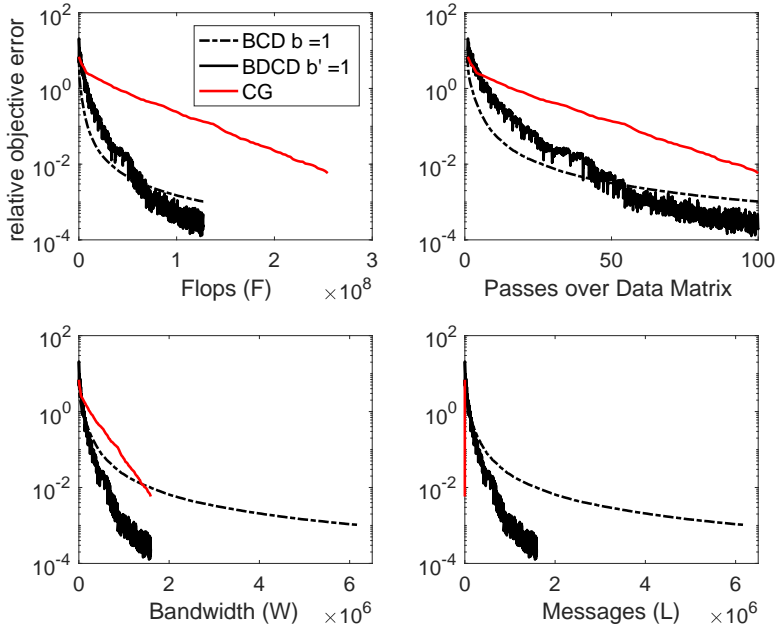}
%
\caption{Comparison of convergence behavior against flops, bandwidth, latency, and data matrix passes of Conjugate Gradients (CG), BCD (with $b = 1$) and BDCD (with $b' = 1$). Convergence is reported in terms of the relative objective error and the experiments are performed on the news20 dataset ($d = 62061$, $n = 15935$, $nnz(X) = 1272569$) obtained from LIBSVM \cite{cc01}. We fix the number of CG iterations to $k = 100$, BCD iterations to $H = 100d$ and BDCD iterations to $H' = 100n$ so that each algorithm performs $100$ passes over $X$.}
\label{fig:compare}
\end{figure}
\subsection{Survey of Regularized Least Squares Methods}\label{sec:lsqbg}
The regularized least-squares problem can be written as the following optimization problem:
\begin{equation}\label{eq:primal}
\argmin_{w\in\mathbb{R}^d} \frac{\lambda}{2}\|w\|^2_2 + \frac{1}{2n}\left\|X^Tw - y\right\|^2_2
\end{equation}
where $X\in\mathbb{R}^{d \times n}$ is the data matrix whose rows are features and columns are data points, $y\in\mathbb{R}^n$ are the labels, $w \in \mathbb{R}^d$ are the weights, and $\lambda>0$ is a regularization parameter. The unregularized ($\lambda = 0$) and regularized ($\lambda > 0$) least squares problems have been well-studied in literature from directly solving the normal equations to other matrix factorization (QR via Gram-Schmidt or Householder, LU, Cholesky, etc.) approaches \cite{demmel08,bjorck96} to Krylov \cite{bjorck96, Carson15, saad03} and (primal and dual) block coordinate descent methods \cite{bottou10, cocoa, shalev13, takac15, Wright15}. Table \ref{tbl:bcdflops} summarizes the parallel algorithm costs of the various algorithms just described. Note that we assume that the data matrix, $X$, is dense for simplicity.

We briefly summarize the difference between the BCD and BDCD algorithms, but defer the derivations to Section \ref{sec:deriv}. The BCD algorithm solves the primal minimization problem \eqref{eq:primal}, whereas, the BDCD algorithm solves the dual minimization problem:
\begin{align}\label{eq:6}
\argmin_{\alpha\in\mathbb{R}^n}&  \frac{\lambda}{2}\left\|\frac{1}{\lambda n} X\alpha\right\|^2_2  + \frac{1}{2 n}\left\|\alpha + y\right\|^2_2
\end{align}
where $\alpha \in \mathbb{R}^n$ is the dual solution vector. The dual problem \cite{shalev13} can be obtained by deriving the convex conjugate of \eqref{eq:primal} and has the following primal-dual solution relationship:
\begin{align}\label{eq:3}
w &= -\frac{1}{\lambda n} X\alpha.
\end{align}
 \begin{table}[t!]
 \begin{center}
 \small
 \begin{tabular}{c|c|c|c}
 \hline
 \multicolumn{4}{c}{Relative Objective Error Comparison}\\ \hline \hline
 CG iteration & CG error & BDCD error & BCD error\\ \hline
 0 & 6.8735 & 6.8735 & 6.8735\\ \hline
 1 & 4.5425 &  7.8231 & 1.2826\\ \hline
 25 & 0.5115 & 0.0441 & 0.0104\\ \hline
 50 & 0.1326 & 0.0043 & 0.0031\\ \hline
 75 & 0.0283 & 5.0779e-04 & 0.0016\\ \hline
 100 & 0.0058 & 1.9346e-04 & 0.0010 \\ \hline
 \end{tabular}
 \end{center}
 \caption{Comparison of CG iterations and relative objective error of CG, BDCD ($b' = 1$) and BCD ($b = 1$). We normalize the BDCD and BCD iterations to match each CG iteration reported. If $k$ is the CG iteration, then BCD performs $H = kd$ and BDCD performs $H' = kn$.}
 \label{tbl:errorcmp}
 \end{table}
Figure \ref{fig:compare} illustrates the tradeoff between convergence behavior and theoretical flops, number of passes over $X$, bandwidth and latency costs of CG, BCD and BDCD based on the costs in Table \ref{tbl:bcdflops}. We plot the sequential flops cost and ignore the $\log P$ factor for latency. We allow each algorithm to make $100$ passes over $X$ and plot the relative objective error, $\frac{f(X, w_{opt}, y) - f(X, w_{alg}, y)}{f(X,w_{opt},y)}$, where $f(X,w,y) = \frac{1}{2n}\|X^Tw - y\|_2^2 + \frac{\lambda}{2}\|w\|_2^2$. $w_{opt}$ is computed {\it a priori} from CG with a tolerance of $10^{-15}$, and $w_{alg}$ is the solution obtained from each iteration of CG, BCD or BDCD. 
Since $X$ is not symmetric, CG requires two matrix-vector products at each iteration (one with $X$ and another with $X^T$). Therefore, the flops cost of CG is twice that of BCD or BDCD. We assume that the two matrix-vector products can be computed with a single pass over $X$. Ignoring communication, we observe that BCD and BDCD converge faster than CG to $10^{-3}$ accuracy before stagnating. If low-accuracy suffices, then BCD and BDCD converge faster in terms of flops and passes over $X$. Since we measure the relative error in the primal objective (rather than the dual objective), the convergence of BDCD does not decrease monotonically. Table \ref{tbl:errorcmp} summarizes the comparison of objective error progress of CG, BCD and BDCD normalized for CG iterations.

When considering communication, CG is more bandwidth efficient than BCD (but not BDCD) and is {\it orders of magnitude} more latency efficient than BCD and BDCD. This suggests that reducing the latency cost of BCD and BDCD is an important step in making these algorithms competitive. In this paper, we focus on the design, numerical stability and performance of these communciation-avoiding variants and leave the design space exploration of choosing the best algorithm for future work.

\section{Communication-Avoiding Primal and Dual Block Coordinate Descent}\label{sec:deriv}

%
In this section, we re-derive the block coordinate descent (BCD) (in section \ref{sec:bcdderiv}) and block dual coordinate descent (BDCD) (in section \ref{sec:bdcdderiv}) algorithms starting from the respective minimization problems. The derivation of BCD and BDCD lead to recurrences which can be unrolled to derive communication-avoiding versions of BCD and BDCD, which we will refer to as CA-BCD and CA-BDCD respectively.

\subsection{Derivation of Block Coordinate Descent}\label{sec:bcdderiv}
The minimization problem in \eqref{eq:primal} can be solved by block coordinate descent with the $b$-dimensional update
\begin{equation}\label{eq:wupdate}
w_{h} = w_{h-1} + \mathbb{I}_h\Delta w_h
\end{equation}
where $w_h \in \mathbb{R}^d$ and $\mathbb{I}_h = \begin{bmatrix}e_{i_1}, e_{i_2}, \ldots, e_{i_b}\end{bmatrix} \in \mathbb{R}^{d \times b}$, $\Delta w_h \in \mathbb{R}^b$, and $i_k \in [d]$ for $k = 1, 2, \ldots, b$. By substitution in \eqref{eq:primal} we obtain the minimization problem
\begin{equation*}
\argmin_{\Delta w_h\in\mathbb{R}^b} \frac{\lambda}{2}\|w_{h-1} + \mathbb{I}_h\Delta w_h\|_2^2 + \frac{1}{2n} \|X^Tw_{h-1} + X^T\mathbb{I}_h\Delta w_h - y\|_2^2
\end{equation*}
with the closed-form solution
\begin{equation}
\Delta w_h = \left(\frac{1}{n}\mathbb{I}_h^TXX^T\mathbb{I}_h + \lambda \mathbb{I}_h^T\mathbb{I}_h\right)^{-1}\left(-\lambda \mathbb{I}_h^Tw_{h-1} - \frac{1}{n} \mathbb{I}_h^TXX^Tw_{h-1} + \frac{1}{n}\mathbb{I}_h^TXy\right).
\end{equation}

\begin{algorithm}[t!]
\caption{Block Coordinate Descent (BCD) Algorithm}\label{alg:bcdres}
\begin{algorithmic}[1]
\State \textbf{Input}: $X \in \mathbb{R}^{d \times n}, y \in \mathbb{R}^n$, $H>1$, $w_0 \in\mathbb{R}^{d}$, $b  \in \mathbb{Z}_+$ s.t. $b \leq d$
\For {$h=1,2,\cdots,H$}
\State choose $\{i_m \in [d] | m = 1, 2, \ldots, b \}$ uniformly at random without replacement\vspace{0.1cm}
\State $\mathbb{I}_h = \left[e_{i_1}, e_{i_2}, \cdots, e_{i_b}\right]$\vspace{0.1cm}
\State $\Gamma_h = {\frac{1}{n} \mathbb{I}_h^TXX^T\mathbb{I}_h + \lambda}\mathbb{I}_h^T\mathbb{I}_h$
\State $\Delta w_h = \Gamma_h^{-1}\left(-\lambda \mathbb{I}_h^Tw_{h-1} - \frac{1}{n} \mathbb{I}_h^TXz_{h-1} + \frac{1}{n}\mathbb{I}_h^TXy\right)$\label{ln:primdw}
\State $w_{h} = w_{h-1} +\mathbb{I}_h\Delta w_h$\label{ln:primwh}
\State $z_{h} = z_{h-1} +X^T\mathbb{I}_h\Delta w_h$\label{ln:primah}
\EndFor
\State \textbf{Output} $w_H$
\end{algorithmic}
\end{algorithm}
The closed-from solution requires a matrix-vector multiply using the entire data matrix to compute $\frac{1}{n} \mathbb{I}_h^TXX^Tw_{h-1}$. However, this can be avoided by introducing the auxiliary variable:
\begin{equation*}
z_h = X^Tw_h
\end{equation*}
which, by substituting \eqref{eq:wupdate}, can be re-arranged into a vector update of the form
\begin{align}
z_h &= X^Tw_{h-1} + X^T\mathbb{I}_h\Delta w_h\nonumber\\
&= z_{h-1} + X^T\mathbb{I}_h\Delta w_h\label{eq:primah}
\end{align}
and the closed-form solution can be written in terms of $z_{h-1}$,
\begin{align}
\Delta w_h = \left(\frac{1}{n}\mathbb{I}_h^TXX^T\mathbb{I}_h + \lambda \mathbb{I}_h^T\mathbb{I}_h\right)^{-1}\left(-\lambda \mathbb{I}_h^Tw_{h-1} - \frac{1}{n} \mathbb{I}_h^TXz_{h-1} + \frac{1}{n}\mathbb{I}_h^TXy\right).\label{eq:primres-orig}
\end{align}
In order to make the communication-avoiding BCD derivation easier, let us define
\begin{equation*}
\Gamma_h = \frac{1}{n}\mathbb{I}_h^TXX^T\mathbb{I}_h + \lambda \mathbb{I}_h^T\mathbb{I}_h.
\end{equation*}
Then \eqref{eq:primres-orig} can be re-written as
\begin{align}
\Delta w_h = \Gamma_h^{-1}\left(-\lambda \mathbb{I}_h^Tw_{h-1} - \frac{1}{n} \mathbb{I}_h^TXz_{h-1} + \frac{1}{n}\mathbb{I}_h^TXy\right).\label{eq:primres}
\end{align}
This re-arrangement leads to the Block Coordinate Descent (BCD) method shown in Algorithm \ref{alg:bcdres}.
\begin{algorithm}[t]
\caption{Communication-Avoiding Block Coordinate Descent (CA-BCD) Algorithm}\label{alg:cabcd}
\begin{algorithmic}[1]
\State \textbf{Input}: $X \in \mathbb{R}^{d \times n}, y \in \mathbb{R}^n$, $ H>1$, $w_0 \in \mathbb{R}^d$, $b  \in \mathbb{Z}_+$ s.t. $b \leq d$\vspace{0.1cm}
\For {$k=0,1,\cdots, \frac{H}{s}$} \vspace{0.1cm}
\For {$j = 1,2,\cdots, s$}
\State choose $\{i_m \in [d] | m = 1, 2, \ldots, b \}$ uniformly at random without replacement\vspace{0.1cm}
\State $\mathbb{I}_{sk + j} = \left[e_{i_1}, e_{i_2}, \cdots, e_{i_b}\right]$\vspace{0.1cm}
\EndFor
\State let $Y = \begin{bmatrix}\mathbb{I}_{sk + 1}, \mathbb{I}_{sk + 2}, \cdots, \mathbb{I}_{sk + s}\end{bmatrix}^TX$.
\State compute the Gram matrix, $G = \frac{1}{n}YY^T + \lambda I$.
\For {$j = 1,2,\cdots, s$}
\State $\Gamma_{sk + j}$ are the $b \times b$ diagonal blocks of $G$.
\State $\Delta w_{sk + j} = \Gamma_{sk + j}^{-1}\biggr(-\lambda \mathbb{I}_{sk + j}^Tw_{sk} -\lambda \sum_{t = 1}^{j-1}\left(\mathbb{I}_{sk + j}^T\mathbb{I}_{sk + t}\Delta w_{sk + t}\right) - \frac{1}{n} \mathbb{I}_{sk + j}^TXz_{sk}$
\Statex \indent\indent\indent\indent\indent$ - \frac{1}{n} \sum_{t = 1}^{j - 1}\left(\mathbb{I}_{sk + j}^TXX^T\mathbb{I}_{sk + t}\Delta w_{sk + t}\right) + \frac{1}{n}\mathbb{I}_{sk + j}^TXy\biggr)$
\State $w_{sk + j} = w_{sk+j-1} + \mathbb{I}_{sk + j}\Delta w_{sk + j}$
\State $z_{sk+ j} = z_{sk+j-1} + X^T\mathbb{I}_{sk + j}\Delta w_{sk + j}$
\EndFor
\EndFor
\State \textbf{Output} $w_{H}$
\end{algorithmic}
\end{algorithm}
The recurrence in lines \ref{ln:primdw}, \ref{ln:primwh}, and \ref{ln:primah} of Algorithm \ref{alg:bcdres} allow us to unroll the BCD recurrences and avoid communication. We begin by changing the loop index from $h$ to $sk + j$ where $k$ is the outer loop index, $s$ is the recurrence unrolling parameter and $j$ is the inner loop index. Assume that we are at the beginning of iteration $sk+ 1$ and $w_{sk}$ and $z_{sk}$ were just computed. Then $\Delta w_{sk+ 1}$ can be computed by
\begin{align*}
\Delta w_{sk + 1} = \Gamma_{sk + 1}^{-1}\left(-\lambda \mathbb{I}_{sk + 1}^Tw_{sk} - \frac{1}{n} \mathbb{I}_{sk + 1}^TXz_{sk} + \frac{1}{n}\mathbb{I}_{sk + 1}^TXy\right).
\end{align*}
By unrolling the recurrence for $w_{sk  +1}$ and $z_{sk + 1}$ we can compute $\Delta w_{sk + 2}$ in terms of $w_{sk}$ and $z_{sk}$
\begin{multline*}
\Delta w_{sk + 2} = \Gamma_{sk + 2}^{-1}\biggr(-\lambda \mathbb{I}_{sk + 2}^Tw_{sk} -\lambda \mathbb{I}_{sk + 2}^T\mathbb{I}_{sk + 1}\Delta w_{sk + 1}\\ - \frac{1}{n} \mathbb{I}_{sk + 2}^TXz_{sk} - \frac{1}{n} \mathbb{I}_{sk + 2}^TXX^T\mathbb{I}_{sk + 1}\Delta w_{sk + 1} + \frac{1}{n}\mathbb{I}_{sk + 2}^TXy\biggr).
\end{multline*}
By induction we can show that $\Delta w_{sk + j}$ can be computed using $w_{sk}$ and $z_{sk}$
\begin{multline}
\Delta w_{sk + j} = \Gamma_{sk + j}^{-1}\biggr(-\lambda \mathbb{I}_{sk + j}^Tw_{sk} -\lambda \sum_{t = 1}^{j-1}\left(\mathbb{I}_{sk + j}^T\mathbb{I}_{sk + t}\Delta w_{sk + t}\right)\\ - \frac{1}{n} \mathbb{I}_{sk + j}^TXz_{sk} - \frac{1}{n} \sum_{t = 1}^{j - 1}\left(\mathbb{I}_{sk + j}^TXX^T\mathbb{I}_{sk + t}\Delta w_{sk + t}\right) + \frac{1}{n}\mathbb{I}_{sk + j}^TXy\biggr)\label{eq:recur}.
\end{multline}
for $j = 1, 2, \ldots, s$. Due to the recurrence unrolling we can defer the updates to $w_{sk}$ and $z_{sk}$ for $s$ steps. Notice that the first summation in \eqref{eq:recur} computes the intersection between the coordinates chosen at iteration $sk + j$ and $sk + t$ for $t = 1, \ldots, j-1$ via the product $\mathbb{I}_{sk + j}^T\mathbb{I}_{sk + t}$. Communication can be avoided in this term by initializing all processors to the same seed for the random number generator. The second summation in \eqref{eq:recur} computes the Gram-like matrices $\mathbb{I}^T_{sk + j}XX^T\mathbb{I}_{sk + t}$ for $t = 1, \ldots, j-1$. Communication can be avoided in this computation by computing the $sb \times sb$ Gram matrix $G = \left(\frac{1}{n}\begin{bmatrix}\mathbb{I}_{sk + 1}, \mathbb{I}_{sk + 2}, \cdots, \mathbb{I}_{sk + s}\end{bmatrix}^TXX^T\begin{bmatrix} + \mathbb{I}_{sk + 1}, \mathbb{I}_{sk + 2}, \cdots, \mathbb{I}_{sk + s}\end{bmatrix} + \lambda I\right)$ once before the inner loop and redundantly storing it on all processors. Finally, at the end of the $s$ inner loop iterations we can perform the vector updates
\begin{align}
w_{sk + s} &= w_{sk} + \sum_{t = 1}^s\left(\mathbb{I}_{sk + t}\Delta w_{sk + t}\right),\\
z_{sk + s}  &= z_{sk} + X^T\sum_{t = 1}^s\left(\mathbb{I}_{sk + t}\Delta w_{sk + t}\right).
\end{align}
The resulting communication-avoiding BCD (CA-BCD) algorithm is shown in Algorithm \ref{alg:cabcd}.

\subsection{Derivation of Block Dual Coordinate Descent}\label{sec:bdcdderiv}
The solution to the primal problem \eqref{eq:primal} can also be obtained by solving the dual minimization problem shown in \eqref{eq:6} with the primal-dual relationship shown in \eqref{eq:3}.
\begin{algorithm}[t]
\caption{Block Dual Coordinate Descent (BDCD) Algorithm}\label{sdca_res}
\begin{algorithmic}[1]
\State \textbf{Input}: $X = \left[x_1, x_2, \ldots x_n\right] \in \mathbb{R}^{d \times n}, y \in \mathbb{R}^n$, $H'>1$, $\alpha_0 \in\mathbb{R}^{n}$, $b'  \in \mathbb{Z}_+$ s.t. $b' \leq n$  \vspace{0.1cm}
\State \textbf{Initialize}: $w_{0} \leftarrow \frac{-1}{\lambda n} X \alpha_{0}$\vspace{0.1cm}
\For {$h=1,2,\cdots,H'$} \vspace{0.1cm}
\State choose $\{i_m \in [n] | m = 1, 2, \ldots, b' \}$ uniformly at random without replacement\vspace{0.1cm}
\State $\mathbb{I}_h = \left[e_{i_1}, e_{i_2}, \cdots, e_{i_{b'}}\right]$\vspace{0.1cm}
\State $\Theta_h = {\frac{1}{\lambda n^2} \mathbb{I}_h^TX^T X\mathbb{I}_h + \frac{1}{n}}\mathbb{I}_h^T\mathbb{I}_h$
\State $\Delta \alpha_h= -\frac{1}{n} \Theta_h^{-1}\left(-\mathbb{I}_h^TX^Tw_{h-1} + \mathbb{I}_h^T\alpha_{h-1} + \mathbb{I}_h^Ty\right)$\label{ln:ah}
\State $\alpha_h = \alpha_{h-1} + \mathbb{I}_h\Delta \alpha_h$\label{ln:alph}
\State $w_h = w_{h-1} - \frac{1}{\lambda n} X\mathbb{I}_h\Delta \alpha_h$\label{ln:wh}
\EndFor
\State \textbf{Output} $\alpha_H'$ and $w_H'$
\end{algorithmic}
\end{algorithm}
The dual problem \eqref{eq:6} can be solved using block coordinate descent which iteratively solves a subproblem in $\mathbb{R}^{b'}$, where $1 \leq b' \leq n$ is a tunable block-size parameter. Let us first define the dual vector update for $\alpha_h \in \mathbb{R}^n$
\begin{align}
\alpha_h &= \alpha_{h-1} + \mathbb{I}_h\Delta \alpha_h.\label{eq:alp}
\end{align}
Where $h$ is the iteration index, $\mathbb{I}_h  = \begin{bmatrix}e_{i_1}, e_{i_2}, \ldots e_{i_{b'}}\end{bmatrix} \in \mathbb{R}^{n \times {b'}}$, $i_k \in [n] ~\text{for}~k = 1,2, \ldots b'$ and $\Delta \alpha_h \in \mathbb{R}^{b'}$. By substitution in \eqref{eq:6}, $\Delta \alpha_h$ is the solution to a minimization problem in $\mathbb{R}^{b'}$ as desired:
\begin{align}
\argmin_{\Delta \alpha_h\in\mathbb{R}^{b'}}  \frac{1}{2\lambda n^2}\left\| X\alpha_{h-1} + X\mathbb{I}_{h}\Delta \alpha_{h}\right\|^2_2  + \frac{1}{2 n}\left\|\alpha_{h-1} + \mathbb{I}_h \Delta \alpha_h + y\right\|^2_2.\label{eq:op2}
\end{align}
Finally, due to \eqref{eq:3} we obtain the primal vector update for $w_h \in \mathbb{R}^d$
\begin{align}
w_h = w_{h - 1} - {1 \over {\lambda n}} X\mathbb{I}_h\Delta \alpha_h\label{eq:wup}.
\end{align}
From \eqref{eq:alp}, \eqref{eq:op2}, and \eqref{eq:wup} we obtain a block coordinate descent algorithm which solves the dual minimization problem. Henceforth, we refer to this algorithm as block dual coordinate descent (BDCD). Note that by setting $b' = 1$ we obtain the SDCA algorithm \cite{shalev13} with the least-squares loss function.

The optimization problem \eqref{eq:op2} which computes the solution along the chosen coordinates has the closed-form
\begin{equation}
\Delta \alpha_h =-\left({\frac{1}{\lambda n^2} \mathbb{I}_h^TX^T X\mathbb{I}_h + \frac{1}{n}}\mathbb{I}_h^T\mathbb{I}_h\right)^{-1}\left(\frac{1}{\lambda n^2} \mathbb{I}_h^TX^T X\alpha_{h-1} + \frac{1}{n}\mathbb{I}_h^T\alpha_{h-1} + \frac{1}{n}\mathbb{I}_h^Ty\right).
\end{equation}
Let us define $\Theta_h \in \mathbb{R}^{b' \times b'}$ such that
\begin{equation*}\label{eq:gam}
\Theta_h = \left({\frac{1}{\lambda n^2} \mathbb{I}_h^TX^T X\mathbb{I}_h + \frac{1}{n}}\mathbb{I}_h^T\mathbb{I}_h\right). \quad
\end{equation*}

\begin{algorithm}[t]
\caption{Communication-Avoiding Block Dual Coordinate Descent (CA-BDCD) Algorithm}\label{casdca}
\begin{algorithmic}[1]
\State \textbf{Input}: $X = \left[x_1, x_2, \ldots x_n\right] \in \mathbb{R}^{d \times n}, y \in \mathbb{R}^n$, $H'>1$, $\alpha_0\in\mathbb{R}^{n}$, $b'  \in \mathbb{Z}_+$ s.t. $b' \leq n$  \vspace{0.1cm}
\State \textbf{Initialize}: $w_0 \leftarrow \frac{-1}{\lambda n} X \alpha_0$\vspace{0.1cm}
\For {$k=0,1,\cdots, \frac{H'}{s}$} \vspace{0.1cm}
\For {$j = 1,2,\cdots, s$}
\State choose $\{i_m \in [n] | m = 1, 2, \ldots, b' \}$ uniformly at random without replacement\vspace{0.1cm}
\State $\mathbb{I}_{sk + j} = \left[e_{i_1}, e_{i_2}, \cdots, e_{i_{b'}}\right]$\vspace{0.1cm}
\EndFor
\State let $Y = X\left[{\mathbb{I}_{sk + 1}}, {\mathbb{I}_{sk + 2}}, \ldots, {\mathbb{I}_{sk + s}}\right]$.
\State compute the Gram matrix, $G' = {1 \over {\lambda n^2}}Y^TY + {1 \over n} I$.
\For {$j = 1,2,\cdots, s$}
\State $\Theta_{sk + j}$ are the $b' \times b'$ diagonal blocks of $G'$.
\State $\Delta \alpha_{sk + j} = -\frac{1}{n}\Theta_{sk + j}^{-1}\biggr(-\mathbb{I}_{sk + j}^T X^T w_{sk} +  \frac{1}{\lambda n}\sum_{t = 1}^{j-1}\left(\mathbb{I}_{sk + j}^TX^TX\mathbb{I}_{sk + t}\Delta \alpha_{sk + t}\right)$
\Statex \indent\indent\indent\indent\indent\indent\indent $ + \mathbb{I}_{sk + j}^T\alpha_{sk} + \sum_{t = 1}^{j-1}\left(\mathbb{I}_{sk + j}^T\mathbb{I}_{sk + t}\Delta \alpha_{sk + t}\right) +  \mathbb{I}_{sk + j}^Ty\biggr)$
\State $\alpha_{sk + j} = \alpha_{sk+ j - 1} + \mathbb{I}_{sk + j}\Delta \alpha_{sk + j}$
\State $w_{sk + j} = w_{sk + j - 1} - \frac{1}{\lambda n}X\mathbb{I}_{sk + j}\Delta \alpha_{sk + j}$
\EndFor
\EndFor
\State \textbf{Output} $\alpha_{ H'}$ and $w_{ H'}$
\end{algorithmic}
\end{algorithm}

From this we have that at iteration $h$, we compute the solution along the $b'$ coordinates of the linear system
\begin{align}
\Delta \alpha_h  = -\frac{1}{n}\Theta_h^{-1} \left(-\mathbb{I}_h^TX^Tw_{h-1} +  \mathbb{I}_h^T\alpha_{h-1} + \mathbb{I}_h^Ty\right)\label{eq:ah}
\end{align}
and obtain the BDCD algorithm shown in Algorithm \ref{sdca_res}. The recurrence in lines \ref{ln:ah}, \ref{ln:alph}, and \ref{ln:wh} of Algorithm \ref{sdca_res} allow us to unroll the BDCD recurrences and avoid communication. We begin by changing the loop index from $h$ to $sk + j$ where $k$ is the outer loop index, $s$ is the recurrence unrolling parameter and $j$ is the inner loop index. Assume that we are at the beginning of iteration $sk+ 1$ and $w_{sk}$ and $\alpha_{sk}$ were just computed. Then $\Delta \alpha_{sk + 1}$ can be computed by
\begin{align*}
\Delta \alpha_{sk + 1} = -\frac{1}{n}\Theta_{sk + 1}^{-1}\left(-\mathbb{I}_{sk + 1}^T X^T w_{sk} + \mathbb{I}_{sk + 1}^T\alpha_{sk} +  \mathbb{I}_{sk + 1}^Ty\right).
\end{align*}
Furthermore, by unrolling the recurrences for $w_{sk+1}$ and $\alpha_{sk + 1}$ we can analogously to \eqref{eq:recur} show by induction that
\begin{multline}
\Delta \alpha_{sk + j} = -\frac{1}{n}\Theta_{sk + j}^{-1}\biggr(-\mathbb{I}_{sk + j}^T X^T w_{sk} +  \frac{1}{\lambda n}\sum_{t = 1}^{j-1}\left(\mathbb{I}_{sk + j}^TX^TX\mathbb{I}_{sk + t}\Delta \alpha_{sk + t}\right) \\+ \mathbb{I}_{sk + j}^T\alpha_{sk} + \sum_{t = 1}^{j-1}\left(\mathbb{I}_{sk + j}^T\mathbb{I}_{sk + t}\Delta \alpha_{sk + t}\right) +  \mathbb{I}_{sk + j}^Ty\biggr)
\end{multline}
for $j = 1, 2, \ldots, s$. Note that due to unrolling the recurrence we can compute $\Delta \alpha_{sk + j}$ from $w_{sk}$ and $\alpha_{sk}$ which are the primal and dual solution vectors from the previous outer iteration. Since the solution vector updates require communication, the recurrence unrolling allows us to defer those updates for $s$ iterations at the expense of additional computation. The solution vectors can be updated at the end of the inner iterations by
\begin{align}
w_{sk + s} &= w_{sk} - \frac{1}{\lambda n} X\sum_{t = 1}^s\left(\mathbb{I}_{sk + t} \Delta \alpha_{sk + t}\right),\\
\alpha_{sk + s} &= \alpha_{sk} + \sum_{t = 1}^s \left(\mathbb{I}_{sk + t} \Delta \alpha_{sk + t}\right).
\end{align}
The resulting communication-avoiding BDCD (CA-BDCD) algorithm is shown in Algorithm \ref{casdca}.

\section{Analysis of Algorithms}\label{sec:anal}
%
From the derivations in Section \ref{sec:deriv}, we can observe that the primal and dual block coordinate descent algorithms perform computations on $XX^T$ and $X^TX$, respectively. This implies that, along with the convergence rates, the shape of $X$ is a key factor in choosing between the two methods. Furthermore, the data partitioning scheme used to distribute $X$ between processors may cause one method to have a lower communication cost than the other. In this section we analyze the cost of BCD and BDCD under two data partitioning schemes: 1D-block row (feature partitioning) and 1D-block column (data point partitioning). In both cases, we derive the associated computation, storage, and communication costs in order to compare the classical algorithms to our communication-avoiding variants. We describe the tradeoffs between the choice of data partitioning scheme and its effect on the communication cost of the BCD and BDCD algorithms. We assume that the matrix $X \in \mathbb{R}^{d \times n}$ is sparse with $fdn$ uniformly distributed non-zeros where $0 < f \leq 1$ is the density of $X$. We assume that the computed Gram matrices and residual and solution vectors are dense and that vectors in $\mathbb{R}^{n}$ are partitioned and vectors in $\mathbb{R}^{d}$ are replicated for 1D-block column. The reverse holds if $X$ is stored in a 1D-block row layout. Since $X$ is sparse the analysis of the computational cost includes passes over the sparse data structure instead of just the floating-point operations associated with the sparse matrix - sparse matrix multiplication (i.e. Gram matrix computation). Therefore, our analysis gives bounds on the local operations for each processor. We begin in Section \ref{sec:costbdcd} with the analysis of the BCD and BDCD algorithms and then analyze our new, communication-avoiding variants in Section \ref{sec:costcabcd}.

\subsection{Classical Algorithms}\label{sec:costbdcd}
We begin with the analysis of the BCD algorithm with $X$ stored in a 1D-block column layout and show how to extend this proof to BDCD with $X$ in a 1D-block row layout.
\begin{theorem}\label{thm:bcd1dcol}
$H$ iterations of the Block Coordinate Descent (BCD) algorithm with the matrix $X \in \mathbb{R}^{d \times n}$ stored in 1D-block column partitions with a block size $b$, on $P$ processors along the critical path costs
\begin{align*}
F = O\left(\frac{Hb^2fn}{P} + Hb^3\right)~ops,~M = O\left(\frac{fdn + n}{P} + b^2 + d\right)~words~of~memory.
\end{align*}
Communication costs
\begin{align*}
 W = O\left(Hb^2\log P\right)~words~moved,~L = O\left(H\log P\right)~messages.
\end{align*}
\end{theorem}
\begin{proof}
The BCD algorithm computes a $b \times b$ Gram matrix, $\Gamma_h$, solves a $b \times b$ linear system to obtain $\Delta w_h$, and updates the vectors $w_h$ and $z_h$. Computing the Gram matrix requires that each processor locally compute a $b \times b$ block of inner-products and then perform an all-reduce (a reduction and broadcast) to sum the partial blocks. Since the $b \times n$ sub-matrix $\mathbb{I}_h^TX$ has $bfn$ non-zeros, the parallel Gram matrix computation ($\mathbb{I}_h^TXX^T\mathbb{I}_h$) requires $O(\frac{b^2fn}{P})$ operations (there are $b^2$ elements of the Gram matrix each of which depend on $fn$ non-zeros) and communicates $O\left(b^2\log P\right)$ words, with $O\left(\log P\right)$ messages. In order to solve the subproblem redundantly on all processors, a local copy of the residual is required. Computing the residual requires $O\left(\frac{bfn}{P}\right)$ operations, and communicates $O\left(b \log P\right)$ words, in $O\left(\log P\right)$ messages. Once the residual is computed the subproblem can be solved redundantly on each processor in $O\left(b^3\right)$ flops. Finally, the vector updates to $w_h$ and $z_h$ can be computed without any communication in $O\left(b + \frac{bfn}{P}\right)$ flops on each processor. The critical path costs of $H$ iterations of this algorithm are $O\left(\frac{Hb^2fn}{P} + Hb^3\right)$ flops, $O\left(Hb^2\log P\right)$ words, and $O\left(H\log P\right)$ messages. Each processor requires enough memory to store $w_h$, $\Gamma_h$, $\Delta w$, $\mathbb{I}_h$ and $\frac{1}{P}$-th of $X, z_h$, and $y$. Therefore the memory cost of each processor is $d + b^2 + 2b + \frac{fdn + 2n}{P} = O\left(\frac{fdn + n}{P} + b^2 + d\right)$ words per processor.
\end{proof}
Note that if $\frac{fn}{P} > b$, then the Gram matrix computation cost dominates the cost of solving the subproblem. Furthermore, the (distributed) storage cost of $X$ dominates the cost of storing the $b \times b$ Gram matrix.
\begin{theorem}\label{thm:bdcd1drow}
$H'$ iterations of the Block Dual Coordinate Descent (BDCD) algorithm with the matrix $X \in \mathbb{R}^{d \times n}$ stored in 1D-block row partitions with a block size $b'$, on $P$ processors along the critical path costs
\begin{align*}
F = O\left(\frac{H'{b'}^2fd}{P} + H'{b'}^3\right)~ops,~M = O\left(\frac{fdn + d}{P} + {b'}^2 + n\right)~words~of~memory.
\end{align*}
Communication costs
\begin{align*}
W = O\left(H'{b'}^2 \log P\right)~words~moved,~L = O\left(H'\log P\right)~messages.
\end{align*}
\end{theorem}
\begin{proof}
The BDCD algorithm computes a $b' \times b'$ Gram matrix, $\Theta_h$, solves a $b' \times b'$ linear system to obtain $\Delta \alpha_h$, and updates the vectors $\alpha_h$ and $w_h$. Since $\Theta_h$ requires inner-products between columns of $X$, a 1D-block row partitioning scheme ensures that all processors contribute to each entry of $\Theta_h$. A similar cost analysis to the one used in Theorem \ref{thm:bcd1dcol} proves this theorem.
\end{proof}
Note that if $\frac{fd}{P} > b$, then the Gram matrix computation cost dominates the cost of solving the subproblem. Furthermore, the (distributed) storage cost of $X$ dominates the cost of storing the $b \times b$ Gram matrix.

If $X$ is stored in a 1D-block row layout, then each processor stores a disjoint subset of the features of $X$. Since BCD selects $b$ features at each iteration, 1D-block row partitioning could lead to load imbalance. In order to avoid load imbalance we re-partition the chosen $b$ features into 1D-block column layout and proceed by using the 1D-block column BCD algorithm. Re-partitioning the $b$ features requires communication, so we begin by bounding the maximum number of features assigned to a single processor\footnote{the bandwidth cost of re-partitioning is bounded by the processor with maximum load (i.e. maximum number of features).}. These bounds only holds with high probability since the features are chosen uniformly at random. To attain bounds on the bandwidth cost we assume that each sampled row of $X$ has $fn$ non-zeros.
\begin{lemma}\label{lem:lb}
Given a matrix $X \in \mathbb{R}^{d \times n}$ and $P$ processors such that each processor stores $\Theta\left(\left\lfloor{d \over P}\right\rfloor\right)$ features, if $b$ features are chosen uniformly at random, then the worst case maximum number of features, $\eta(b,P)$, assigned to a single processor w.h.p. is:
\begin{align*}
\eta(b,P) =
\begin{cases}
  O\left(\frac{b}{P} + \sqrt{\frac{b \log P}{P}}\right) & \text{if}~ b > P \log P,\\
  O\left(\frac{\log b}{\log \log b}\right) & \text{if}~ b = P,\\
  O\left(\frac{\log P}{\log \frac{P}{b}}\right) & \text{if}~ b < \frac{P}{\log P}.
 \end{cases}
 \end{align*}
\end{lemma}
\begin{proof}
This is the well-known generalization of the balls and bins problem introduced by Gonnet \cite{gonnet81} and later extended by Mitzenmacher \cite{mitzenmacher96} and Raab et. al. \cite{raab98}.
\end{proof}
Note that a similar result holds for the BDCD algorithm with $X$ stored in a 1D-block column layout.
\begin{theorem}\label{thm:bcd1drow}
$H$ iterations of the Block Coordinate Descent (BCD) algorithm with the matrix $X \in \mathbb{R}^{d \times n}$ stored in 1D-block row partitions with a block size $b$, on $P$ processors along the critical path costs
\begin{align*}
F = O\left(\frac{Hb^2fn}{P} + Hb^3\right)~ops,~M = O\left(\frac{fdn + n}{P} + b^2 + d\right)~words~of~memory.
\end{align*}
For small messages, communication costs w.h.p.
\begin{align*}
&W = O\left(\left(b^2 + {\eta(b,P)fn}\right)H\log P\right)~words~moved,~L = O\left(H\log P\right)~messages.
\end{align*}
For large messages, communication costs w.h.p.
\begin{align*}
 &W = O\left(Hb^2\log P + {H\eta(b,P)fn}\right)~words~moved,~L = O\left(HP\right)~messages.
\end{align*}
\end{theorem}
\begin{proof}
The 1D-block row partitioning scheme implies that the $b \times b$ Gram matrix, $\Gamma_h$, computation may be load imbalanced. Since we randomly select $b$ rows, some processors may hold multiple rows while others hold none. In order to balance the computational load we perform an all-to-all to convert the $b \times n$ sampled matrix into the 1D-block column layout. The amount of data moved is bounded by the max-loaded processor, which from Lemma \ref{lem:lb}, stores $O\left(\eta(b,P)\right)$ rows w.h.p. in the worst-case. This requires $W = O\left({\eta(b,P)fn}\log P\right)$ and $L = O\left(\log P\right)$ for small messages or $W = O\left({\eta(b,P)fn}\right)$ and $L = O\left(HP\right)$ for large messages. The all-to-all requires additional storage on each processor of $M = O\left(\frac{bfn}{P}\right)$ words. Once the sampled matrix is converted, the BCD algorithm proceeds as in Theorem \ref{thm:bcd1dcol}. By combining the cost of the all-to-all over $H$ iterations and the costs from Theorem \ref{thm:bcd1dcol}, we obtain the costs for the BCD algorithm with $X$ stored in a 1D-block row layout.
\end{proof}
Note that the additional storage for the all-to-all does not dominate since $b < d$ by definition.
\begin{theorem}
$H'$ iterations of the Block Dual Coordinate Descent (BDCD) algorithm with the matrix $X \in \mathbb{R}^{d \times n}$ stored in 1D-block column partitions with a block size $b'$, on $P$ processors along the critical path costs w.h.p.
\begin{align*}
F = O\left(\frac{H'{b'}^2fd}{P} + H'{b'}^3\right)~ops,~M = O\left(\frac{fdn + d}{P} + {b'}^2 + n\right)~words~of~memory.
\end{align*}
For small messages, communciation costs w.h.p.
\begin{align*}
W = O\left(\left({b'}^2 + {\eta(b',P)fd}\right)H'\log P\right)~words~moved,~L = O\left(H'\log P\right)~messages.
\end{align*}
For large messages, communication costs w.h.p.
\begin{align*}
W = O\left(H'{b'}^2\log P + {H'\eta(b',P)fd}\right)~words~moved,~L &= O\left(H'P\right)~messages.
\end{align*}
\end{theorem}
\begin{proof}
The BDCD algorithm computes a ${b'} \times {b'}$ Gram matrix, $\Theta_h$. A 1D-block column partitioning scheme implies that the Gram matrix computation will be load imbalanced and, therefore, requires an all-to-all to convert the sampled matrix into a 1D-block row layout. A similar cost analysis to the one used in Theorem \ref{thm:bcd1drow} proves this theorem.
\end{proof}

\subsection{Communication-Avoiding Algorithms}\label{sec:costcabcd}
In this section, we derive the computation, storage, and communication costs of our communication-avoiding BCD and BDCD algorithm under the 1D-block row and 1D-block column data layouts. In both cases we show that our algorithm reduces the latency  costs by a factor of $s$ over the classical algorithms. We begin with the CA-BCD algorithm in 1D-block column layout and, then show how this proof extends to CA-BDCD in 1D-block row layout.

\begin{theorem}\label{thm:cabcd1dcol}
$H$ iterations of the Communication-Avoiding Block Coordinate Descent (CA-BCD) algorithm with the matrix $X \in \mathbb{R}^{d \times n}$ stored in 1D-block column partitions with a block size $b$, on $P$ processors along the critical path costs
\begin{align*}
F = O\left(\frac{Hb^2sfn}{P} + Hb^3\right)~ops,~M = O\left(\frac{fdn + n}{P} + b^2s^2 + d\right)~words~of~memory.
\end{align*}
Communication costs
\begin{align*}
  W = O\left(Hb^2s \log P\right)~words~moved,~L = O\left(\frac{H}{s}\log P\right)~messages.
\end{align*}
\end{theorem}
\begin{proof}
The CA-BCD algorithm computes the $sb \times sb$ Gram matrix, $G = \frac{1}{n}YY^T + \lambda I$, where $Y = \begin{bmatrix}\mathbb{I}_{sk + 1}, \mathbb{I}_{sk + 2}, \cdots, \mathbb{I}_{sk + s}\end{bmatrix}^TX$, solves $s$ ($b \times b$) linear systems to compute $\Delta w_{sk + j}$ and updates the vectors $w_{sk + s}$ and $z_{sk + s}$. Computing the Gram matrix requires that each processor locally compute a $sb \times sb$ block of inner-products and then perform an all-reduce (a reduction and broadcast) to sum the partial blocks. This operation requires $O\left(\frac{b^2s^2fn}{P}\right)$ operations (there are $s^2b^2$ elements of the Gram matrix each of which depends on $fn$ non-zeros), communicates $O\left(s^2b^2 \log P\right)$ words, and requires $O\left(\log P\right)$ messages. In order to solve the subproblem redundantly on all processors, a local copy of the residual is required. Computing the residual requires $O\left(\frac{bsfn}{P}\right)$ flops, and communicates $O\left(sb \log P\right)$ words, in $O\left(\log P\right)$ messages. Once the residual is computed the subproblem can be solved redundantly on each processor in $O\left(b^3s + b^2s^2\right)$ flops. Finally, the vector updates to $w_{sk + s}$ and $z_{sk + s}$ can be computed without any communication in $O\left(bs + \frac{bsfn}{P}\right)$ flops on each processor. Since the critical path occurs every $\frac{H}{s}$ iterations (every outer iteration), the algorithm costs $O\left(\frac{Hb^2sfn}{P} + Hb^3\right)$ flops, $O\left(Hb^2s \log P\right)$ words, and $O\left(\frac{H}{s}\log P\right)$ messages. Each processor requires enough memory to store $w_{sk + j}$, $G$, $\Delta w_{sk + j}$, $\mathbb{I}_{sk + j}$ and $\frac{1}{P}$-th of $X, z_{sk + j}$, and $y$. Therefore the memory cost of each processor is $d + s^2b^2 + 2sb + \frac{fdn + 2n}{P} = O\left(\frac{fdn + n}{P} + b^2s^2 + d\right)$ words per processor.
\end{proof}

\begin{theorem}\label{thm:cabdcd1drow}
$H'$ iterations of the Communication-Avoiding Block Dual Coordinate Descent (CA-BDCD) algorithm with the matrix $X \in \mathbb{R}^{d \times n}$ stored in 1D-block row partitions with a block size $b'$, on $P$ processors along the critical path costs
\begin{align*}
F = O\left(\frac{H'{b'}^2sfd}{P} + H'{b'}^3\right)~ops,~M &= O\left(\frac{fdn + d}{P} + {b'}^2s^2 + n\right)~words~of~mem.
\end{align*}
Communication costs
\begin{align*}
W = O\left(H'{b'}^2s\log P\right) words~moved,~L = O\left(\frac{H'}{s}\log P\right)~messages.
\end{align*}
\end{theorem}
\begin{proof}
The CA-BDCD algorithm computes the $s{b'} \times s{b'}$ Gram matrix, $G' = \frac{1}{\lambda n^2}Y^TY + \frac{1}{n}I$, where $Y = X\begin{bmatrix}\mathbb{I}_{sk + 1}, \mathbb{I}_{sk + 2}, \cdots, \mathbb{I}_{sk + s}\end{bmatrix}$. The 1D-block column partitioning layout ensures that each processor computes a partial $s{b'} \times s{b'}$ block of the Gram matrix. A similar cost analysis to Theorem \ref{thm:cabcd1dcol} proves this theorem.
\end{proof}

\begin{theorem}\label{thm:cabcd1drow}
$H$ iterations of the Communication-Avoiding Block Coordinate Descent (CA-BCD) algorithm with the matrix $X \in \mathbb{R}^{d \times n}$ stored in 1D-block row partitions with a block size $b$, on $P$ processors along the critical path costs
\begin{align*}
F = O\left(\frac{Hb^2sfn}{P} + Hb^3\right)~ops,~M = O\left(\frac{(d + bs)fn + n}{P} +  b^2s^2 + d\right) words.
\end{align*}
For small messages, communication costs w.h.p.
\begin{align*}
W = O\left(\left(b^2s + {\eta(sb,P)fn}\right)H\log P\right)~words~moved,L = O\left(\frac{H}{s}\log P\right)~messages.
\end{align*}
For large messages, communication costs w.h.p.
\begin{align*}
W = O\left(Hb^2s\log P + {H\eta(sb,P)fn}\right)~words~moved,~L = O\left(\frac{H}{s}P\right)~messages.
\end{align*}
\end{theorem}
\begin{proof}
The 1D-block row partitioning scheme implies that the $sb \times sb$ Gram matrix computation may be load imbalanced. Since we randomly select $sb$ rows, some processors may hold multiple chosen rows while some hold none. In order to balance the computational load we perform an all-to-all to convert the $sb \times n$ sampled matrix into the 1D-block column layout. The amount of data moved is bounded by the max-loaded processor, which from Lemma \ref{lem:lb}, stores $O\left(\eta(sb,P)\right)$ rows w.h.p. in the worst-case. This requires $W = O\left({\eta(sb,P)fn}\log P\right)$ and $L = O\left(\log P\right)$ for small messages or $W = O\left(\eta(sb,P){fn}\right)$ and $L = O\left(HP\right)$ for large messages. The all-to-all requires additional storage on each processor of $M = O\left(\frac{bsfn}{P}\right)$ words. Once the sampled matrix is converted, the BCD algorithm proceeds as in Theorem \ref{thm:cabcd1dcol}. By combining the cost of the all-to-all over $H$ iterations and the costs from Theorem \ref{thm:cabcd1dcol}, we obtain the costs for the CA-BCD algorithm with $X$ stored in a 1D-block row layout.
\end{proof}
Note that the additional storage for the all-to-all may dominate if $d < bs$. Therefore, $b$ and $s$ must be chosen carefully.

\begin{theorem}
$H$ iterations of the Communication-Avoiding Block Dual Coordinate Descent (CA-BDCD) algorithm with the matrix $X \in \mathbb{R}^{d \times n}$ stored in 1D-block column partitions with a block size $b'$, on $P$ processors along the critical path costs
\begin{align*}
F = O\left(\frac{H'{b'}^2sfd}{P} + H'{b'}^3\right)~ops, M = O\left(\frac{(n + {b'}s)fd + d}{P} +  {b'}^2s^2 + n\right)~words  .
\end{align*}
For small messages, communication costs w.h.p.
\begin{align*}
&W = O\left(\left({b'}^2s+ {\eta(s{b'},P)fd}\right)H'\log P\right)~words~moved, L = O\left(\frac{H'}{s}\log P\right)~msgs.
\end{align*}
For large messages, communication costs w.h.p.
\begin{align*}
W = O\left(H'{b'}^2s\log P + {H'\eta(s{b'},P)fd}\right)~words~moved,~L = O\left(\frac{H'}{s}P\right)~messages.
\end{align*}
\end{theorem}
\begin{proof}
The CA-BDCD algorithm computes a $s{b'} \times s{b'}$ Gram matrix, $G$. A 1D-block column partitioning scheme implies that the Gram matrix computation will be load imbalanced and, therefore, requires an all-to-all to convert the sampled matrix into a 1D-block row layout. A similar cost analysis to the one used in Theorem \ref{thm:cabcd1drow} proves this theorem.
\end{proof}
The communication-avoiding variants that we have derived require a factor of $s$ fewer messages than their classical counterparts, at the cost of more computation, bandwidth and memory. This suggests that $s$ must be chosen carefully to balance the additional computation, bandwidth and memory usage with the reduction in the latency cost. This suggests that if latency is the dominant cost then our communication-avoiding variants can attain a $s$-fold speedup.

\section{Experimental Evaluation}\label{sec:eval}
%
We proved in Section \ref{sec:anal} that the CA-BCD and CA-BDCD algorithms reduce latency (the dominant cost) at the expense of additional bandwidth and computation. The recurrence unrolling we propose may also affect the numerical stability of CA-BCD and CA-BDCD since the sequence of computations and vector updates are different. In Section \ref{sec:numexp} we experimentally show that the communication-avoiding variants are numerically stable (in contrast to some CA-Krylov methods \cite{Carson15, carson14,carson15b, carson13, carson14b, Hoemmen10}) and, in Section \ref{sec:perfexp}, we show that the communication-avoiding variants can lead to large speedups on a Cray XC30 supercomputer using MPI.
\begin{table}
\begin{center}
\footnotesize
\begin{tabular}{l|c|c|c|c|c|l}
\hline
\multicolumn{6}{c}{Summary of datasets}\\
\hline\hline
\multicolumn{1}{c}{Name} &  \multicolumn{1}{|c|}{Features ($d$)} &Data Points ($n$) & \multicolumn{1}{|c|}{NNZ$\%$} & $\sigma_{min}$ &$\sigma_{max}$& Source\\
\hline
news20 & $62,061$ &$15,935$ & $0.13$ & $1.7e{-6}$ & $6.0e{+5}$ & LIBSVM \cite{lang95}\\ \hline
a9a & $123$ & $32,561$ & $11$ & $4.9e{-6}$ & $2.0e{+5}$ & UCI \cite{lichman13}\\ \hline
real-sim & $20,958$ & $72,309$ & $0.24$ & $1.1e{-3}$ & $9.2e{+2}$& LIBSVM \cite{mccallum}\\ \hline
\end{tabular}
\end{center}
\caption{Properties of the LIBSVM datasets used in our experiments. We report the largest and smallest singular values (same as the eigenvalues) of $X^TX$.} 
\label{tbl:dsets}
\end{table}
\subsection{Numerical Experiments}\label{sec:numexp}
The algorithm transformations derived in Section \ref{sec:deriv} require that the CA-BCD and CA-BDCD operate on Gram matrices of size $sb \times sb$ instead of size $b \times b$ every outer iteration. Due to the larger dimensions, the condition number of the Gram matrix increases and may have an adverse affect on the convergence behavior. We explore this tradeoff between convergence behavior, flops, communication and the choices of $b$ and $s$ for the standard and communication-avoiding algorithms. All numerical stability experiments were performed in MATLAB version R2016b on a 2.3 GHz Intel i7 machine with 8GB of RAM with datasets obtained from the LIBSVM repository \cite{cc01}. Datasets were chosen so that all algorithms were tested on a range of shapes, sizes, and condition numbers. Table \ref{tbl:dsets} summarizes the important properties of the datasets tested. For all experiments, we set the regularization parameter to $\lambda = 1000\sigma_{min}$. The regularization parameter reduces the condition numbers of the datasets and allows the BCD and BDCD algorithms to converge faster. In practice, $\lambda$ should be chosen based on metrics like prediction accuracy on the test data (or hold-out data). Smaller values of $\lambda$ would slow the convergence rate and require more iterations, therefore we choose $\lambda$ so that our experiments have reasonable running times. We do not explore tradeoffs among $\lambda$ values, convergence rate and running times in this paper. In order to measure convergence behavior, we plot the relative solution error, $\frac{\|w_{opt} - w_h\|_2}{\|w_{opt}\|_2}$, where $w_h$ is the solution obtained from the coordinate descent algorithms at iteration $h$ and $w_{opt}$ is obtained from conjugate gradients with $tol = 1e{-15}$. We also plot the relative objective error, $\frac{f(X, w_{opt}, y) - f(X, w_h, y)}{f(X, w_{opt}, y)}$, where $f(X, w, y) = \frac{1}{2n}\|X^Tw - y\|_2^2 + \frac{\lambda}{2}\|w\|_2^2$, the primal objective. We use the primal objective to show convergence behavior for BCD, BDCD and their communication-avoiding variants. We explore the tradeoff between the block sizes, $b$ and $b'$, and convergence behavior to test BCD and BDCD stability due to the choice of block sizes. Then, we fix the block sizes and explore the tradeoff between $s$, the recurrence unrolling parameter, and convergence behavior to study the stability of the communication-avoiding variants. Finally, for both sets of experiments we also plot the algorithm costs against convergence behavior to illustrate the theoretical performance tradeoffs due to choice of block sizes and choice of $s$. For the latter experiments we assume that the datasets are partitioned in 1D-block column for BCD and 1D-block row for BDCD. We plot the sequential flops cost for all algorithms, ignore the $\log P$ factor for the number of messages and ignore constants. We obtain the Gram matrix computation cost from the SuiteSparse \cite{davis16} routine \verb+ssmultsym+\footnote{Symbolically executes the sparse matrix - sparse matrix multiplication and reports an estimate of the flops cost (counting multiplications and additions).}.

\begin{figure}[t!]
\begin{subfigure}{.329\textwidth}
\centering
\includegraphics[trim = .5in 2.5in .7in 2.5in,  clip,width=\textwidth]{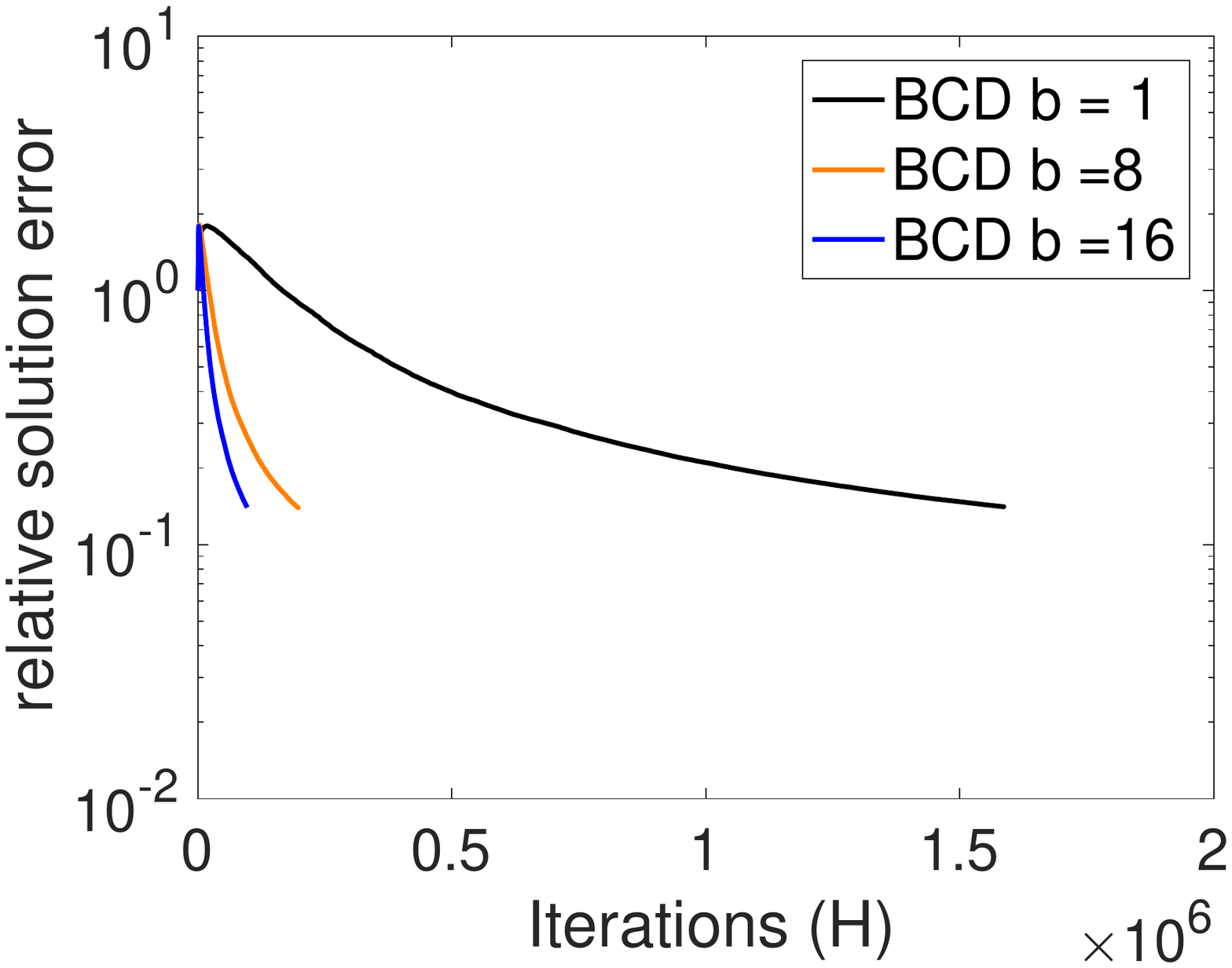}
\caption{news20}
\label{fig:news202}
\end{subfigure}
\begin{subfigure}{.329\textwidth}
\centering
\includegraphics[trim = .5in 2.5in .7in 2.5in,  clip,width=\textwidth]{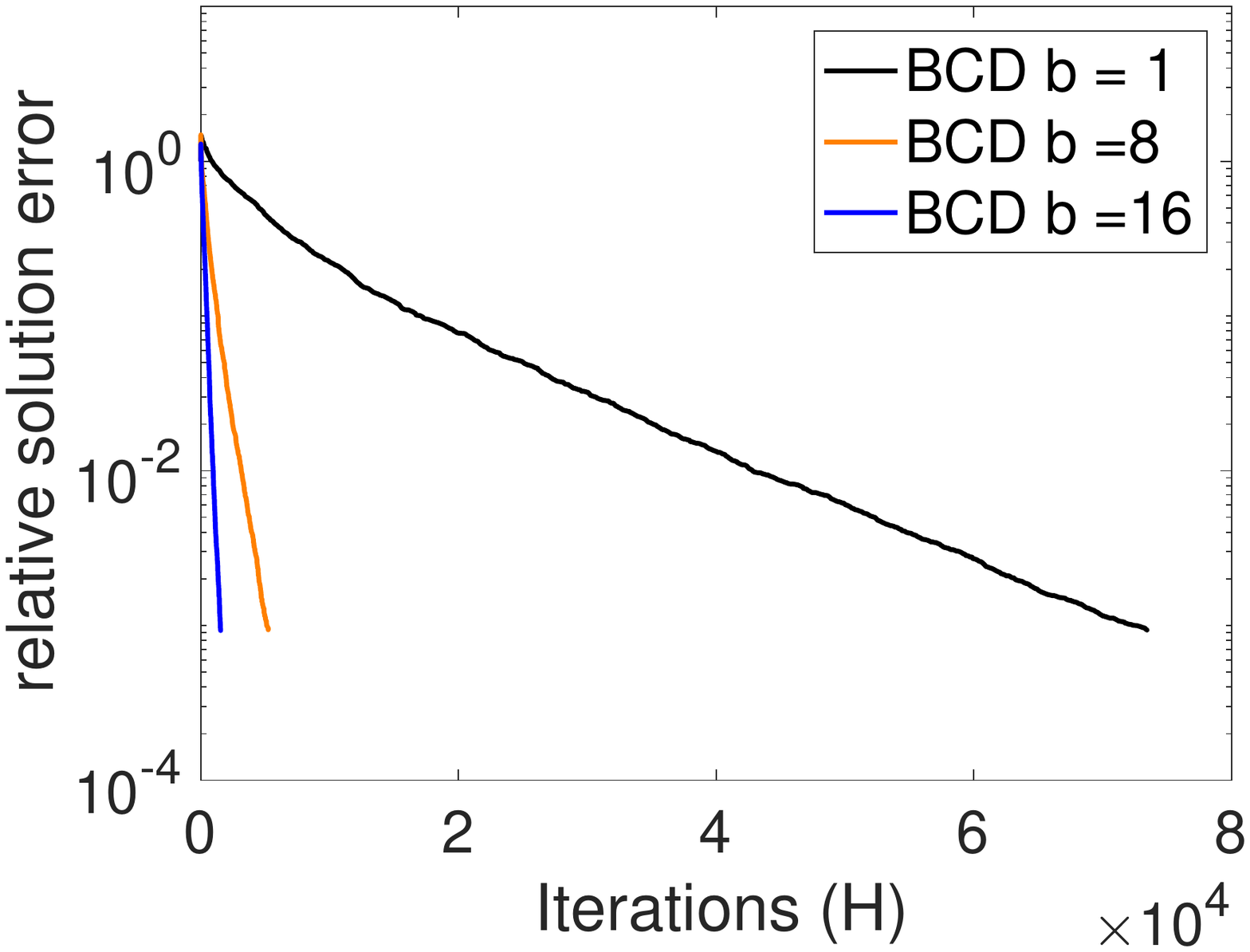}
\caption{a9a}
\label{fig:a9a2}
\end{subfigure}
\begin{subfigure}{.329\textwidth}
\centering
\includegraphics[trim = .5in 2.5in .7in 2.5in,  clip,width=\textwidth]{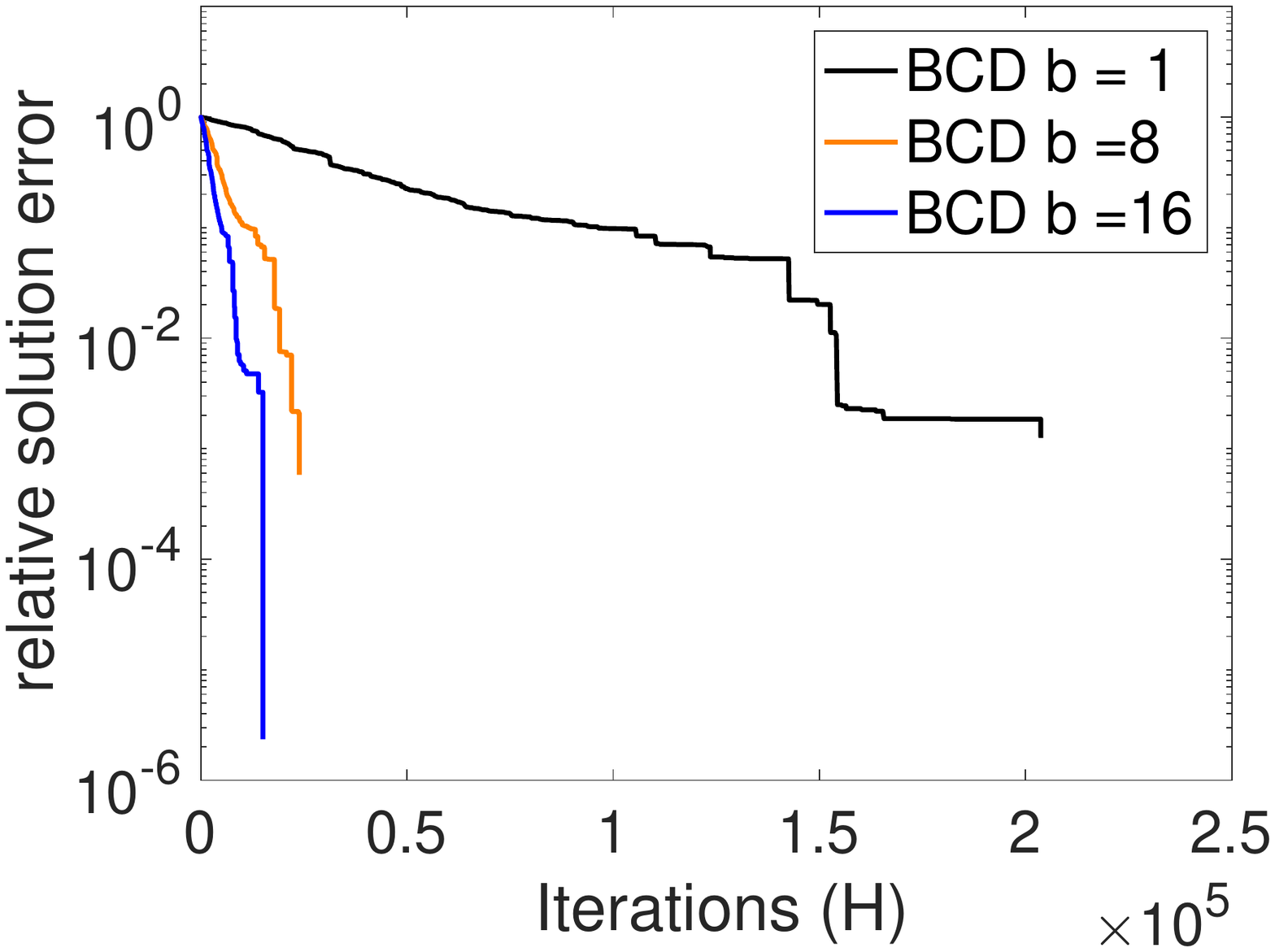}
\caption{real-sim}
\label{fig:realsim2}
\end{subfigure}

\begin{subfigure}{.329\textwidth}
\centering
\includegraphics[trim = .5in 2.5in  .7in 2.5in,  clip,width=\textwidth]{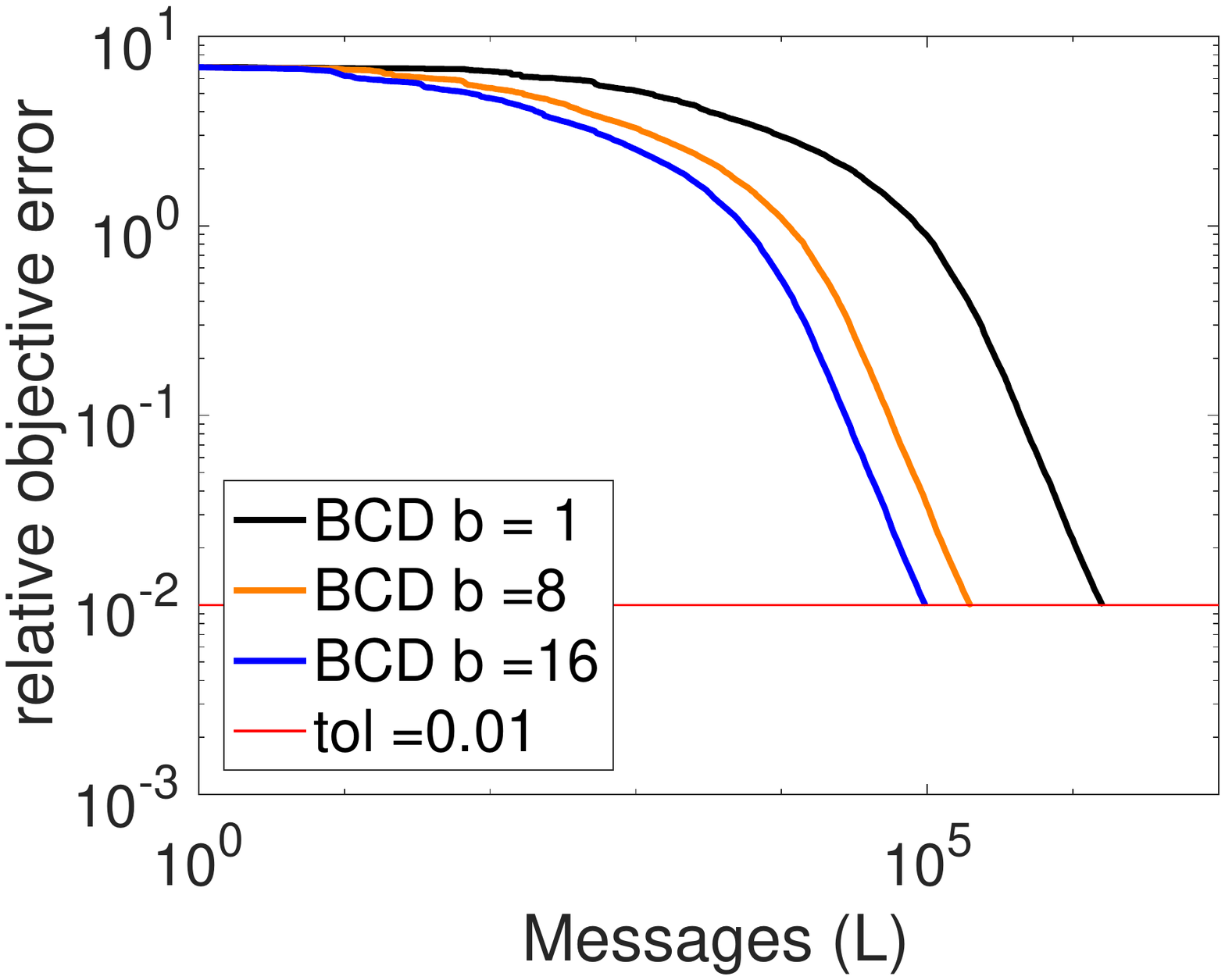}
\caption{news20}
\label{fig:news205}
\end{subfigure}
\begin{subfigure}{.329\textwidth}
\centering
\includegraphics[trim = .5in 2.5in  .7in 2.5in,  clip,width=\textwidth]{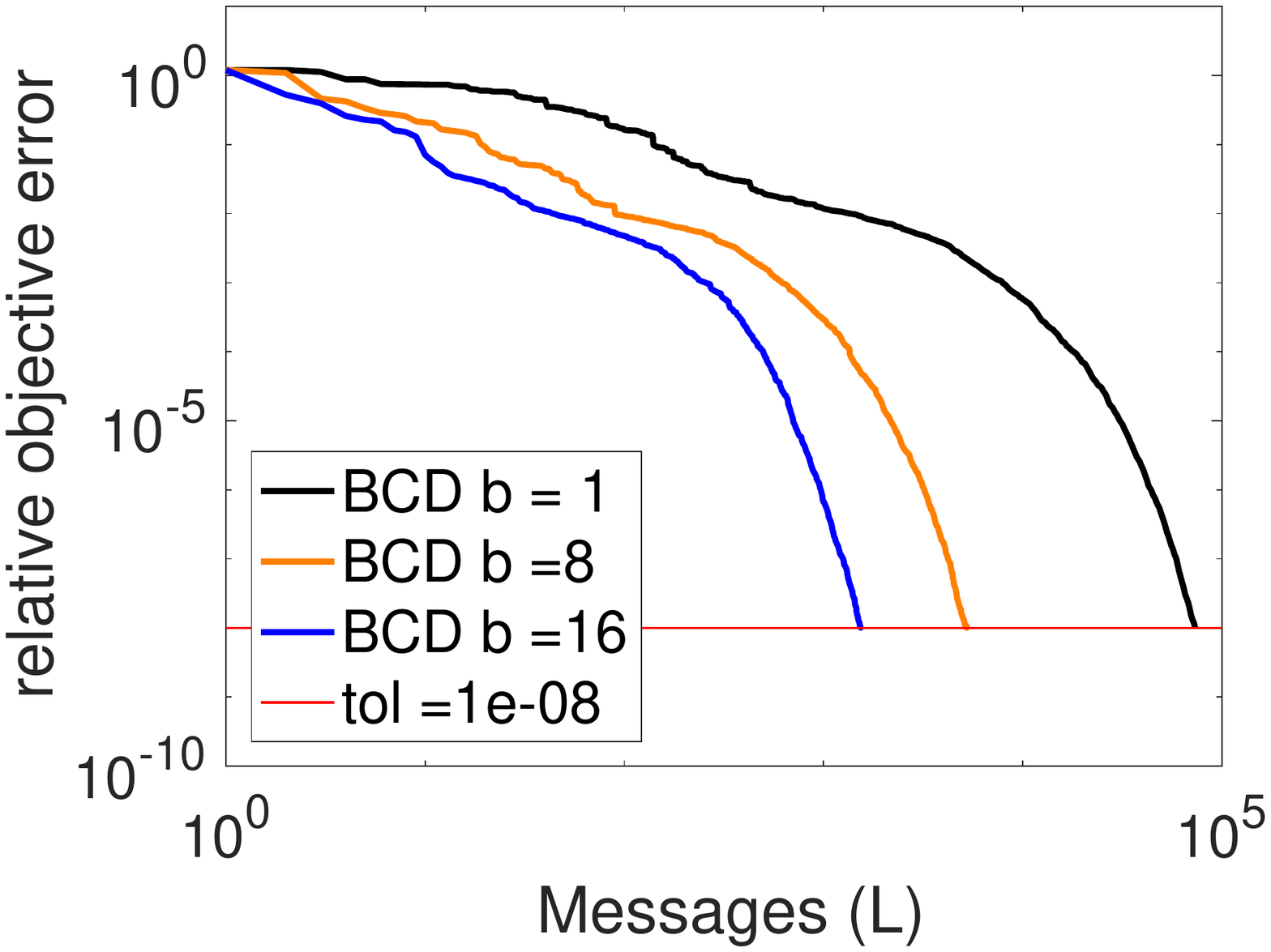}
\caption{a9a}
\label{fig:a9a5}
\end{subfigure}
\begin{subfigure}{.329\textwidth}
\centering
\includegraphics[trim = .5in 2.5in .7in 2.5in,  clip,width=\textwidth]{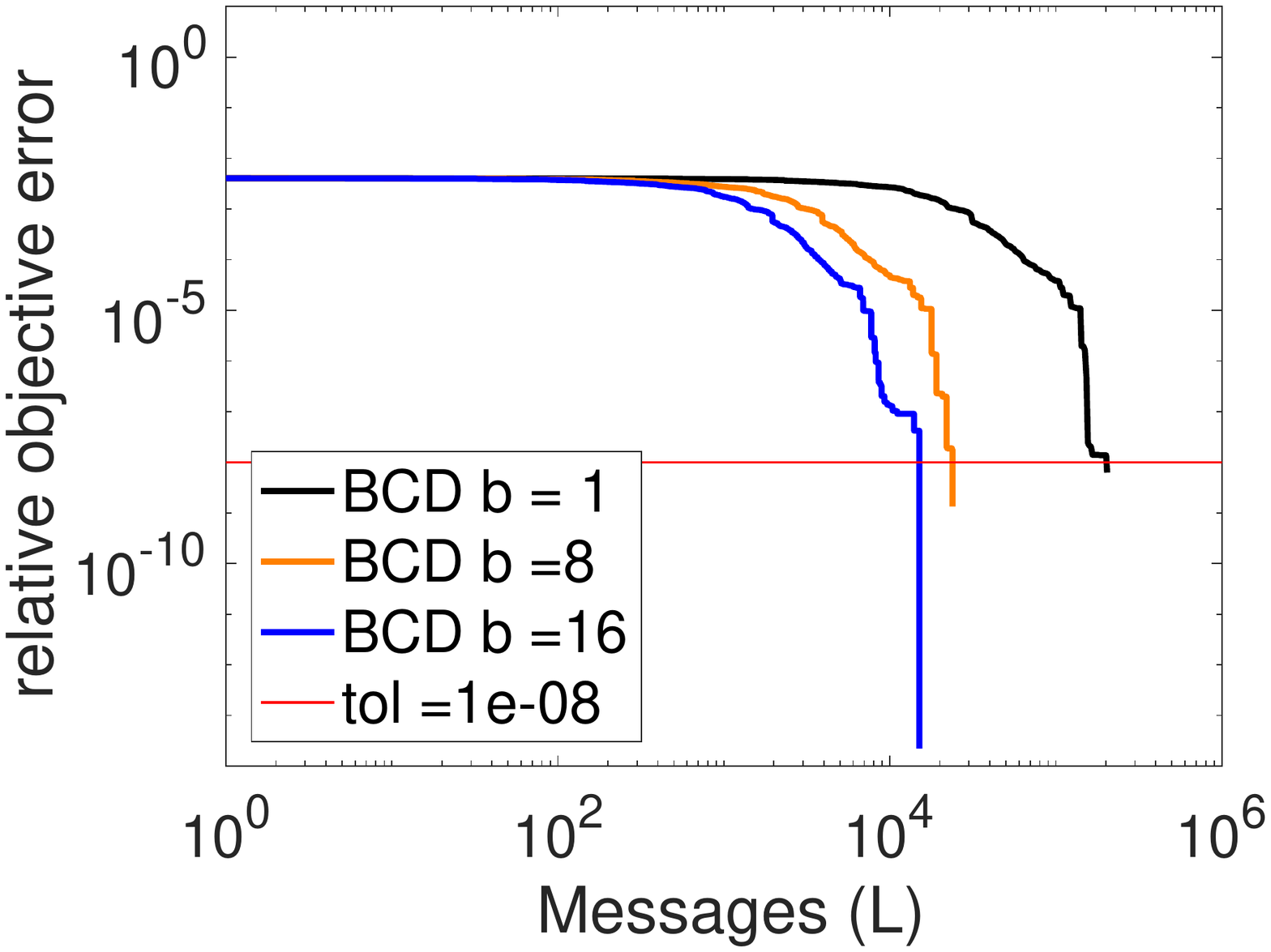}
\caption{real-sim}
\label{fig:realsim5}
\end{subfigure}
\caption{We compare the convergence behavior of BCD for several block sizes, $b$, such that $1 \leq b < d$ on several machine learning datasets. We show relative solution error (top row, Figs. \ref{fig:news202}-\ref{fig:realsim2}) and objective error (bottom row, Fig. \ref{fig:news205}-\ref{fig:realsim5}) convergence plots with $\lambda = 1000\sigma_{min}$. We fix the objective error tolerance for news20 to $1e{-2}$ and $1e{-8}$ for a9a and real-sim. The x-axis for Figures \ref{fig:news205}-\ref{fig:realsim5} show the number of messages required on a $\log_{10}$ scale. Since BCD communicates at every iteration, the x-axis is also equivalent to the number of iterations (modulo $\log_{10}$ scale).}
\label{fig:bcdconv}
\end{figure}

\begin{figure}[t]
%
\begin{subfigure}{.329\textwidth}
\centering
\includegraphics[trim = .5in 2.5in  .7in 2.5in,  clip,width=\textwidth, ]{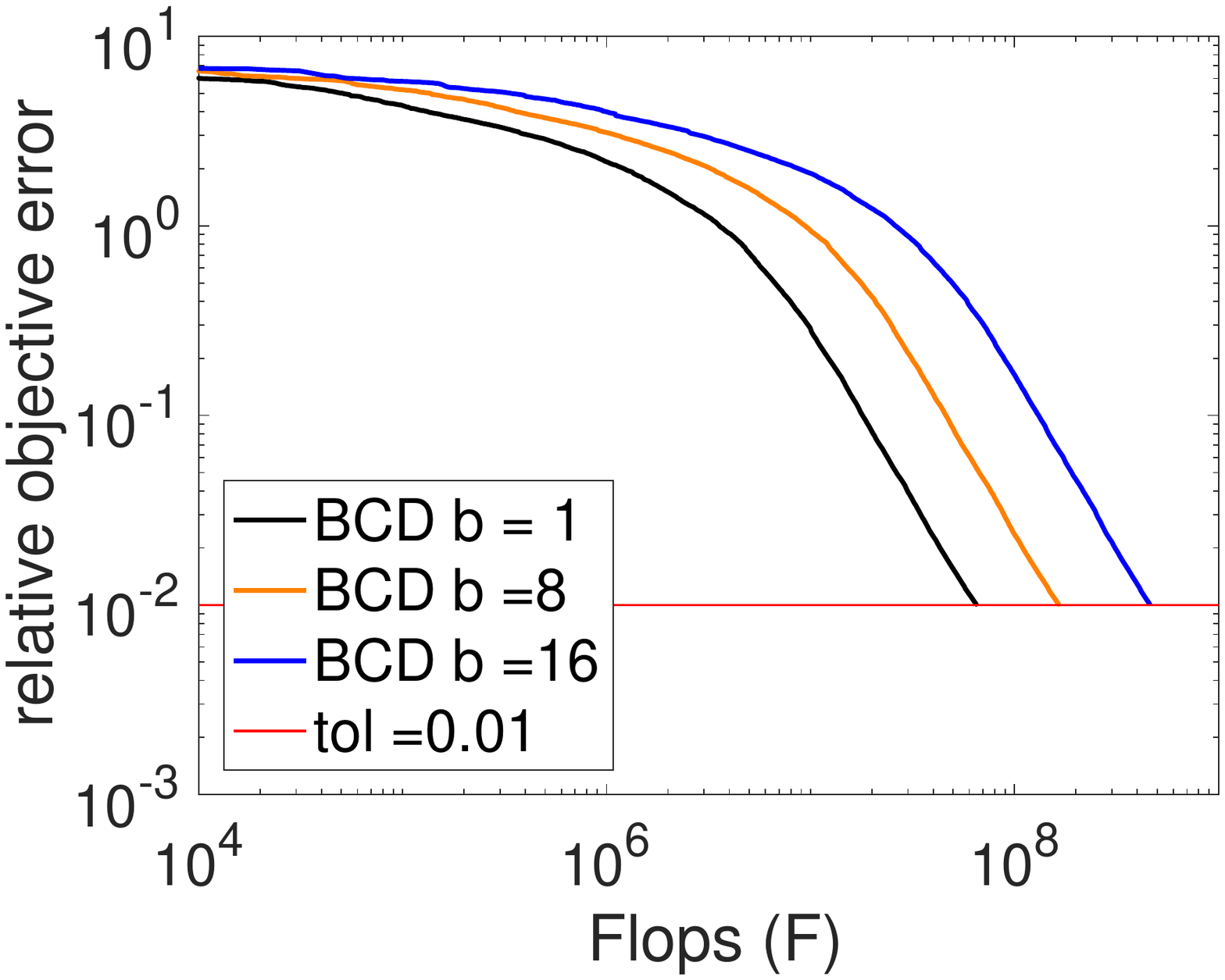}
\caption{news20}
\label{fig:news203}
\end{subfigure}
\begin{subfigure}{.329\textwidth}
\centering
\includegraphics[trim = .5in 2.5in  .7in 2.5in,  clip,width=\textwidth]{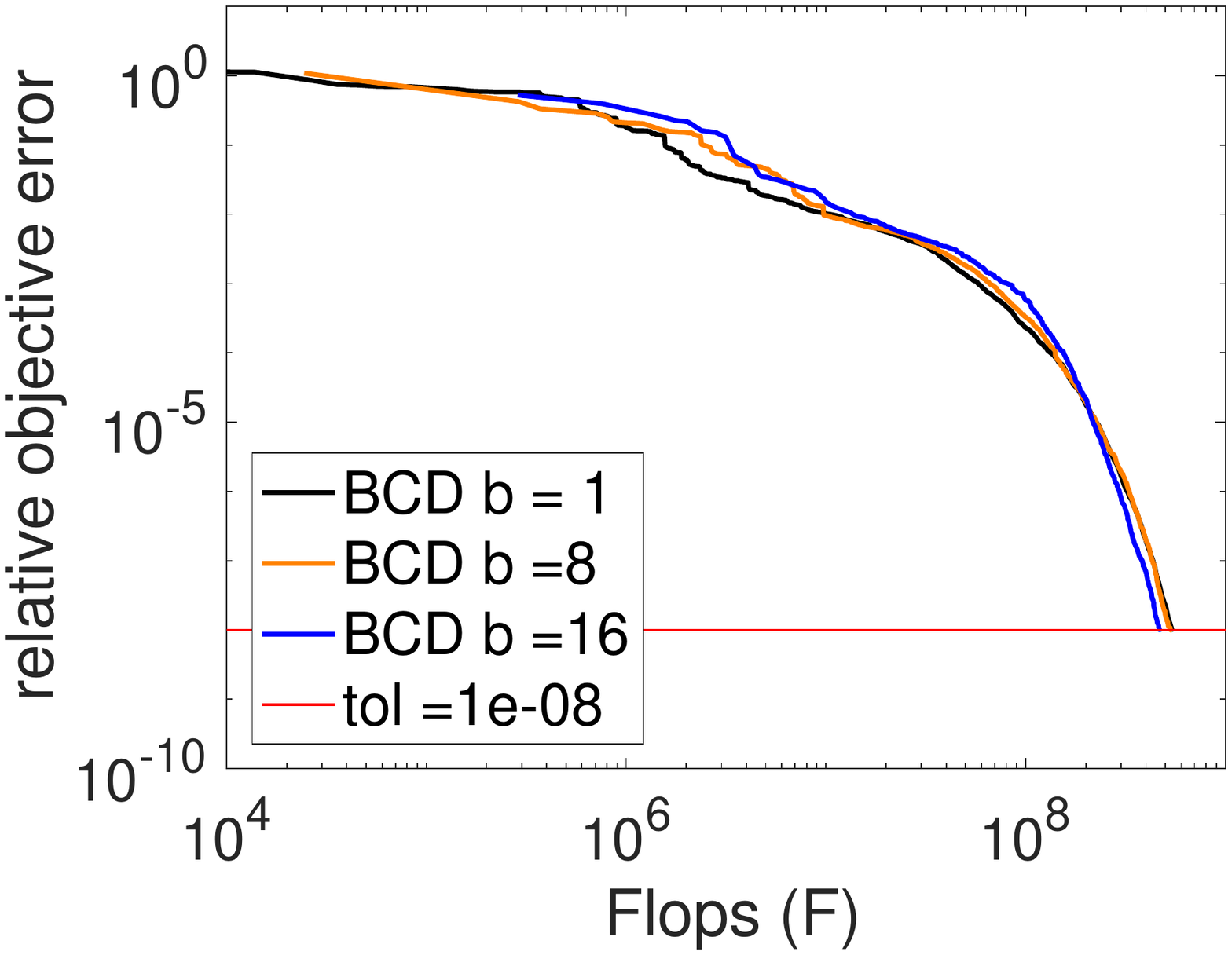}
\caption{a9a}
\label{fig:a9a3}
\end{subfigure}
\begin{subfigure}{.329\textwidth}
\centering
\includegraphics[trim = .5in 2.5in  .7in 2.5in,  clip,width=\textwidth]{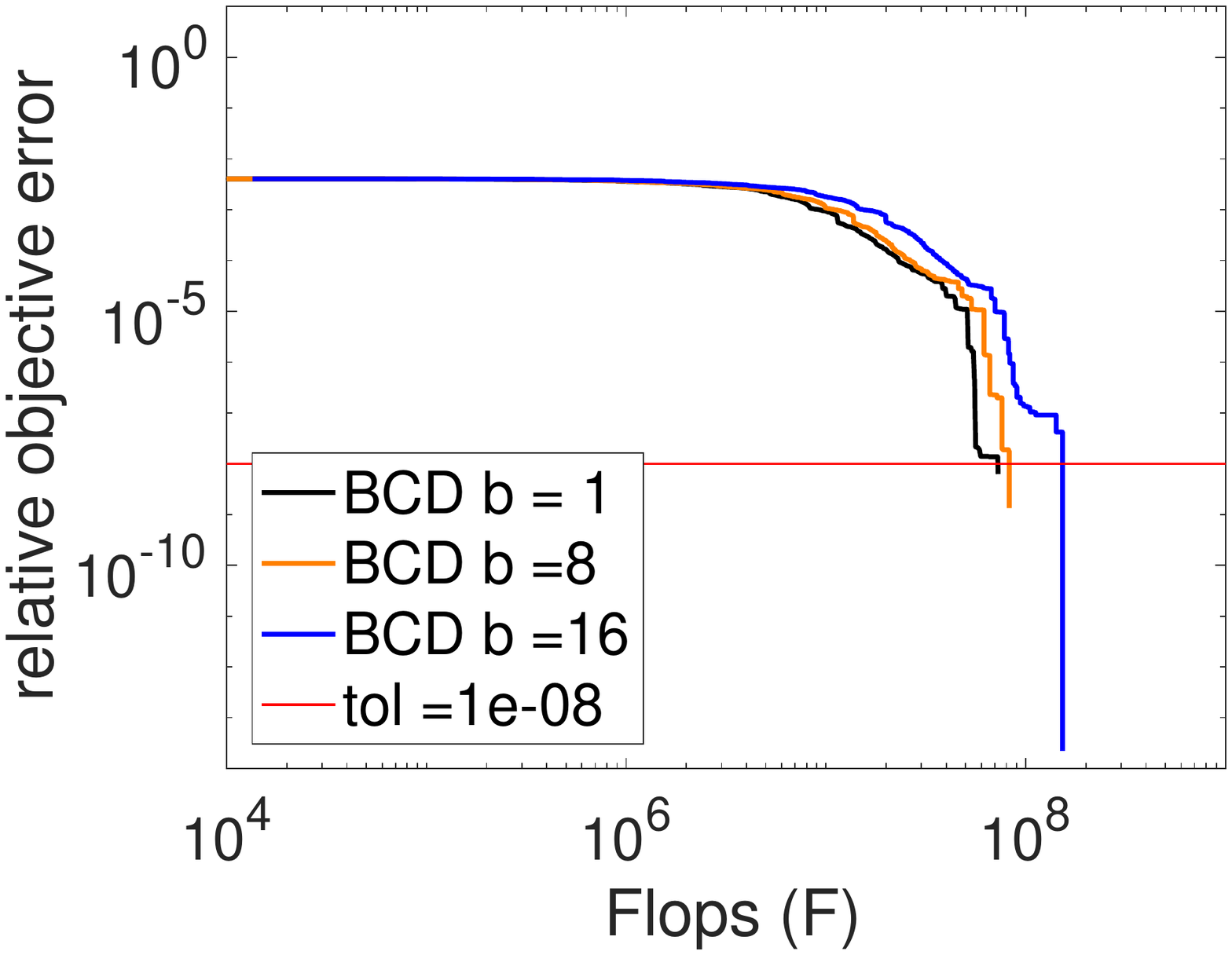}
\caption{real-sim}
\label{fig:realsim3}
\end{subfigure}

\begin{subfigure}{.329\textwidth}
\centering
\includegraphics[trim = .5in 2.5in  .7in 2.5in,  clip,width=\textwidth]{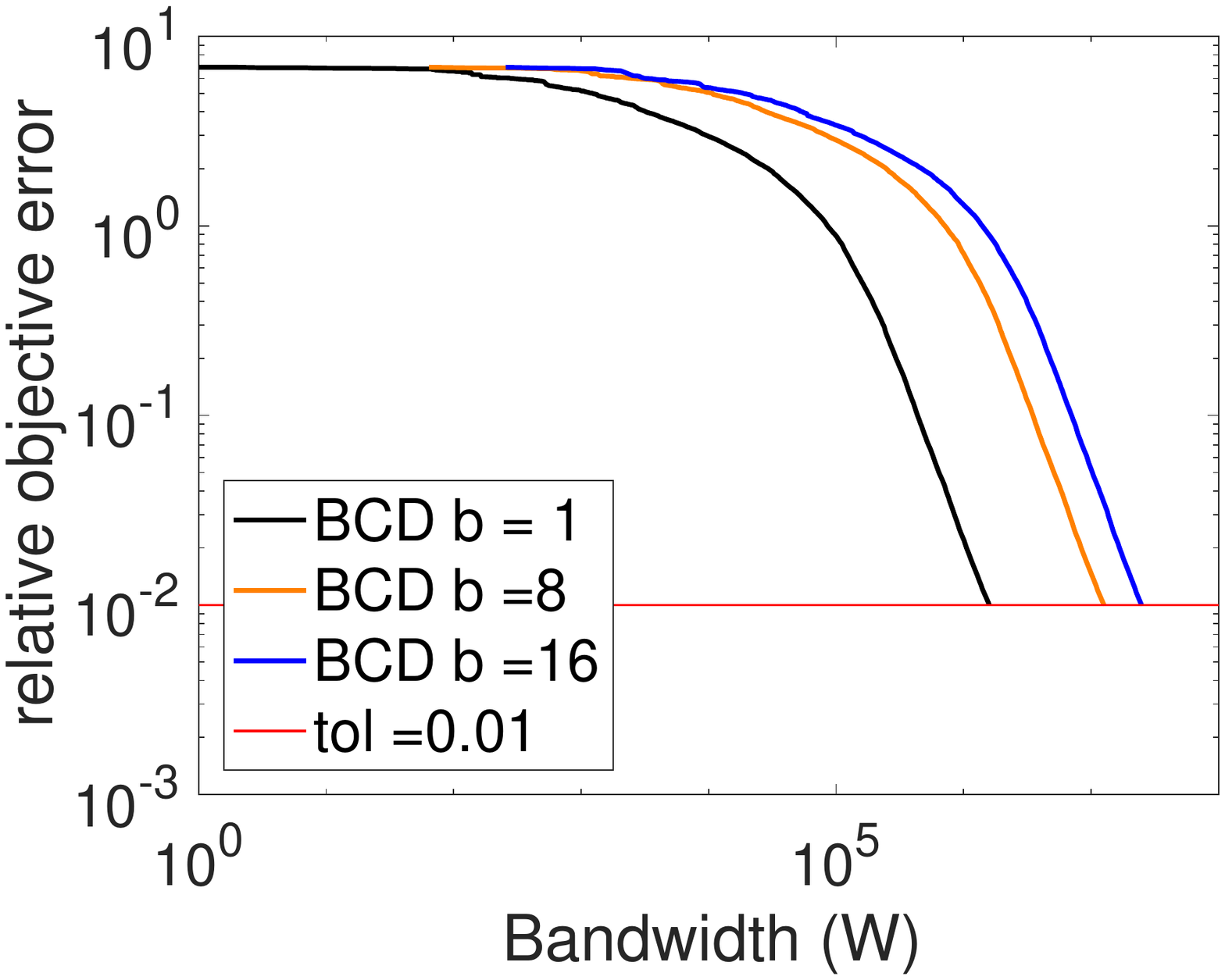}
\caption{news20}
\label{fig:news204}
\end{subfigure}
\begin{subfigure}{.329\textwidth}
\centering
\includegraphics[trim = .5in 2.5in  .7in 2.5in,  clip,width=\textwidth]{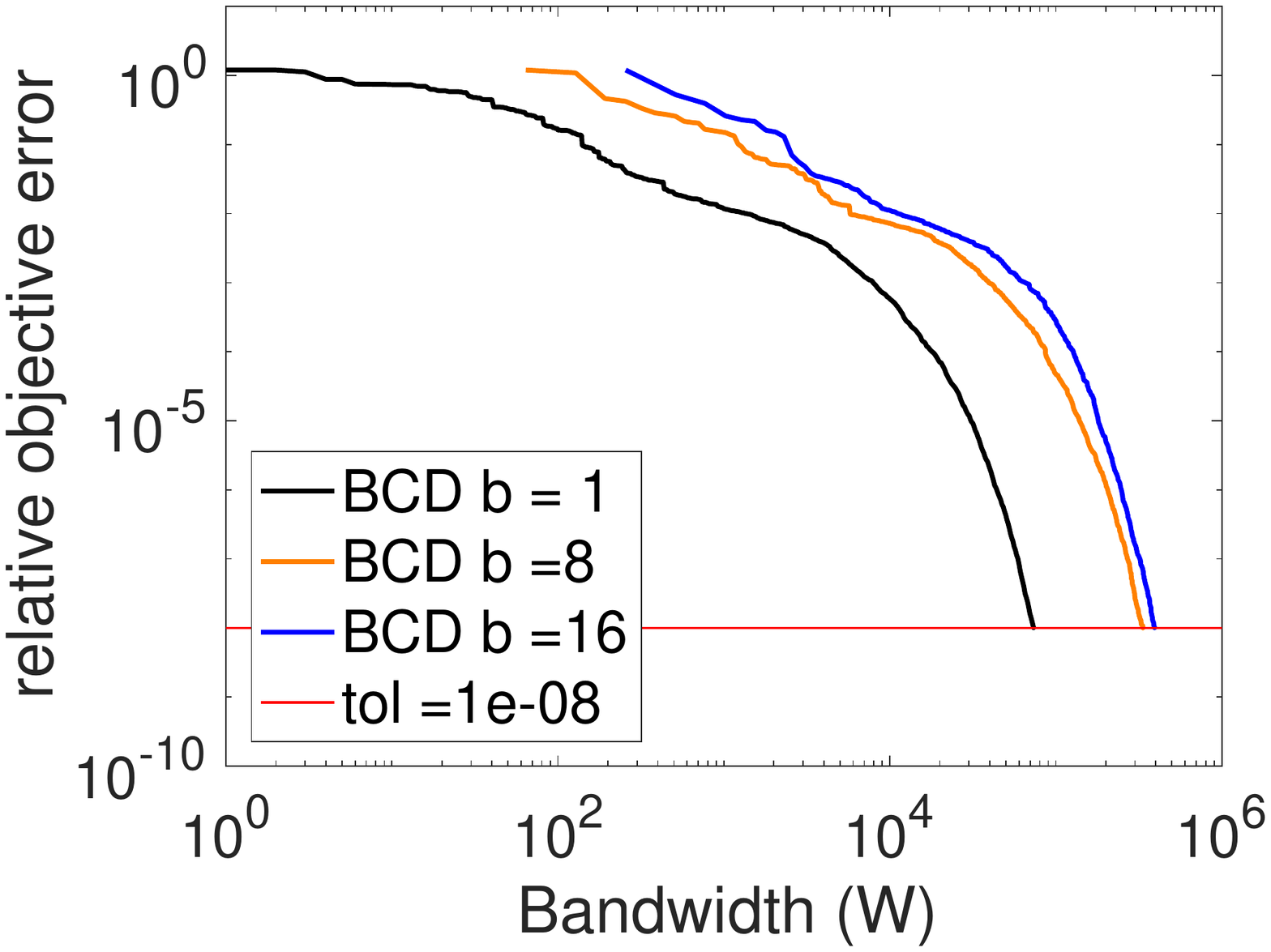}
\caption{a9a}
\label{fig:a9a4}
\end{subfigure}
\begin{subfigure}{.329\textwidth}
\centering
\includegraphics[trim = .5in 2.5in  .7in 2.5in,  clip,width=\textwidth]{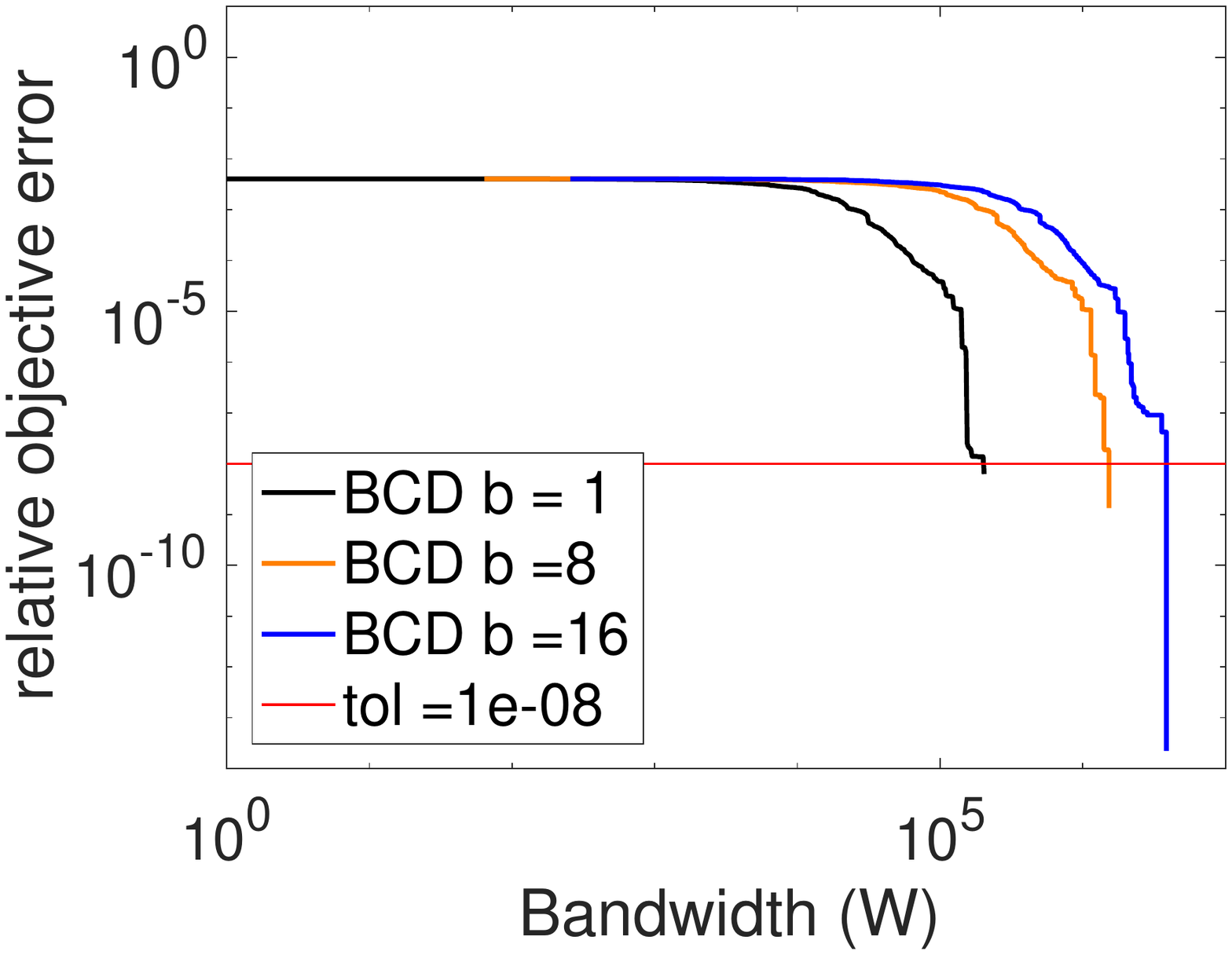}
\caption{real-sim}
\label{fig:realsim4}
\end{subfigure}

\caption{We compare the convergence behavior of BCD for several block sizes, $b$, such that $1 \leq b < d$ on several machine learning datasets. Flops cost (top row, Figs. \ref{fig:news203}-\ref{fig:realsim3}) and bandwidth cost (middle row, Figs. \ref{fig:news204}-\ref{fig:realsim4}) versus convergence with $\lambda = 1000\sigma_{min}$.}\label{fig:bcdcost}
\end{figure}
\subsubsection{Block Coordinate Descent}\label{bcdeval}
\begin{figure}[t!]
%
\begin{subfigure}{.329\textwidth}
\centering
\includegraphics[trim = .5in 2.5in .6in 2.5in,  clip,width=\textwidth, ]{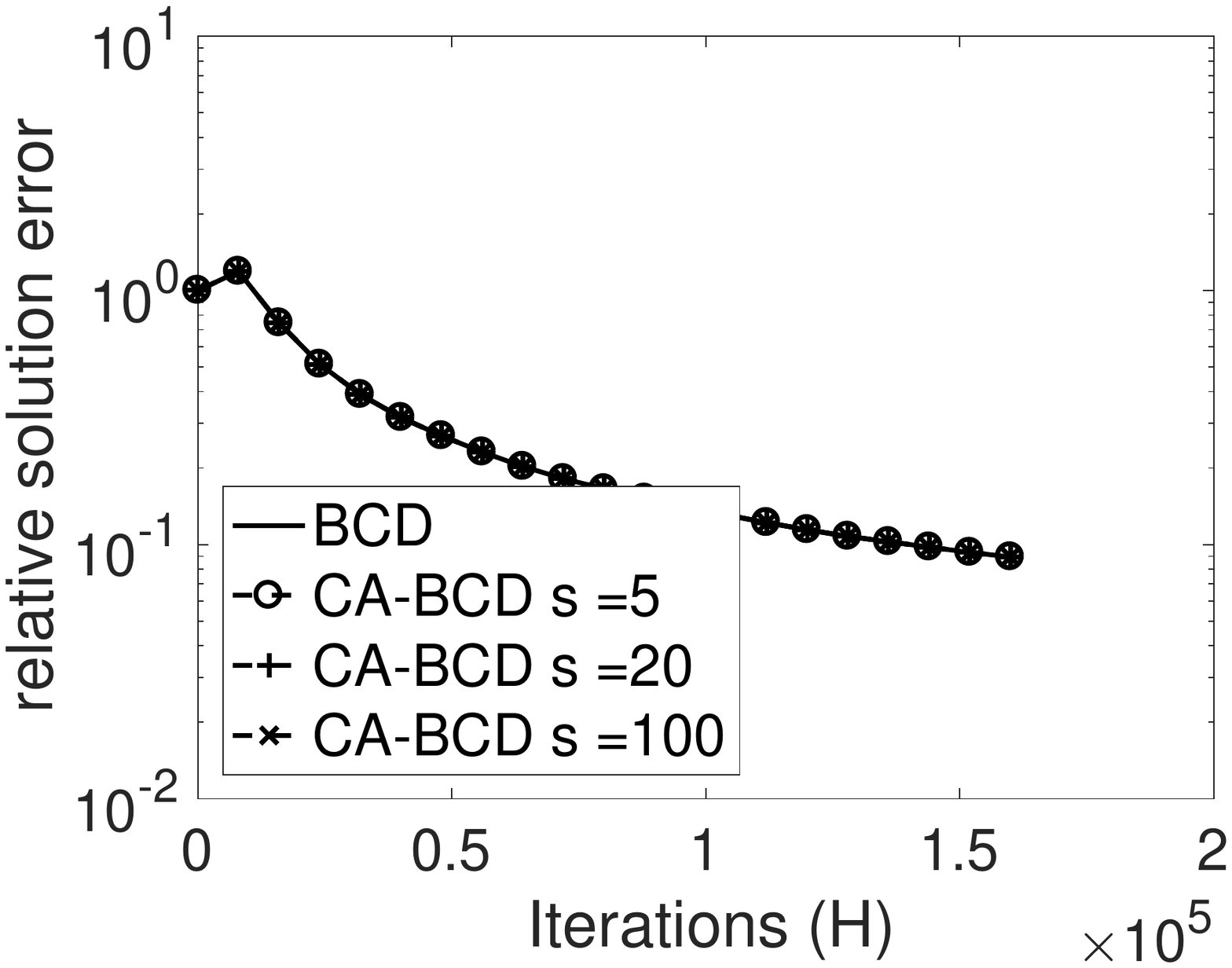}
\caption{news20}
\label{fig:news206}
\end{subfigure}
\begin{subfigure}{.329\textwidth}
\centering
\includegraphics[trim = .5in 2.5in .6in 2.5in,  clip,width=\textwidth, ]{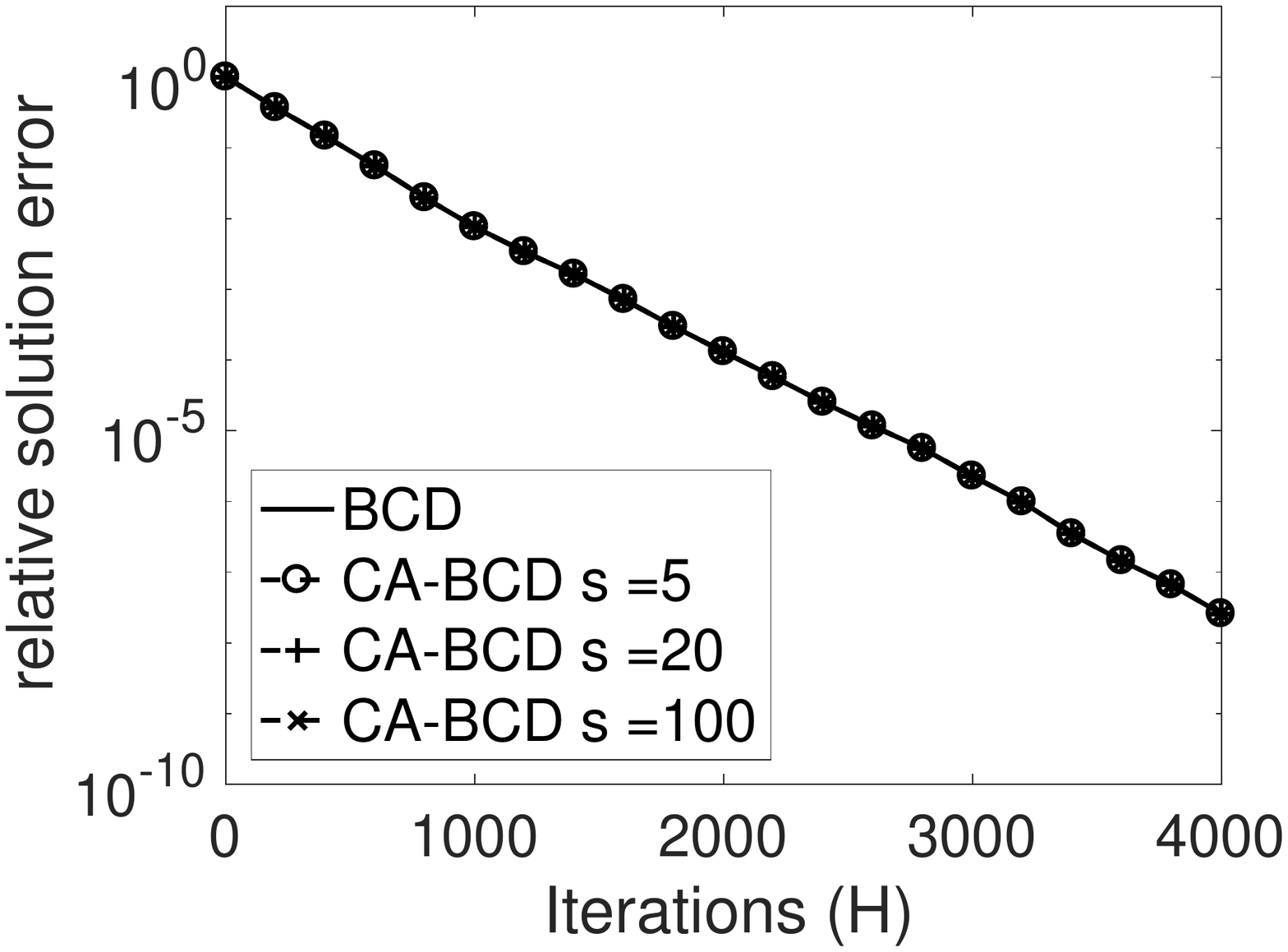}
\caption{a9a}
\label{fig:a9a6}
\end{subfigure}
\begin{subfigure}{.329\textwidth}
\centering
\includegraphics[trim = .5in 2.5in .6in 2.5in,  clip,width=\textwidth, ]{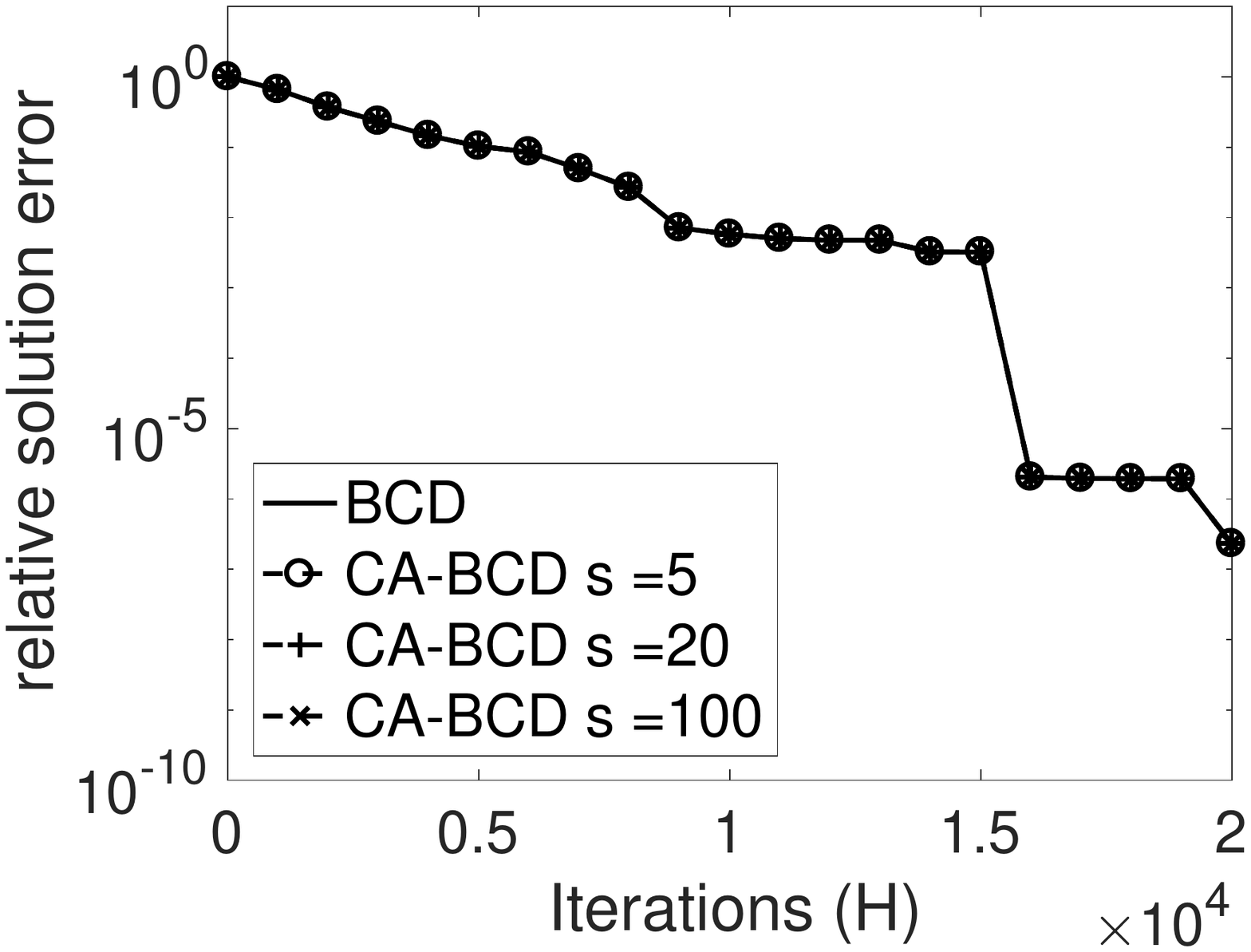}
\caption{real-sim}
\label{fig:realsim6}
\end{subfigure}

\begin{subfigure}{.329\textwidth}
\centering
\includegraphics[trim = .5in 2.5in .6in 2.5in,  clip,width=\textwidth, ]{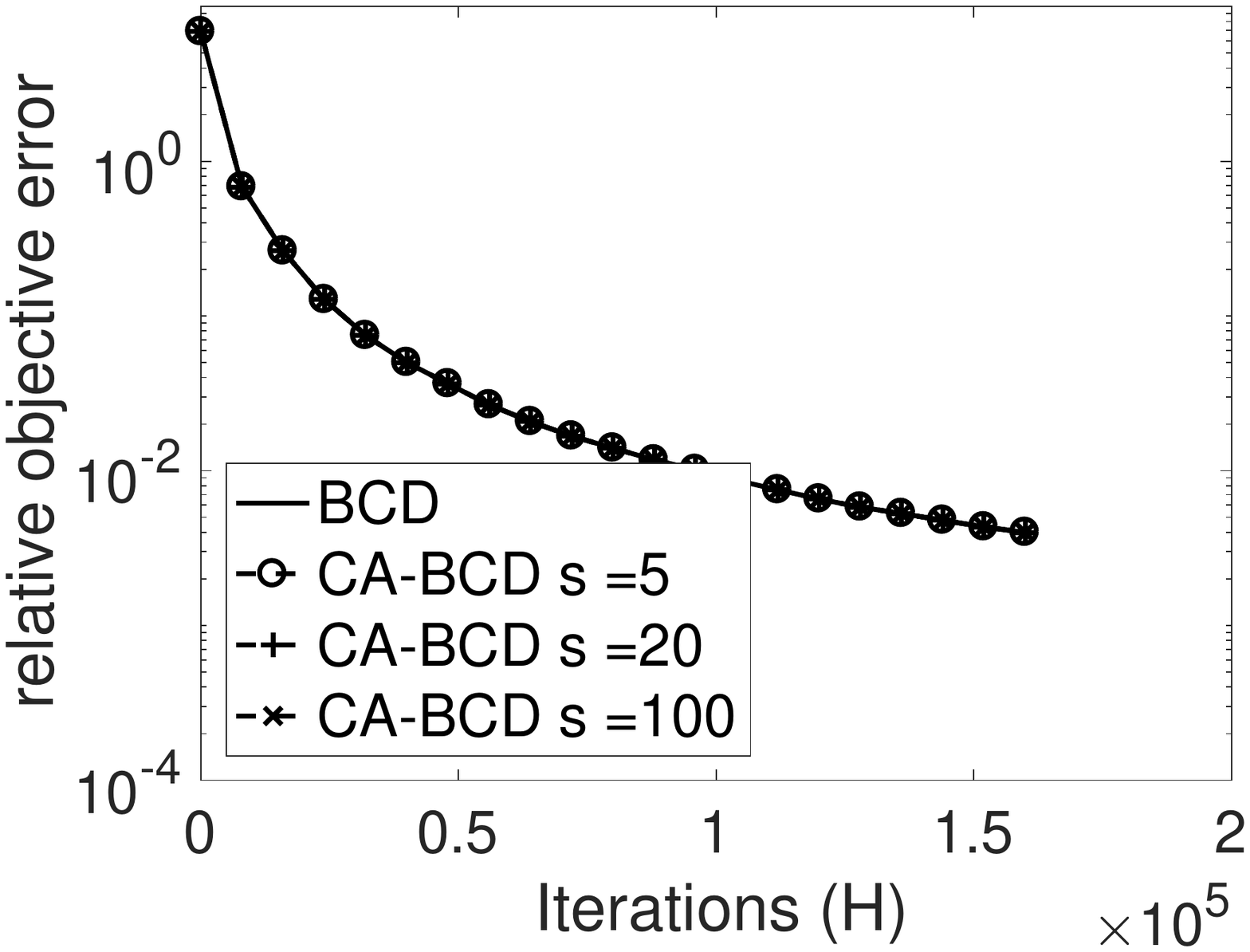}
\caption{news20}
\label{fig:news207}
\end{subfigure}
\begin{subfigure}{.329\textwidth}
\centering
\includegraphics[trim = .5in 2.5in .6in 2.5in,  clip,width=\textwidth, ]{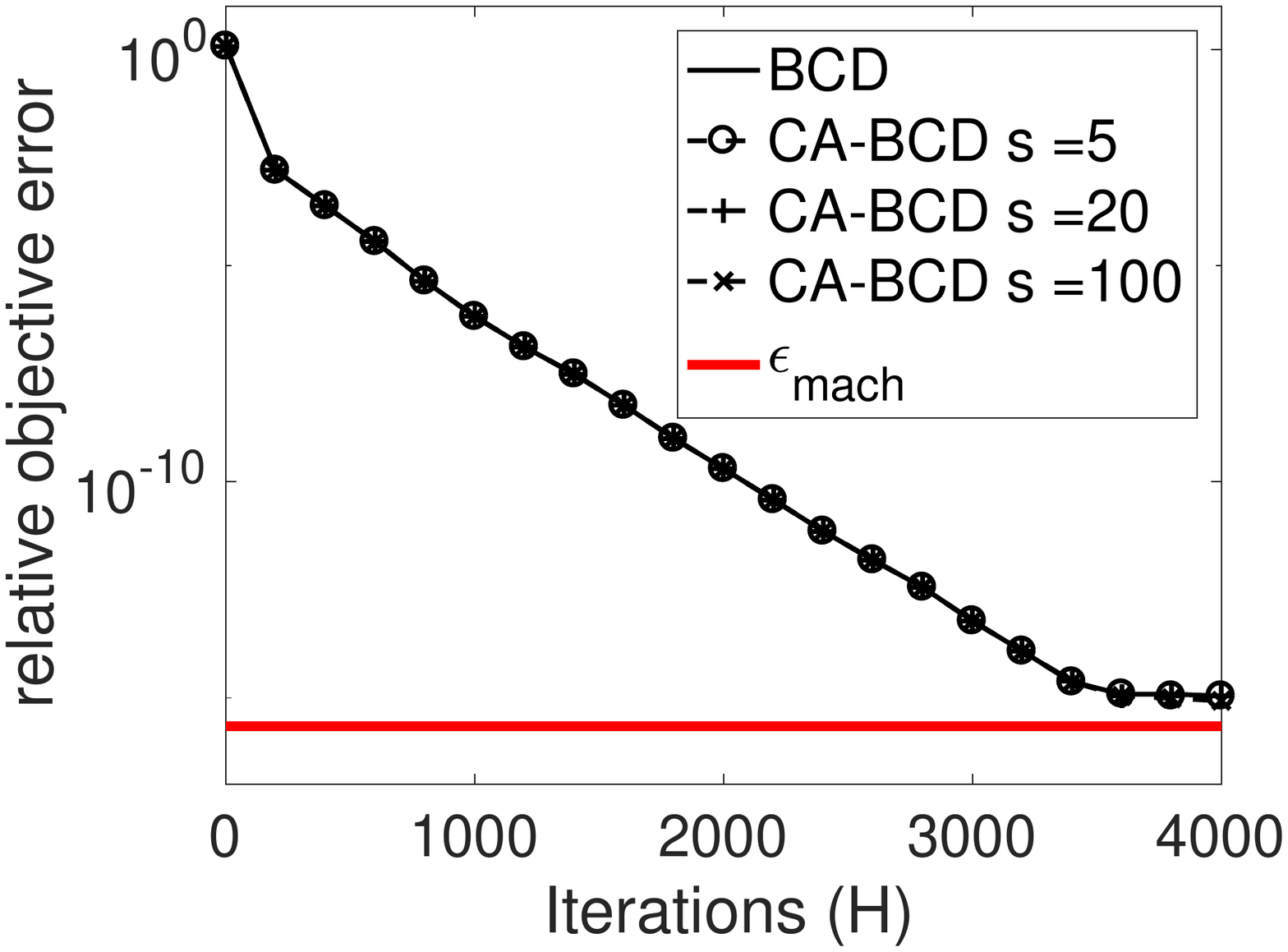}
\caption{a9a}
\label{fig:a9a7}
\end{subfigure}
\begin{subfigure}{.329\textwidth}
\centering
\includegraphics[trim = .5in 2.5in .6in 2.5in,  clip,width=\textwidth, ]{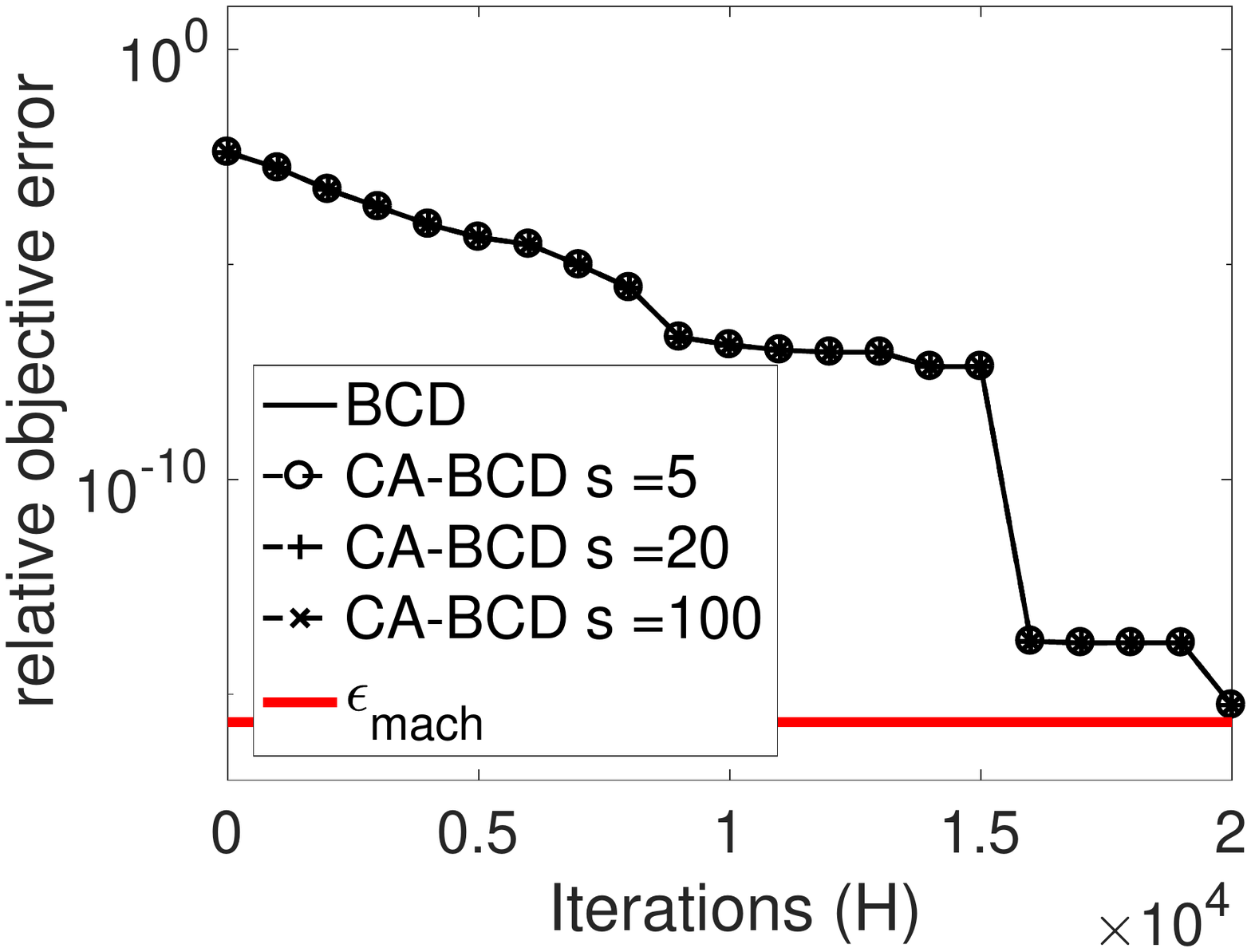}
\caption{real-sim}
\label{fig:realsim7}
\end{subfigure}

\begin{subfigure}{.329\textwidth}
\centering
\includegraphics[trim = .4in 2.5in .7in 2.5in,  clip,width=\textwidth, ]{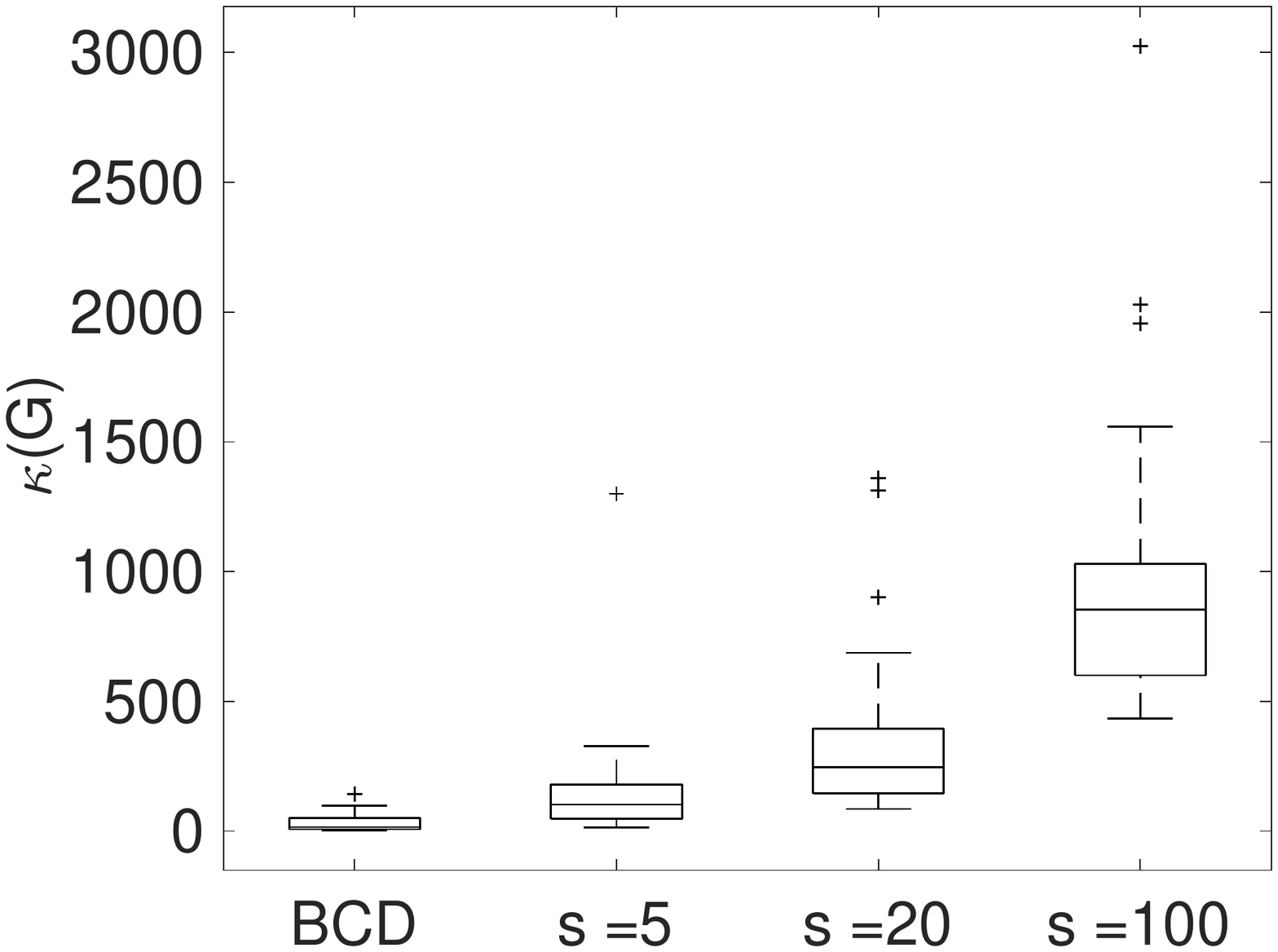}
\caption{news20}
\label{fig:news208}
\end{subfigure}
\begin{subfigure}{.329\textwidth}
\centering
\includegraphics[trim = .4in 2.5in .7in 2.5in,  clip,width=\textwidth, ]{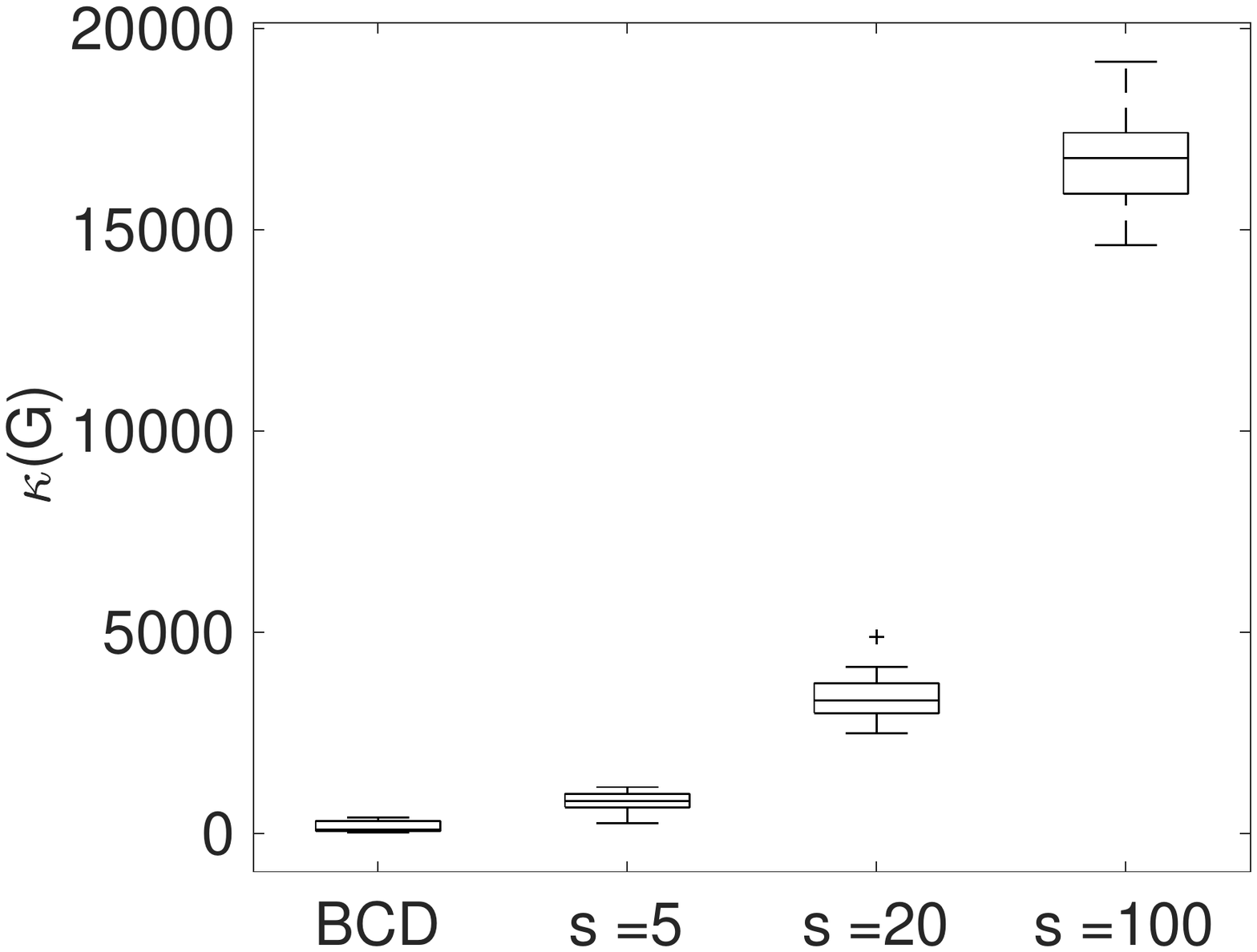}
\caption{a9a}
\label{fig:a9a8}
\end{subfigure}
\begin{subfigure}{.329\textwidth}
\centering
\includegraphics[trim = .4in 2.5in .7in 2.5in,  clip,width=\textwidth, ]{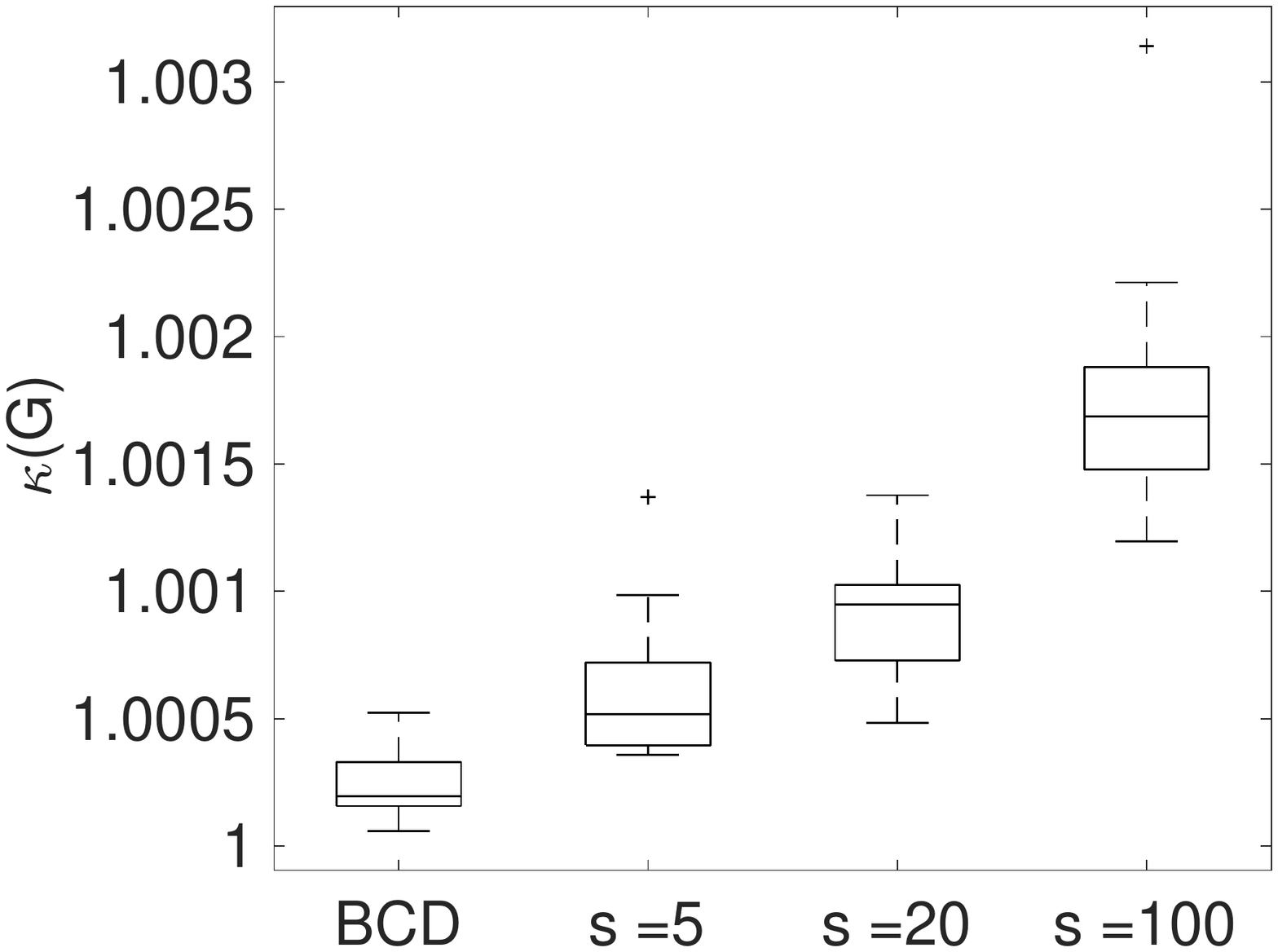}
\caption{real-sim}
\label{fig:realsim8}
\end{subfigure}
\caption{We compare the convergence behavior of BCD and CA-BCD with several values of $s$. Relative solution error (top row, Figs. \ref{fig:news206}-\ref{fig:realsim6}), relative objective error (middle row, Figs. \ref{fig:news207}-\ref{fig:realsim7}), and statistics of the Gram matrix condition numbers (bottom row, Figs. \ref{fig:news208}-\ref{fig:realsim8}) versus convergence. The block size for each dataset is set to $b = 16$. The boxplots (Figs. \ref{fig:news208} - \ref{fig:realsim8}) use standard MATLAB convention \cite{boxplot}.}
\label{fig:cabcdconv}
\end{figure}

Recall that the BCD algorithm computes a $b \times b$ Gram matrix and solves a $b$-dimensional subproblem at each iteration. Therefore, one should expect that as $b$ increases the algorithm converges faster but requires more flops and bandwidth per iteration. So we begin by exploring the block size vs. convergence behavior tradeoff for BCD with $1 \leq b < d$.

Figure \ref{fig:bcdconv} shows the convergence behavior of the datasets in Table \ref{tbl:dsets} in terms of the relative solution error (Figs. \ref{fig:news202}-\ref{fig:realsim2}) and relative objective error (Figs. \ref{fig:news205}-\ref{fig:realsim5}). The x-axis for the latter figures are on $\log_{10}$ scale. Note that the number of messages is equivalent to the number of iterations, since BCD communicates every iteration. We observe that the convergence rates for all datasets improve as the block sizes increase.

Figure \ref{fig:bcdcost} shows the convergence behavior (in terms of the objective error) vs. flops and bandwidth costs for each dataset. From these results, we observe that BCD with $b = 1$ is more flops and bandwidth efficient, whereas $b > 1$ is more latency efficient (from Figs. \ref{fig:news205}-\ref{fig:realsim5}). This indicates the existence of a tradeoff between BCD convergence rate (which depends on the block size) and hardware-specific parameters (like flops rate, memory/network bandwidth and latency).

\begin{figure}[t!]
%
\begin{subfigure}{.329\textwidth}
\centering
\includegraphics[trim = .5in 2.5in 1.in 2.5in,  clip,width=\textwidth, ]{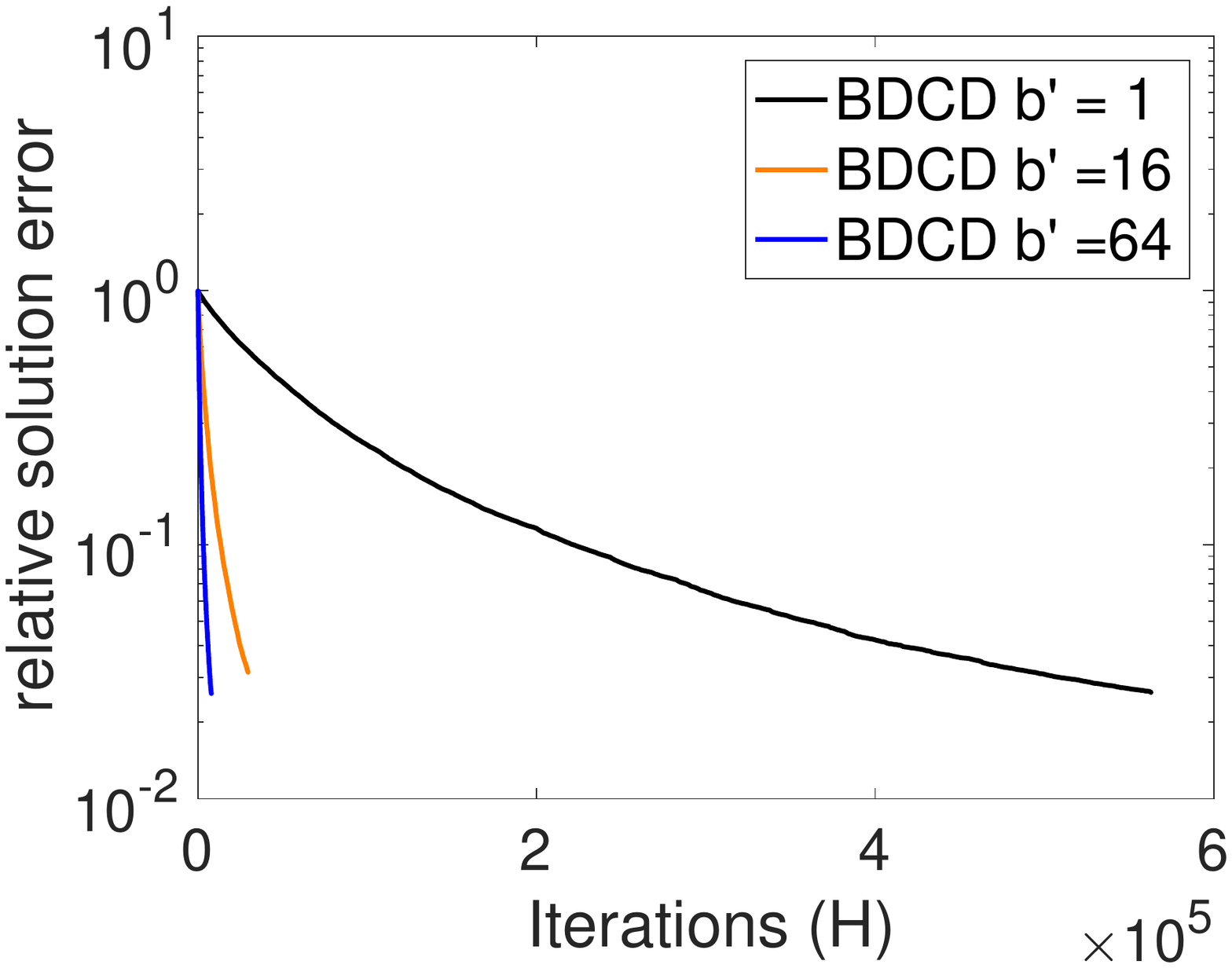}
\caption{news20}
\label{fig:news2012}
\end{subfigure}
\begin{subfigure}{.329\textwidth}
\centering
\includegraphics[trim = .5in 2.5in 1.in 2.5in,  clip,width=\textwidth, ]{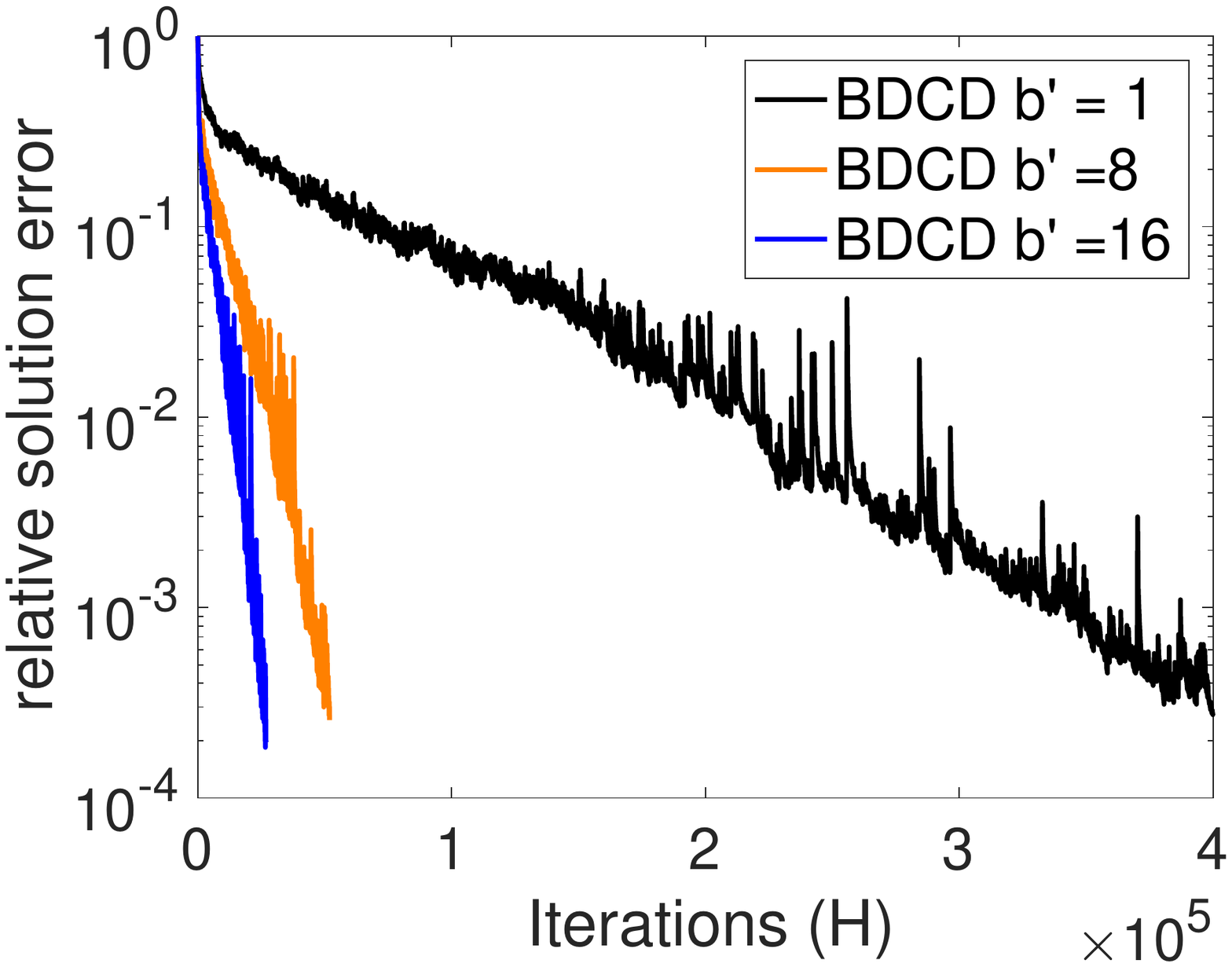}
\caption{a9a}
\label{fig:a9a12}
\end{subfigure}
\begin{subfigure}{.329\textwidth}
\centering
\includegraphics[trim = .5in 2.5in 1.in 2.5in,  clip,width=\textwidth, ]{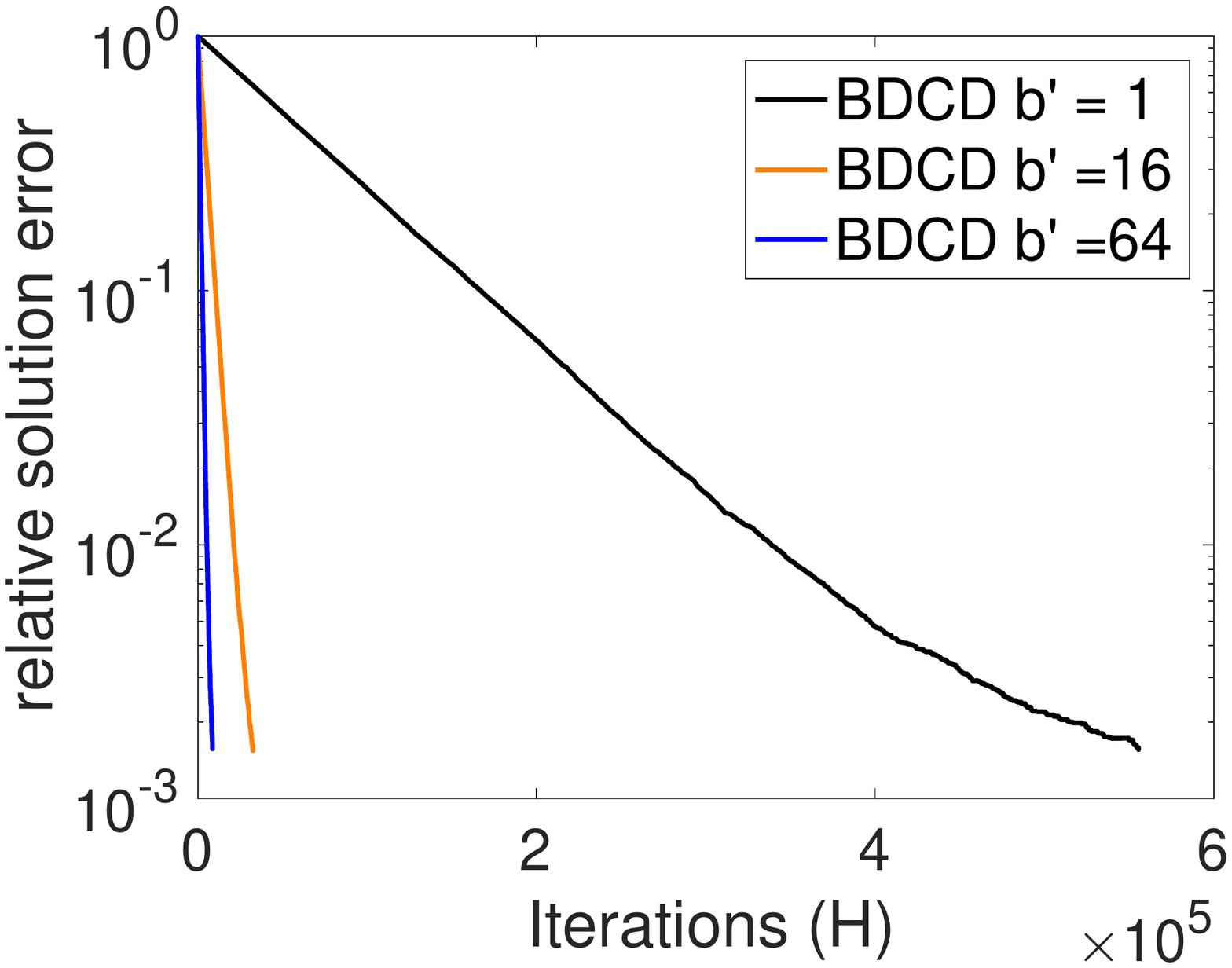}
\caption{real-sim}
\label{fig:realsim12}
\end{subfigure}

\begin{subfigure}{.329\textwidth}
\centering
\includegraphics[trim = .5in 2.5in .7in 2.5in,  clip,width=\textwidth, ]{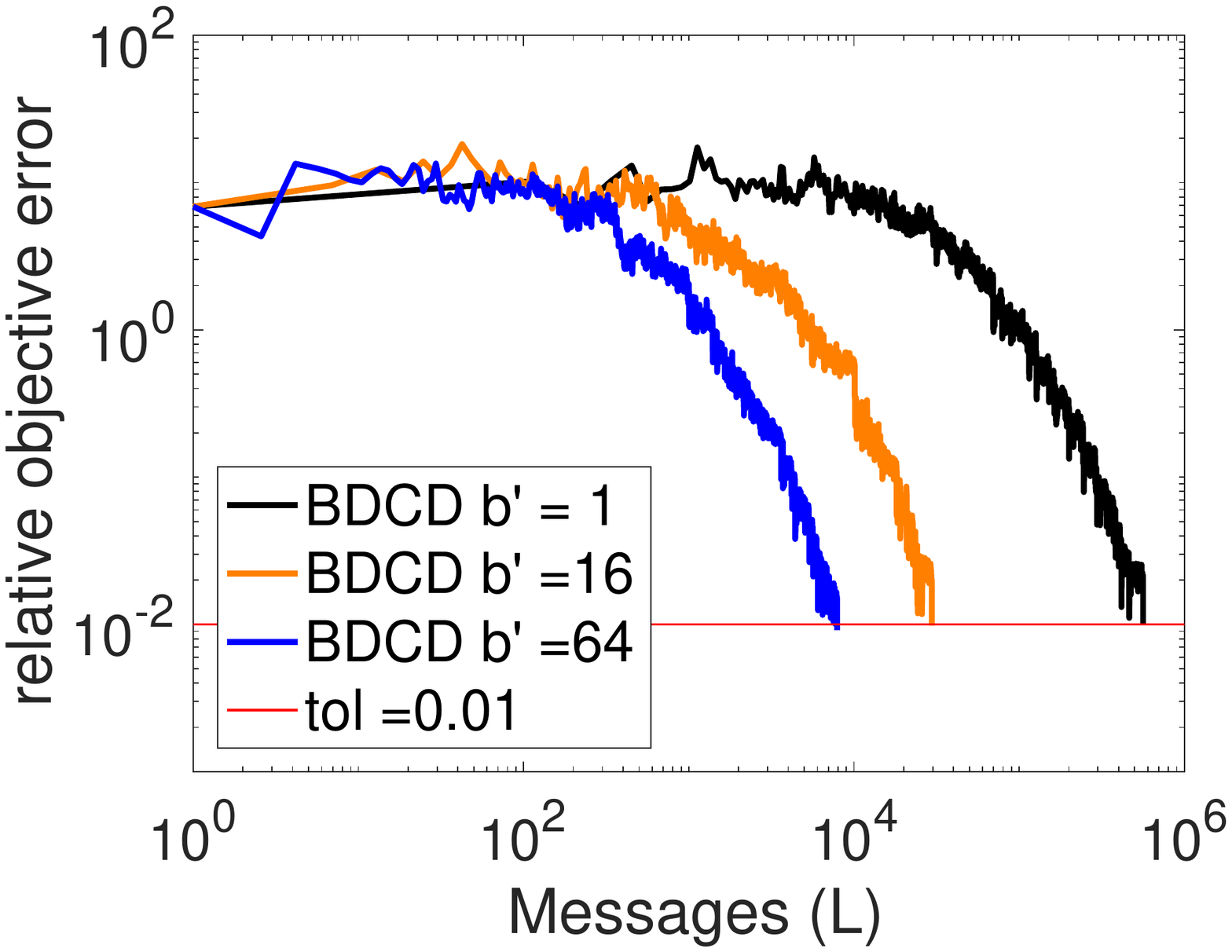}
\caption{news20}
\label{fig:news2016}
\end{subfigure}
\begin{subfigure}{.329\textwidth}
\centering
\includegraphics[trim = .5in 2.5in .7in 2.5in,  clip,width=\textwidth, ]{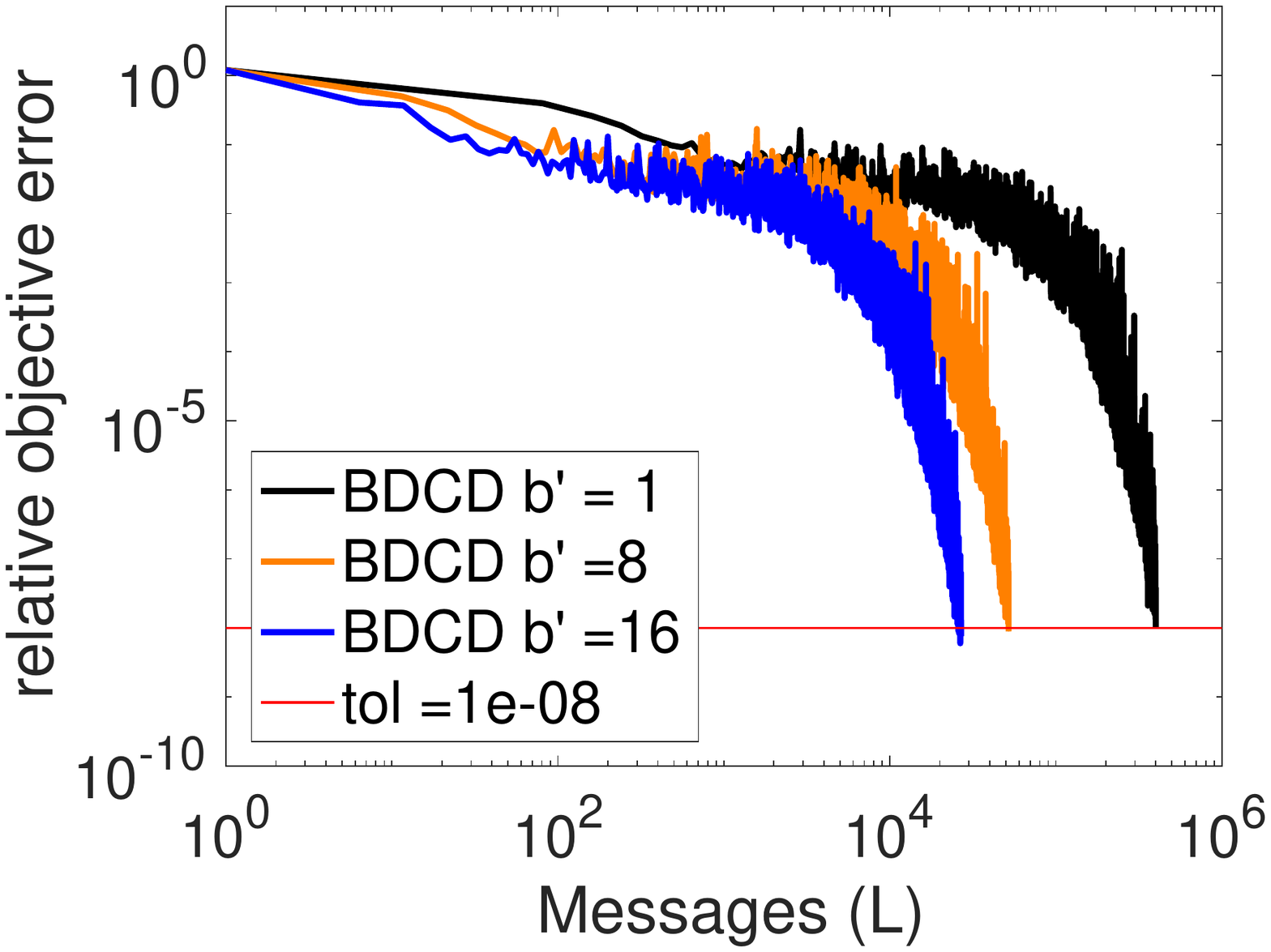}
\caption{a9a}
\label{fig:a9a16}
\end{subfigure}
\begin{subfigure}{.329\textwidth}
\centering
\includegraphics[trim = .5in 2.5in .7in 2.5in,  clip,width=\textwidth, ]{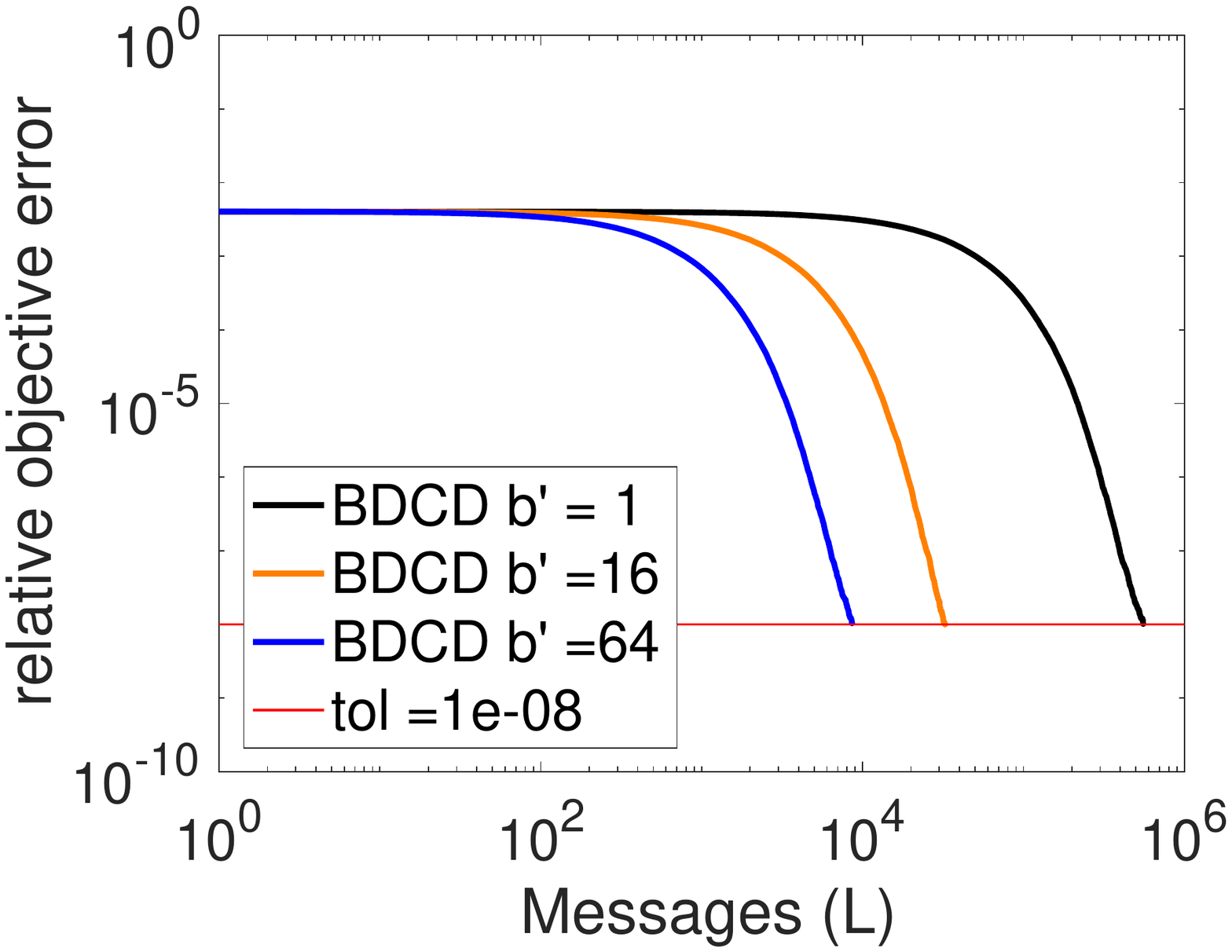}
\caption{real-sim}
\label{fig:realsim16}
\end{subfigure}
\caption{We compare the convergence behavior of BDCD for several block sizes, $b'$, such that $1 \leq b' < n$ on several machine learning datasets. We show relative solution error (top row, Figs. \ref{fig:news2012}-\ref{fig:realsim12}) and objective error (bottom row, Fig. \ref{fig:news2016}-\ref{fig:realsim16}) convergence plots with $\lambda = 1000\sigma_{min}$. The x-axis for Figures \ref{fig:news2016}-\ref{fig:realsim16} show the number of messages required on a $\log_{10}$ scale. Since BDCD communicates at every iteration, the x-axis is also equivalent to the number of iterations (modulo $\log_{10}$ scale).}\label{fig:bdcdconv}
\end{figure}

\subsubsection{Communication-Avoiding Block Coordinate Descent}\label{cabcdeval}
\begin{figure}[t!]
%
\begin{subfigure}{.329\textwidth}
\centering
\includegraphics[trim = .5in 2.5in .7in 2.5in,  clip,width=\textwidth, ]{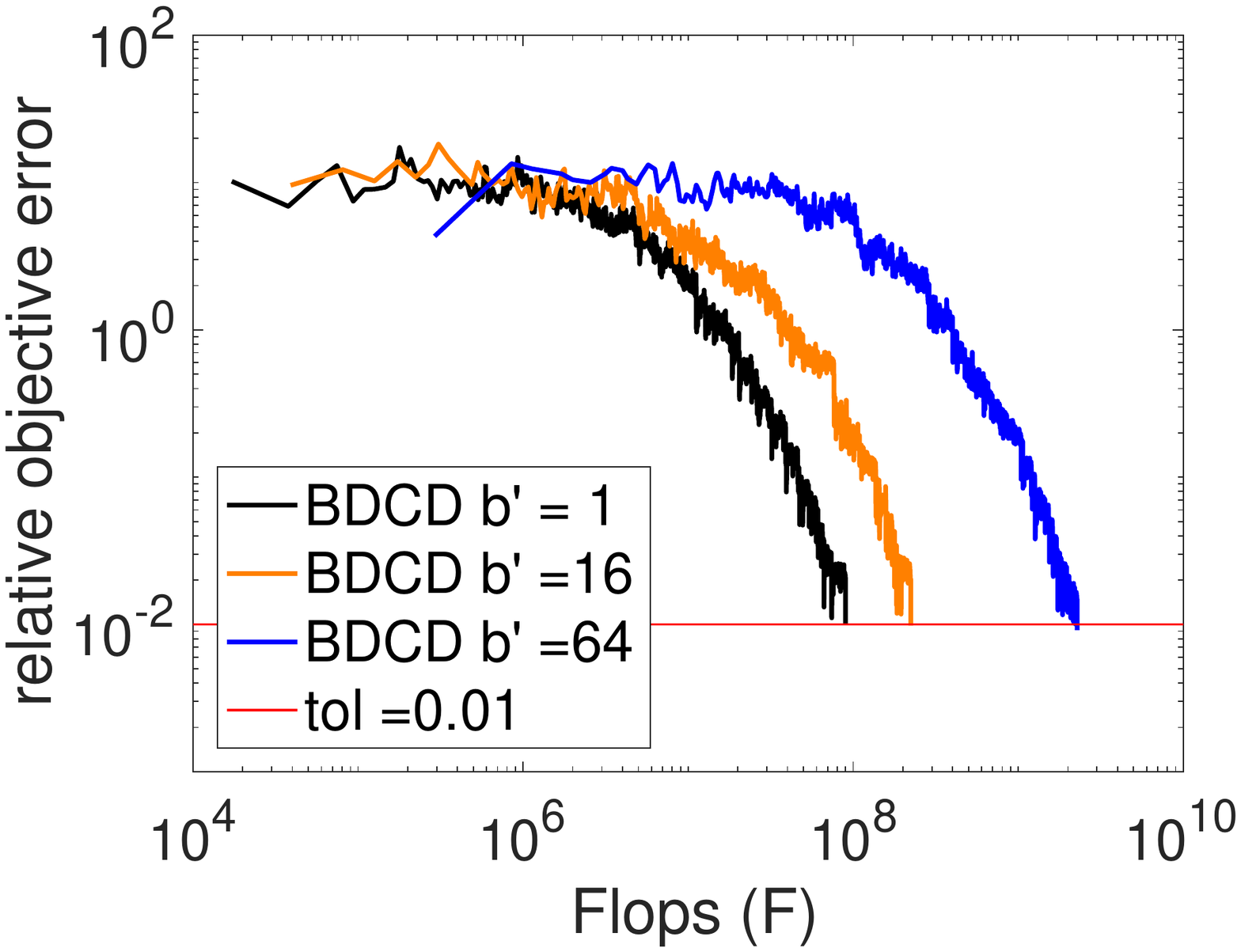}
\caption{news20}
\label{fig:news2014}
\end{subfigure}
\begin{subfigure}{.329\textwidth}
\centering
\includegraphics[trim = .5in 2.5in .7in 2.5in,  clip,width=\textwidth, ]{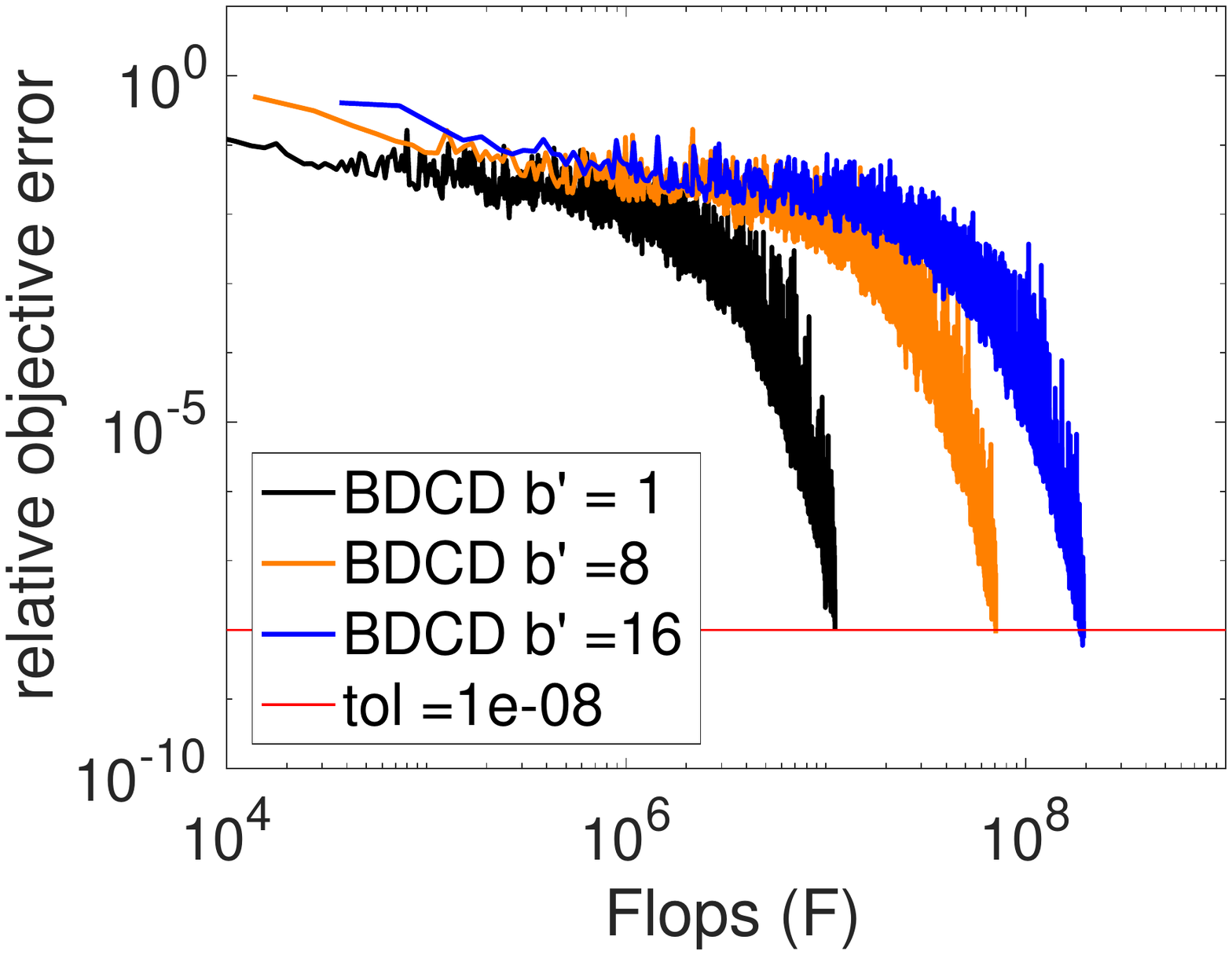}
\caption{a9a}
\label{fig:a9a14}
\end{subfigure}
\begin{subfigure}{.329\textwidth}
\centering
\includegraphics[trim = .5in 2.5in .7in 2.5in,  clip,width=\textwidth, ]{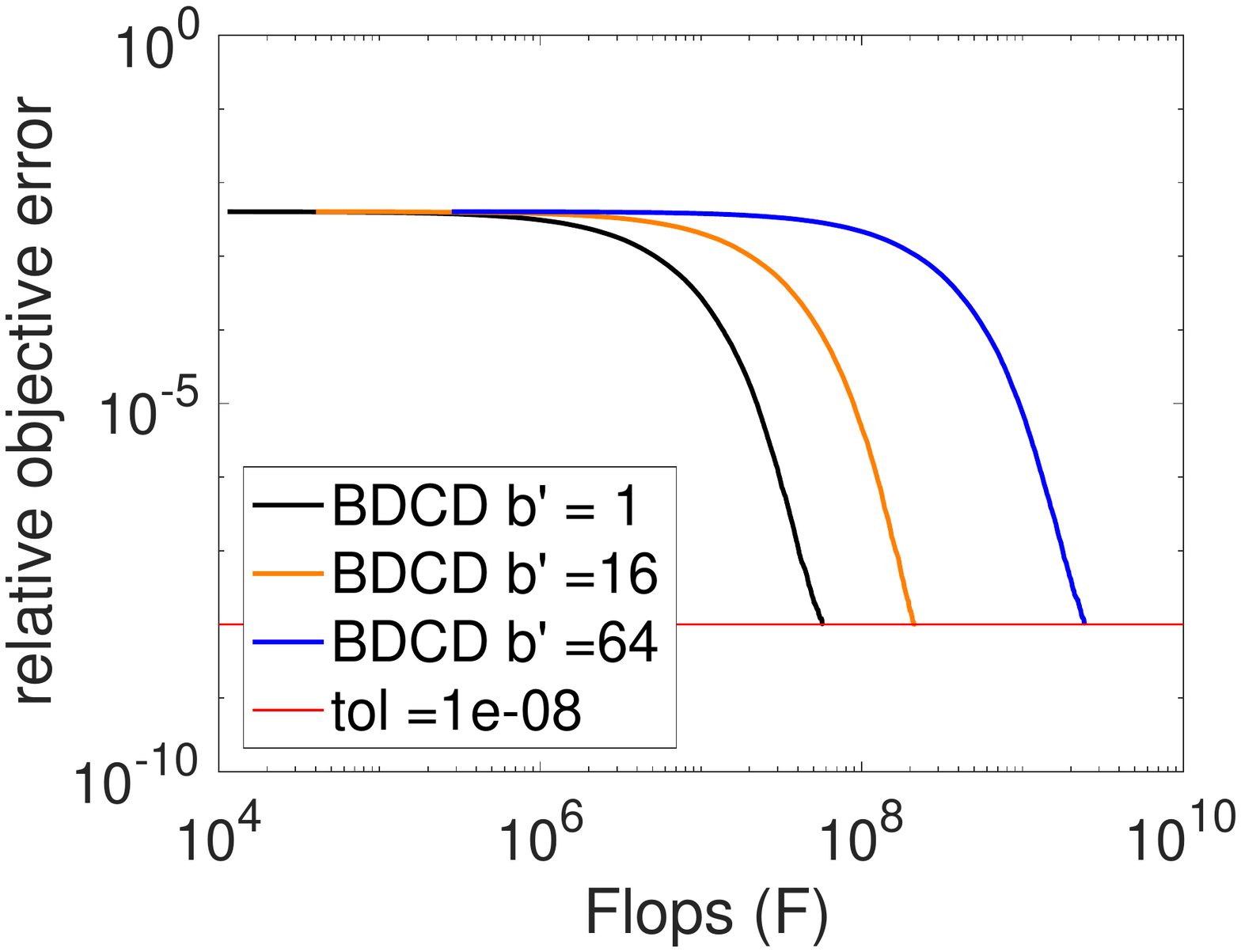}
\caption{real-sim}
\label{fig:realsim14}
\end{subfigure}

\begin{subfigure}{.329\textwidth}
\centering
\includegraphics[trim = .5in 2.5in .7in 2.5in,  clip,width=\textwidth, ]{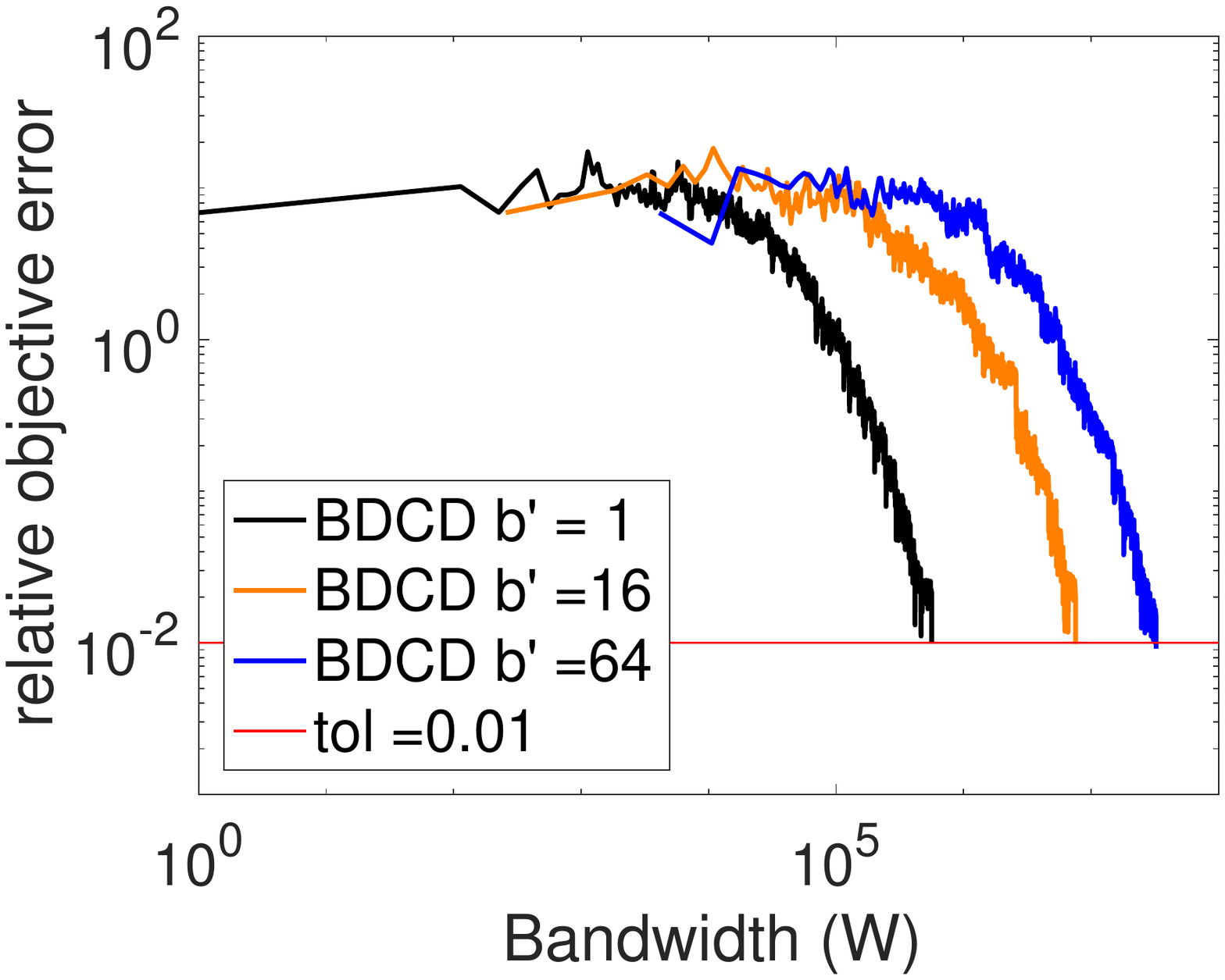}
\caption{news20}
\label{fig:news2015}
\end{subfigure}
\begin{subfigure}{.329\textwidth}
\centering
\includegraphics[trim = .5in 2.5in .7in 2.5in,  clip,width=\textwidth, ]{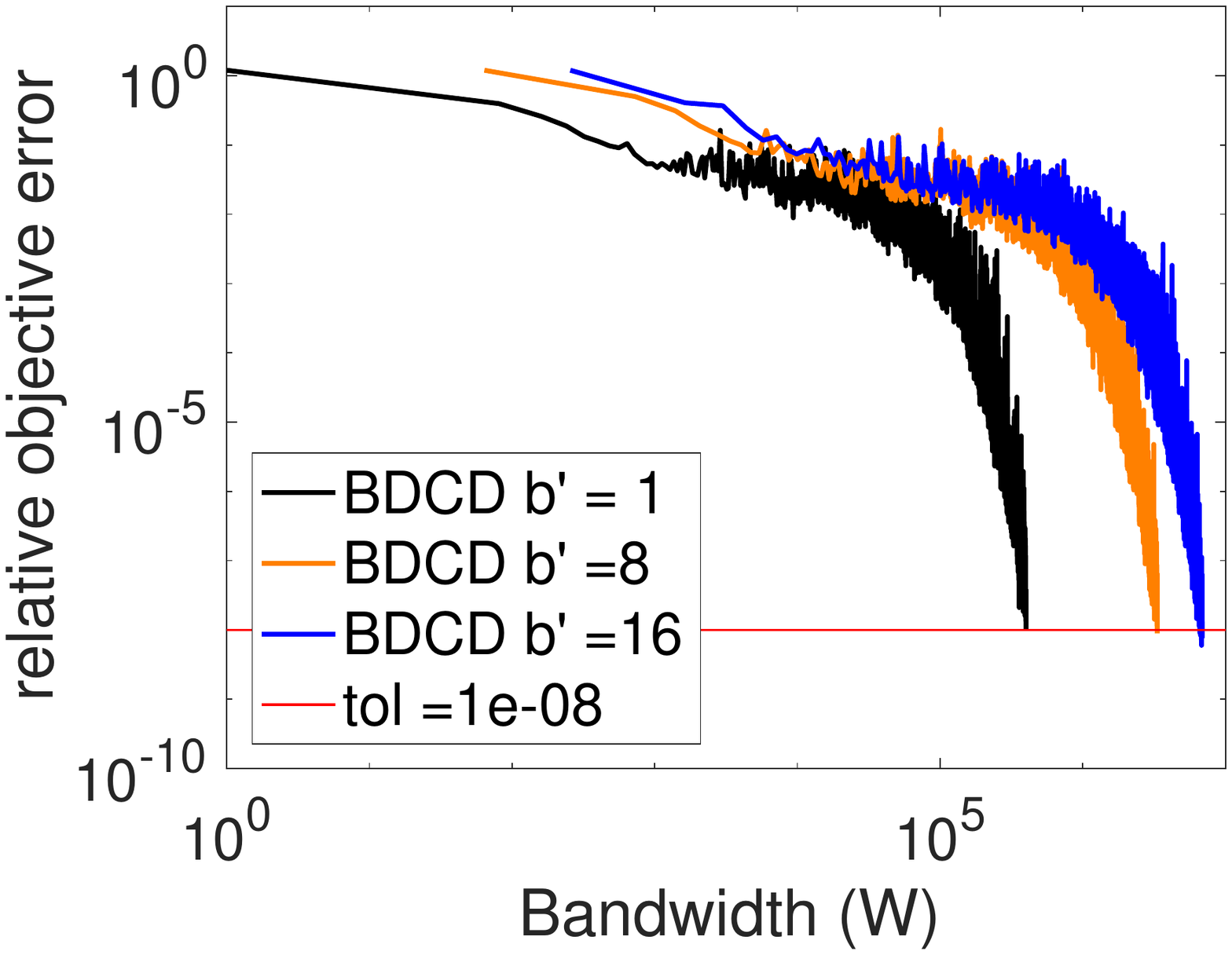}
\caption{a9a}
\label{fig:a9a15}
\end{subfigure}
\begin{subfigure}{.329\textwidth}
\centering
\includegraphics[trim = .5in 2.5in .7in 2.5in,  clip,width=\textwidth, ]{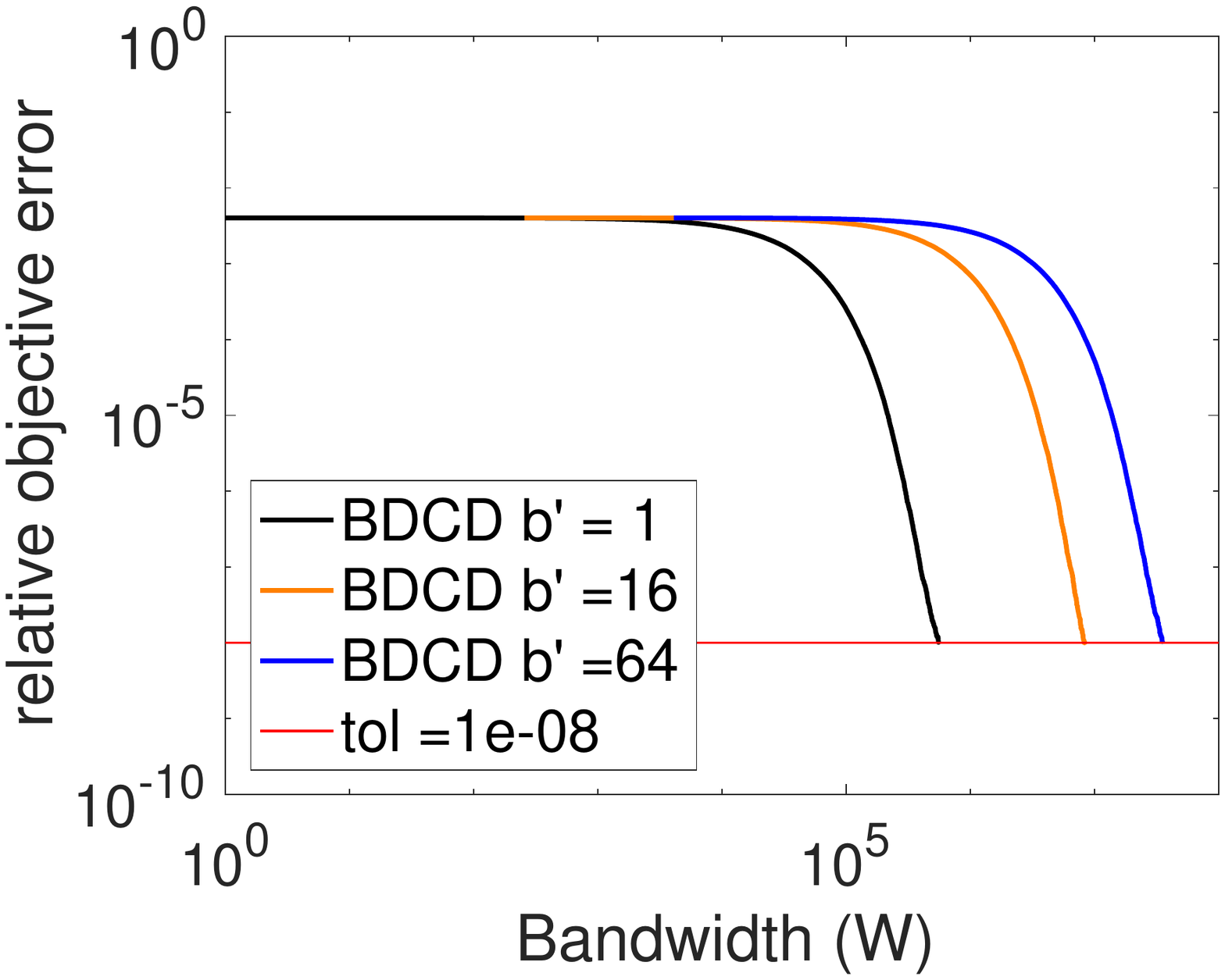}
\caption{real-sim}
\label{fig:realsim15}
\end{subfigure}

\caption{We compare the convergence behavior of BDCD for several block sizes, $b'$, such that $1 \leq b' < n$ on several machine learning datasets. Flops cost (top row, Figs. \ref{fig:news2014}-\ref{fig:realsim14}), and bandwidth cost (middle row, Figs. \ref{fig:news2015}-\ref{fig:realsim15}) versus convergence with $\lambda = 1000\sigma_{min}$.}\label{fig:bdcdcost}
\end{figure}

Our derivation of the CA-BCD algorithm showed that by unrolling the vector update recurrences we can reduce the latency cost of the BCD algorithm by a factor of $s$. However, this comes at the cost of computing a larger $sb \times sb$ Gram matrix whose condition number is larger than the $b \times b$ Gram matrix computed in the BCD algorithm. The larger condition number implies that the CA-BCD algorithm may not be stable for $s >1$ due to round-off error. We begin by experimentally showing the convergence behavior of the CA-BCD algorithm on the datasets in Table \ref{tbl:dsets} with fixed block sizes of $b = 16$ for news20, a9a, and real-sim, respectively.

Figure \ref{fig:cabcdconv} compares the convergence behavior of BCD and CA-BCD for $s > 1$. We plot the relative solution error, relative objective error and statistics of the Gram matrix condition numbers. The convergence plots indicate that CA-BCD shows almost no deviation from the BCD convergence. While the Gram matrix condition numbers increase with $s$ for CA-BCD, those condition numbers are not so large as to significantly alter the numerical stability. Figures \ref{fig:a9a7} and \ref{fig:realsim7} show that the objective error converges very close to $\epsilon_{mach}$. The well-conditioning of the real-sim dataset in addition to the regularization and small block size (relative to $d$) makes the Gram matrices almost perfectly conditioned. Based on these results, it is likely that the factor of $s$ increase in flops and bandwidth will be the primary bottleneck.

\subsubsection{Block Dual Coordinate Descent}\label{dcdeval}

\begin{figure}[t!]
%
\begin{subfigure}{.329\textwidth}
\centering
\includegraphics[trim = .5in 2.5in 1.in 2.5in,  clip,width=\textwidth, ]{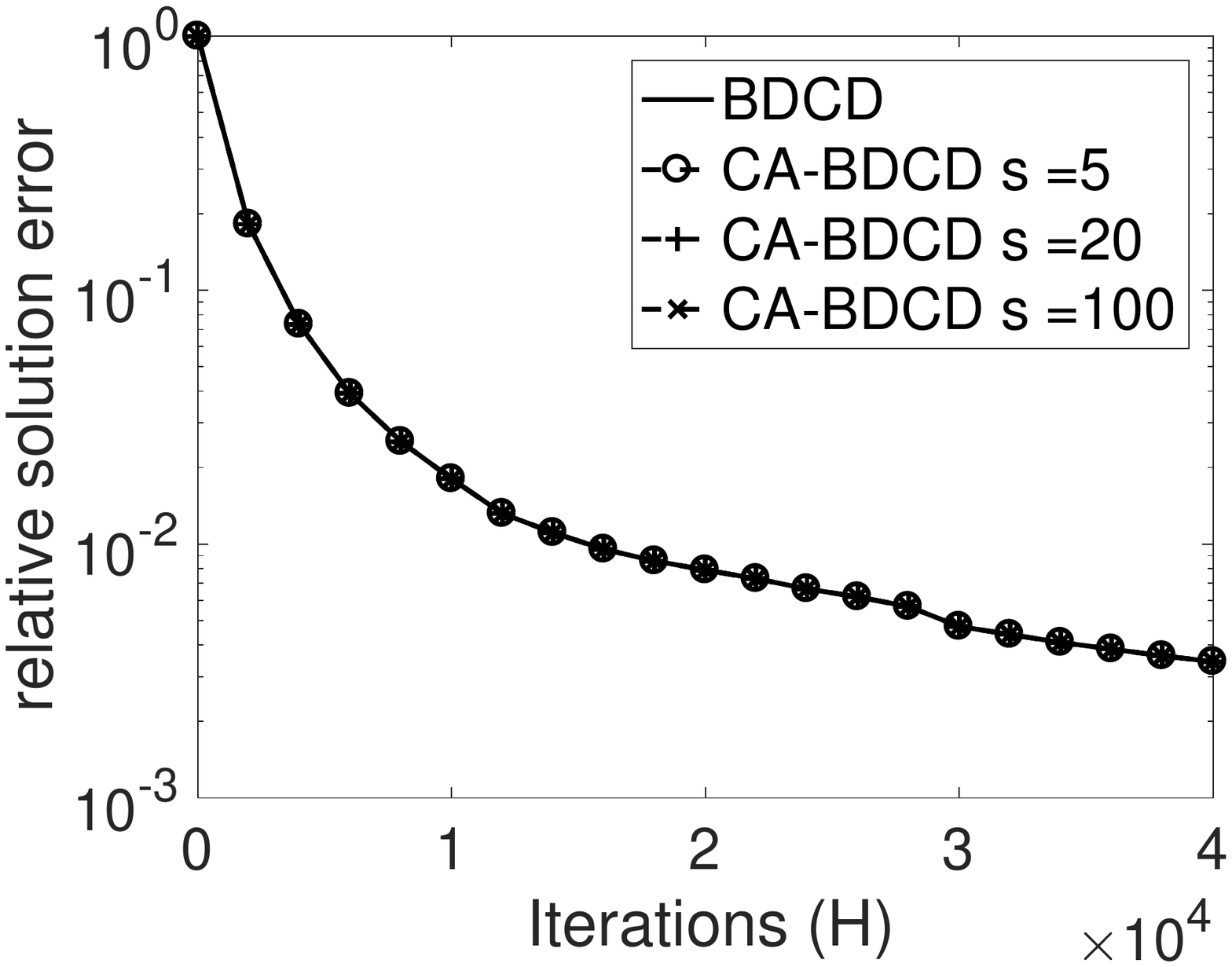}
\caption{news20}
\label{fig:news2018}
\end{subfigure}
\begin{subfigure}{.329\textwidth}
\centering
\includegraphics[trim = .5in 2.5in 1.in 2.5in,  clip,width=\textwidth, ]{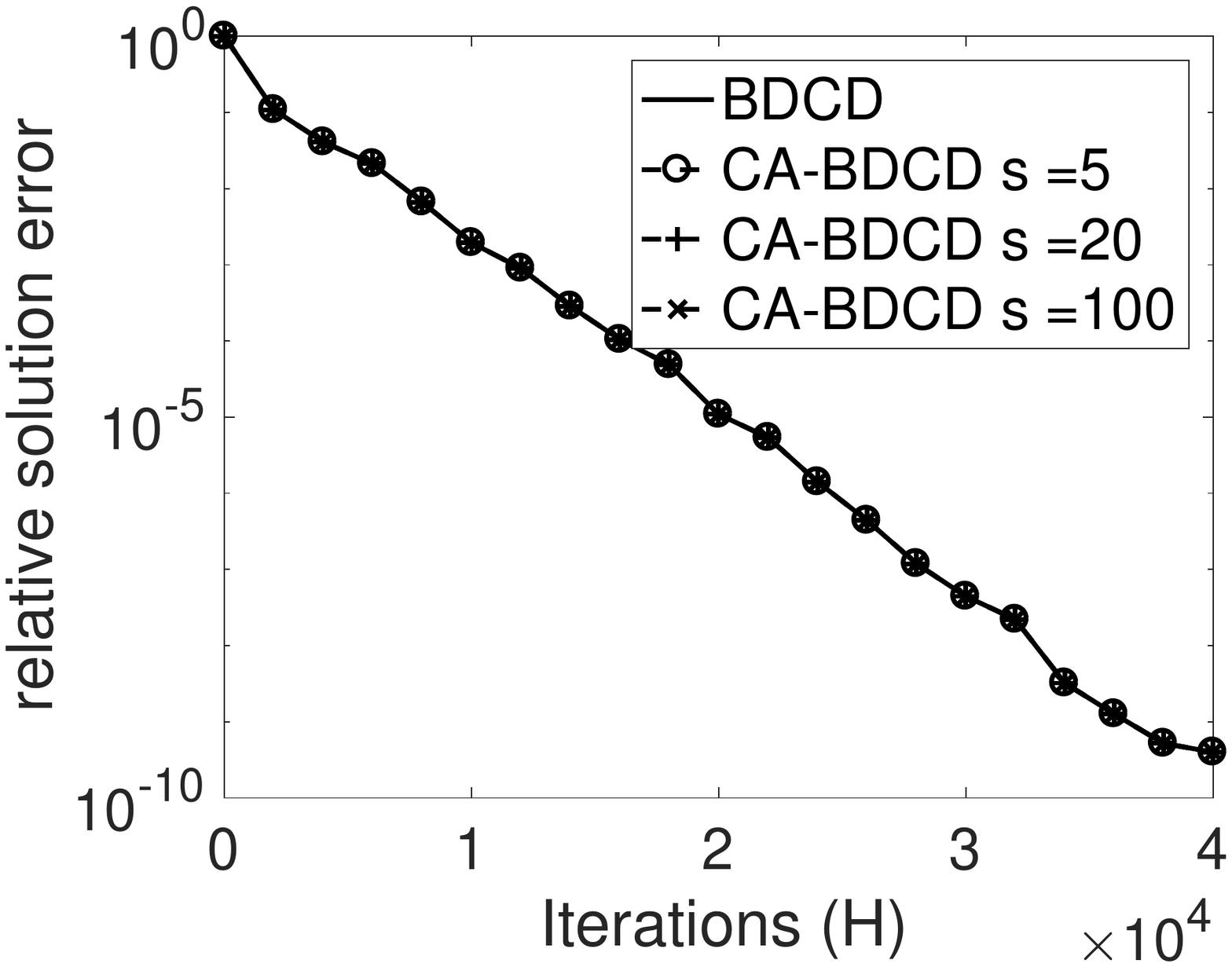}
\caption{a9a}
\label{fig:a9a18}
\end{subfigure}
\begin{subfigure}{.329\textwidth}
\centering
\includegraphics[trim = .5in 2.5in 1.in 2.5in,  clip,width=\textwidth, ]{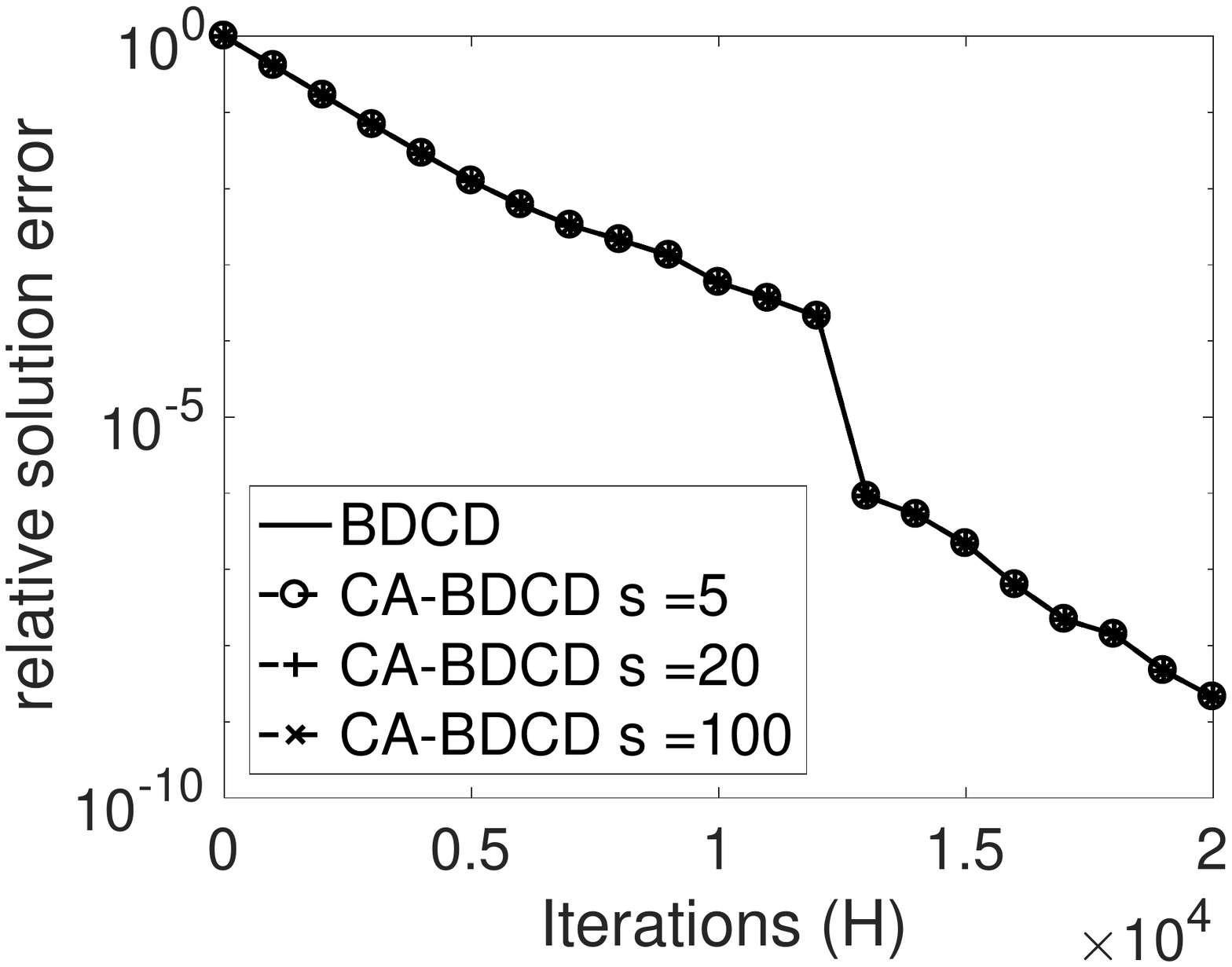}
\caption{real-sim}
\label{fig:realsim18}
\end{subfigure}

\begin{subfigure}{.329\textwidth}
\centering
\includegraphics[trim = .5in 2.5in 1.in 2.5in,  clip,width=\textwidth, ]{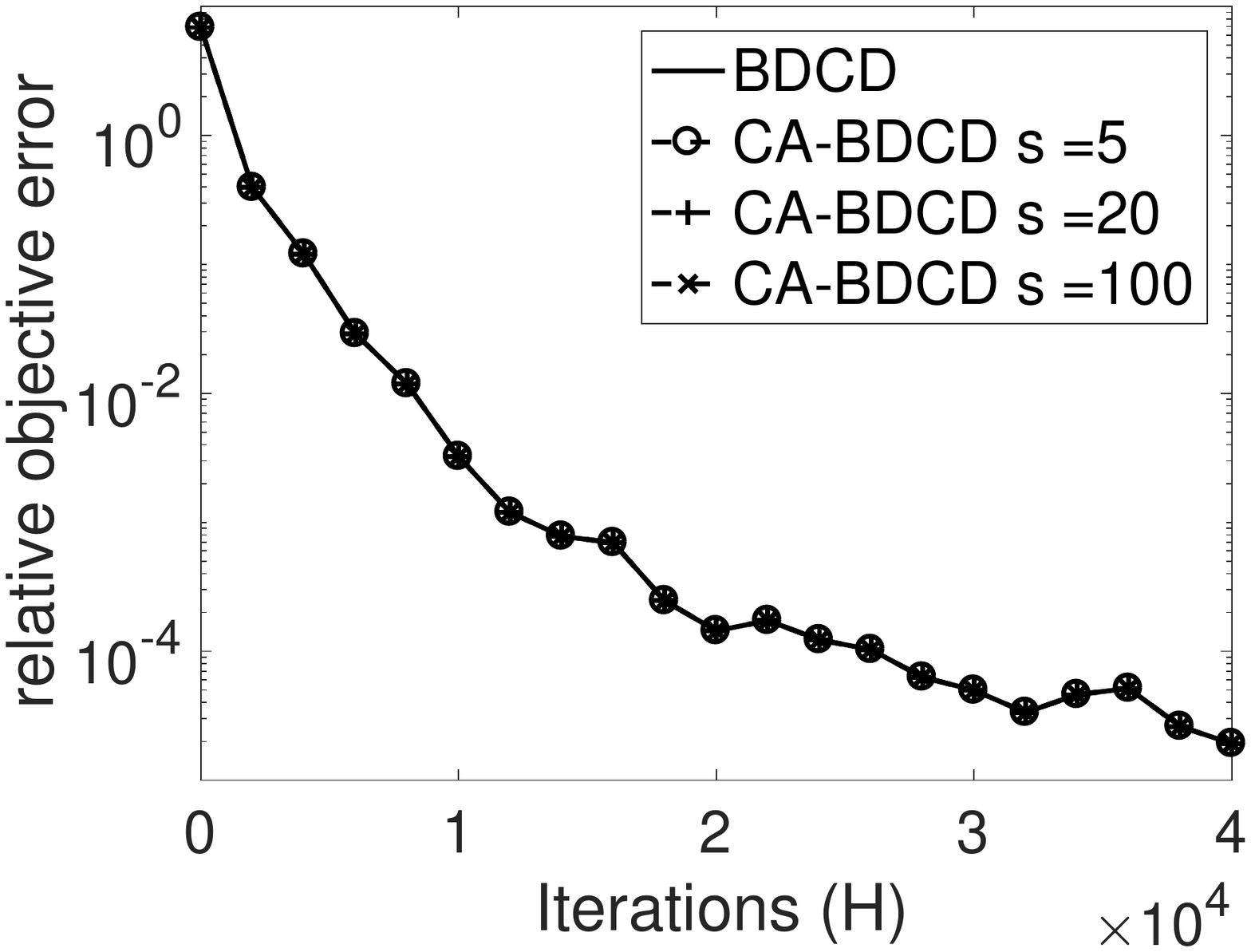}
\caption{news20}
\label{fig:news2017}
\end{subfigure}
\begin{subfigure}{.329\textwidth}
\centering
\includegraphics[trim = .5in 2.5in 1.in 2.5in,  clip,width=\textwidth, ]{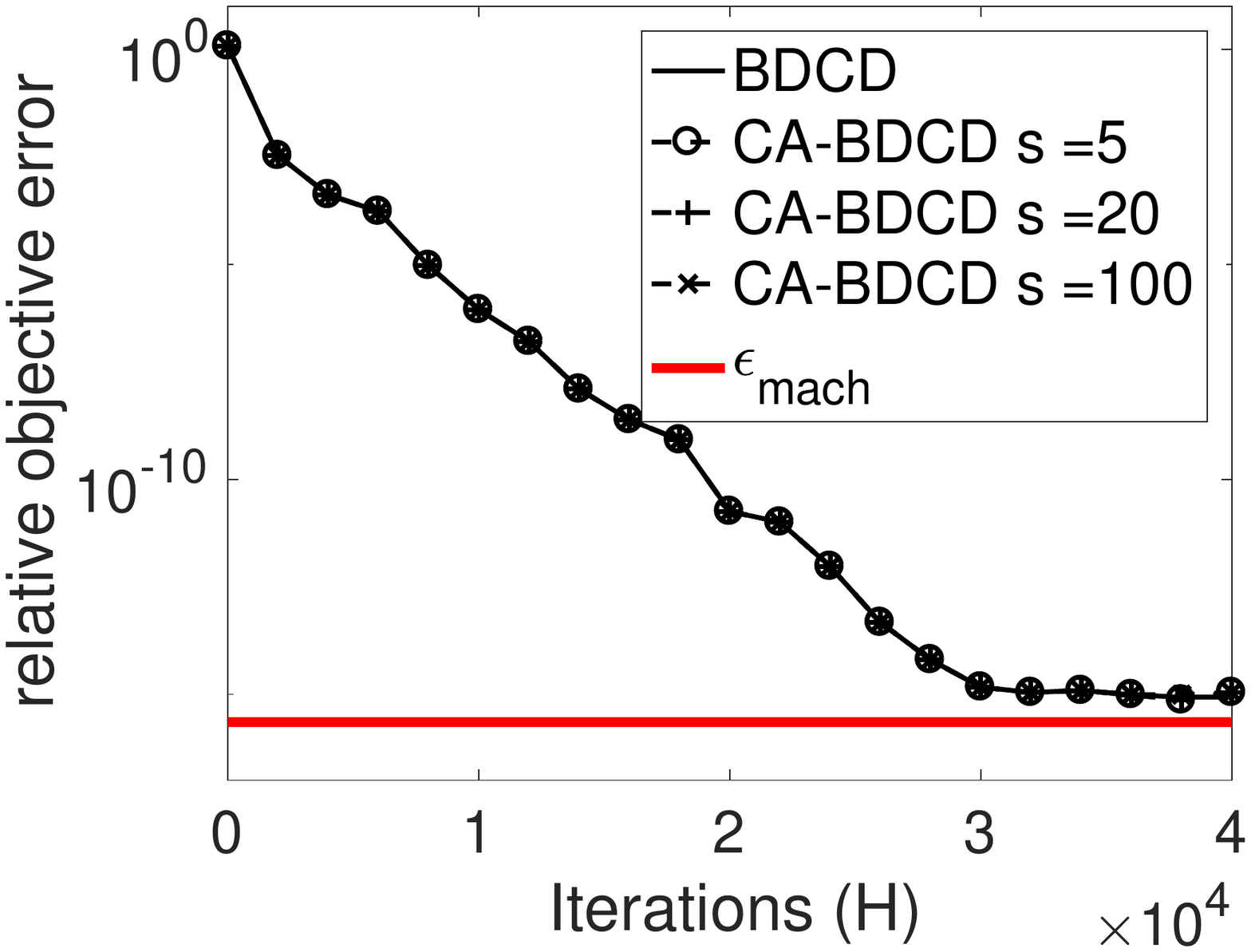}
\caption{a9a}
\label{fig:a9a17}
\end{subfigure}
\begin{subfigure}{.329\textwidth}
\centering
\includegraphics[trim = .5in 2.5in 1.in 2.5in,  clip,width=\textwidth, ]{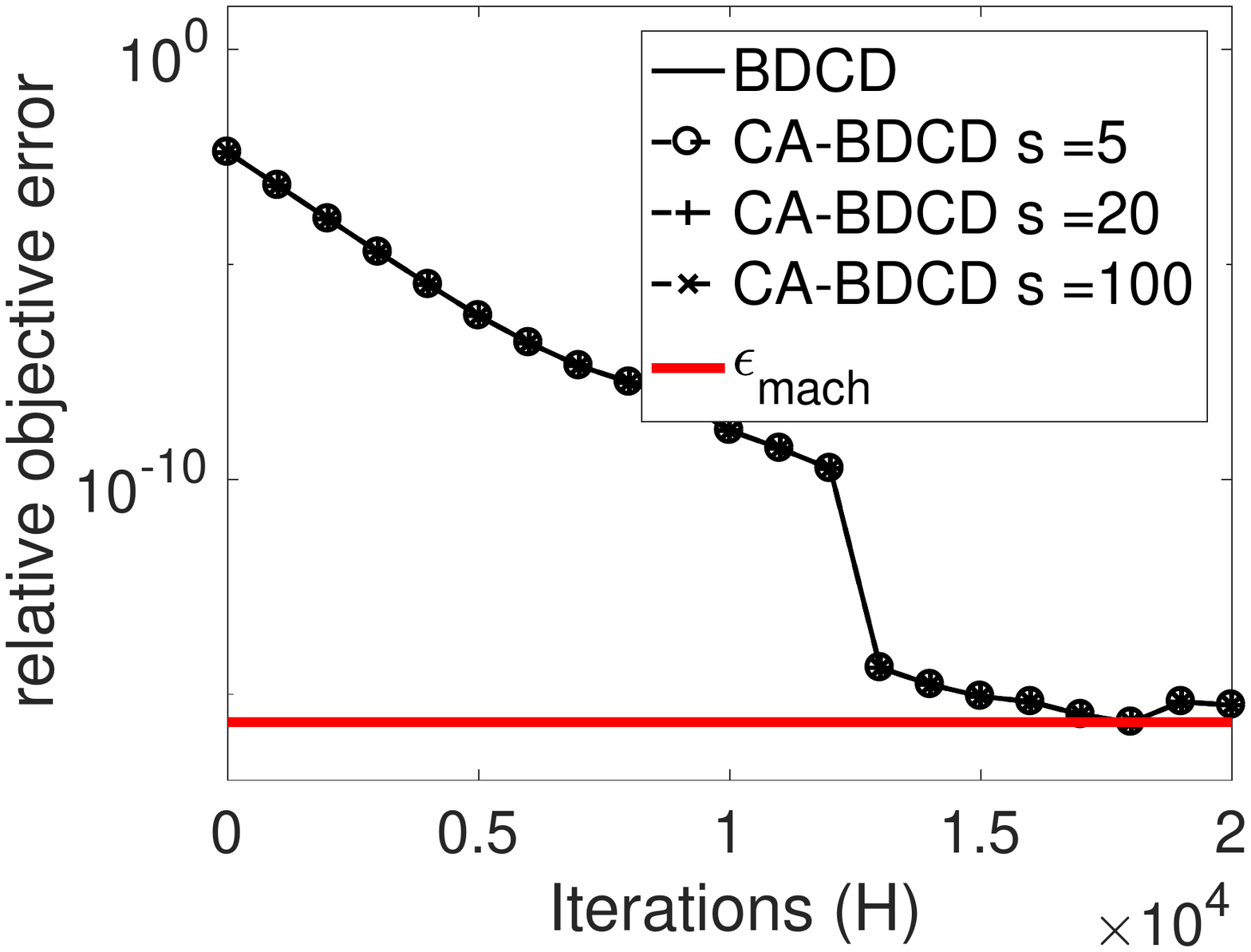}
\caption{real-sim}
\label{fig:realsim17}
\end{subfigure}

\begin{subfigure}{.329\textwidth}
\centering
\includegraphics[trim = .5in 2.5in 1.in 2.5in,  clip,width=\textwidth, ]{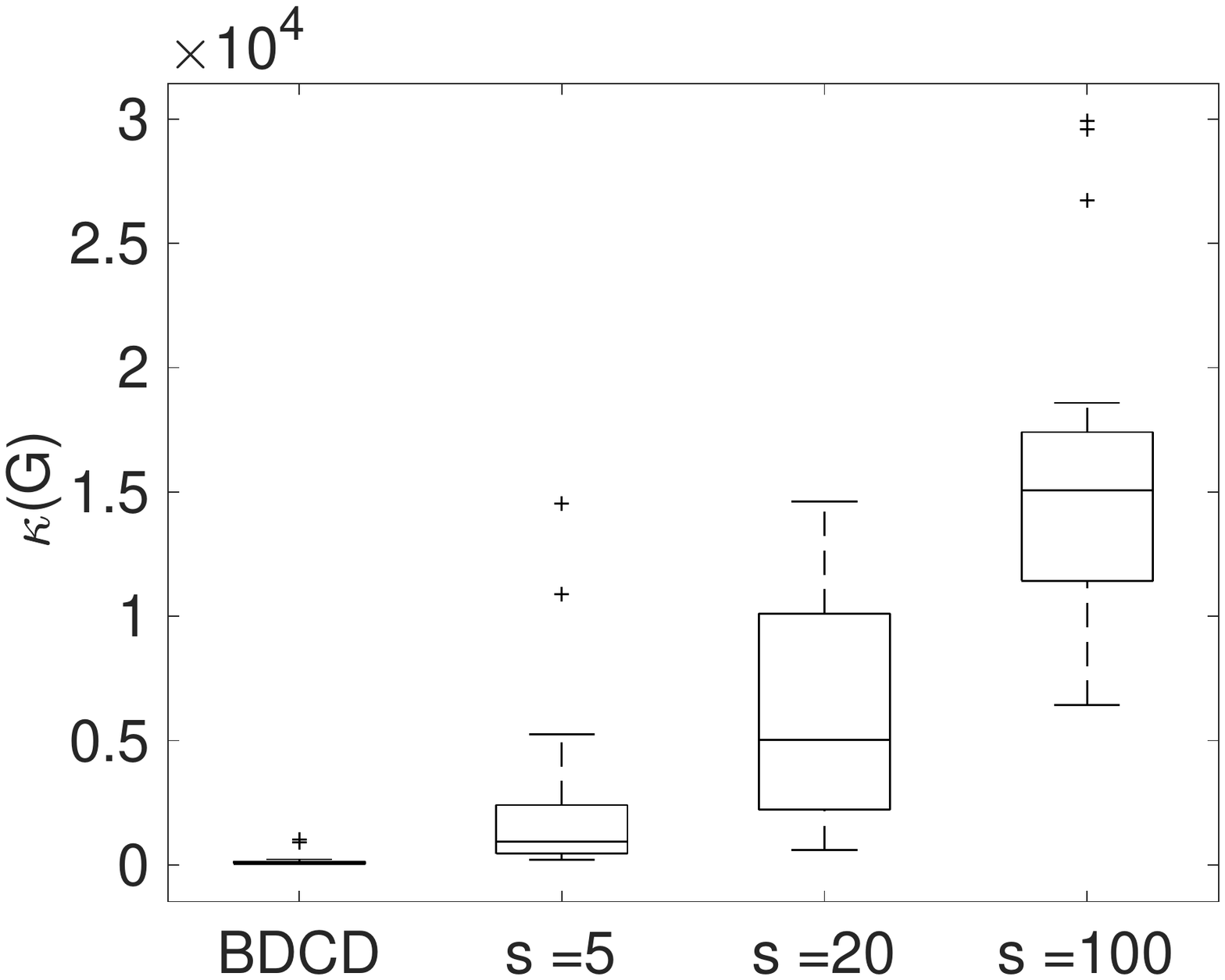}
\caption{news20}
\label{fig:news2019}
\end{subfigure}
\begin{subfigure}{.329\textwidth}
\centering
\includegraphics[trim = .5in 2.5in 1.in 2.5in,  clip,width=\textwidth, ]{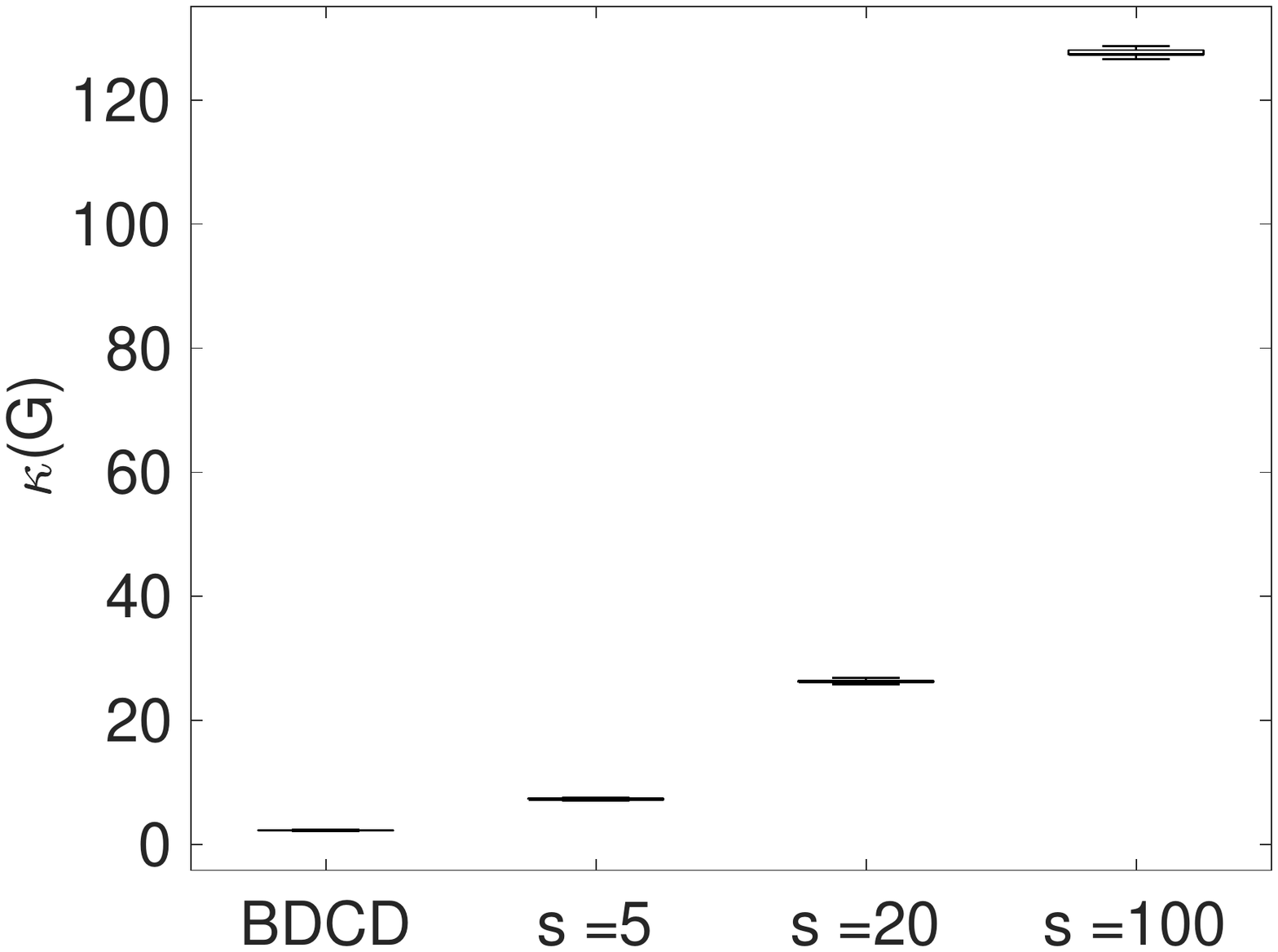}
\caption{a9a}
\label{fig:a9a19}
\end{subfigure}
\begin{subfigure}{.329\textwidth}
\centering
\includegraphics[trim = .5in 2.5in 1.in 2.5in,  clip,width=\textwidth, ]{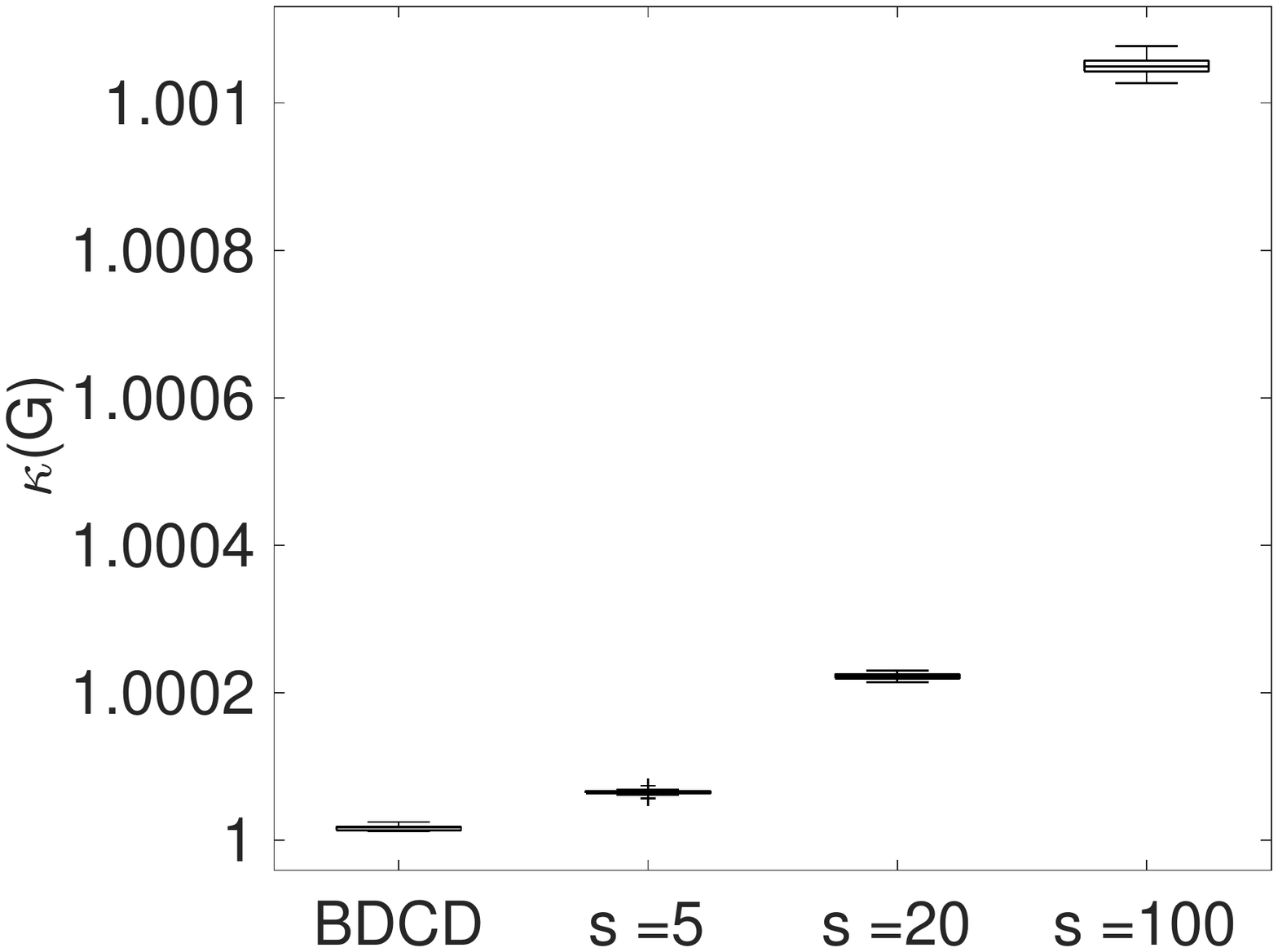}
\caption{real-sim}
\label{fig:realsim19}
\end{subfigure}

\caption{We compare the convergence behavior of BDCD and CA-BDCD with several values of $s$. Relative solution error (top row, Figs. \ref{fig:news2018}-\ref{fig:realsim18}), relative objective error (middle row, Figs. \ref{fig:news2017}-\ref{fig:realsim17}), and statistics of the Gram matrix condition numbers (bottom row, Figs. \ref{fig:news2019}-\ref{fig:realsim19}) versus convergence. The block sizes for each dataset are: news20 with $b' = 64$, a9a with $b' = 16$, and real-sim with $b' = 64$.}
\label{fig:cabdcdconv}
\end{figure}

The BDCD algorithm solves the dual of the regularized least-squares problem by computing a $b' \times b'$ Gram matrix obtained from the columns of $X$ (instead of the rows of $X$ for BCD) and solves a $b'$-dimensional subproblem at each iteration. Similar to BCD, we expect that as $b'$ increases, the BDCD algorithm converges faster at the cost of more flops and bandwidth. We explore this tradeoff space by comparing the convergence behavior (solution error and objective error) and algorithm costs for BDCD with $1 \leq b' < n$.

Figure \ref{fig:bdcdconv} shows the convergence behavior on the datasets in Table \ref{tbl:dsets} for various block sizes and measures the relative solution error (Figs. \ref{fig:news2012}-\ref{fig:realsim12}) and relative objective error (Figs. \ref{fig:news2016}-\ref{fig:realsim16}). Similar to BCD, as the block sizes increase the convergence rates of each dataset improves. However, unlike BCD, the objective error does not immediately decrease for some datasets (news20 and a9a). This is expected behavior since BDCD minimizes the dual objective (see Section \ref{sec:bdcdderiv}) and obtains the primal solution vector, $w_h$, by taking linear combinations of $b'$ columns of $X$ and $w_{h-1}$. This also accounts for the non-monotonic decrease in the primal objective and primal solution errors. 

Figure \ref{fig:bdcdcost} shows the convergence behavior (in terms of the objective error) vs. flops and bandwidth costs of BDCD for the datasets and block sizes tested in Figure \ref{fig:bdcdconv}. We see that small block sizes are more flops and bandwidth efficient while large block sizes are latency efficient (from Figs. \ref{fig:news2016}-\ref{fig:realsim16}). Due to this tradeoff it important to select block sizes that balance these costs based on machine-specific parameters.


\subsubsection{Communication-Avoiding Block Dual Coordinate Descent}\label{cadcdeval}

The CA-BDCD algorithm avoids communication in the dual problem by unrolling the vector update recurrences by a factor of $s$. This allows us to reduce the latency cost by computing a larger $sb' \times sb'$ Gram matrix instead of a $b' \times b'$ Gram matrix in the BDCD algorithm. The larger condition number implies that the CA-BDCD algorithm may not be stable, so we begin by experimentally showing the convergence behavior of the CA-BCD algorithm on the datasets in Table \ref{tbl:dsets}.

Figure \ref{fig:cabdcdconv} compares the convergence behavior of BDCD and CA-BDCD for $s > 1$ with block sizes of $b' = 64, 16, $ and $64$ for the news20, a9a and real-sim datasets, respectively. The results indicate that CA-BDCD is numerically stable for all tested values of $s$ on all datasets. 
While the condition numbers of the Gram matrices increase with $s$, the numerical stability is not significantly affected. The well-conditioning of the real-sim dataset in addition to the regularization and small block size (relative to $n$) make the Gram matrices almost perfectly conditioned.

\subsubsection{Stopping Criterion}\label{sec:stopping}
\begin{figure}[t!]
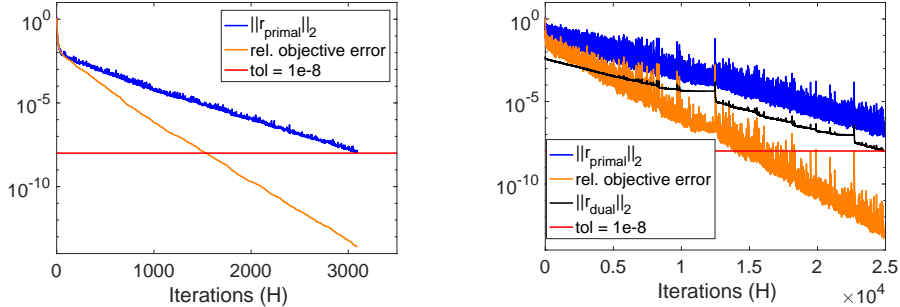

\begin{subfigure}{.49\textwidth}
\centering
\includegraphics[trim = 0in 1.5in 0in 1in, clip,width=\textwidth]{\data/../num/bcd_stopping}
\caption{BCD on a9a dataset ($b = 16$).}
\label{fig:bcdstopping}
\end{subfigure}
\begin{subfigure}{.49\textwidth}
\centering
\includegraphics[trim = 0in 1.5in 0in 1in, clip,width=\textwidth]{\data/../num/bdcd_stopping}
\caption{BDCD on a9a dataset ($b' = 16$).}
\label{fig:bdcdstopping}
\end{subfigure}
\caption{We plot the relative objective error, norm of the primal residual (for BCD Figure \ref{fig:bcdstopping}), and norm of the dual residual (for BDCD Figure \ref{fig:bdcdstopping}) for the a9a dataset with block sizes $b = b' = 16$.}
\label{fig:stopping}
\end{figure}
At iteration $h$ of BCD, we solve a $b$-dimensional subproblem
\begin{align*}
\Delta w_h = \left(\frac{1}{n}\mathbb{I}^T_hXX^T\mathbb{I}_h + \lambda \mathbb{I}_h^T\mathbb{I}_h\right)^{-1}\left(-\lambda\mathbb{I}^T_hw_{h-1} - \frac{1}{n} \mathbb{I}^T_hXz_{h-1} + \frac{1}{n}\mathbb{I}_h^TXy\right).
\end{align*}
Note that the $b$-dimensional vector, $\left(-\lambda\mathbb{I}^T_hw_{h-1} - \frac{1}{n} \mathbb{I}^T_hXz_{h-1} + \frac{1}{n}\mathbb{I}_h^TXy\right)$, is the sub-sampled primal residual vector and is explicitly computed at every iteration. Therefore, a natural stopping criteria is to occasionally compute the full-dimensional residual to check for convergence. Figure \ref{fig:bcdstopping} illustrates the convergence of the residual in comparison to the relative objective error (the optimal objective value is obtained with Conjugate Gradients) for the a9a dataset with $b = 16$. Since the optimal objective value is, in general, unknown the residual can be used as an upper bound on the objective error.

At iteration $h$ of BDCD, we solve the $b'$-dimensional subproblem
\begin{align*}
  \Delta \alpha_h =-\frac{1}{n}\left({\frac{1}{\lambda n^2} \mathbb{I}_h^TX^T X\mathbb{I}_h + \frac{1}{n}}\mathbb{I}_h^T\mathbb{I}_h\right)^{-1}\left(-\mathbb{I}_h^TX^Tw_{h-1} +  \mathbb{I}_h^T\alpha_{h-1} + \mathbb{I}_h^Ty\right)
\end{align*}
This $b'-$dimensional vector, $\left(-\mathbb{I}_h^TX^Tw_{h-1} +  \mathbb{I}_h^T\alpha_{h-1} + \mathbb{I}_h^Ty\right)$, is the sub-sampled dual residual vector. One can similarly compute the full-dimensional dual residual occasionally to check for convergence. Figure \ref{fig:bdcdstopping} illustrates the convergence of the dual residual in comparison to the relative objective error, and the primal residual. We can observe that the dual residual is a lower bound on the primal residual, therefore, the dual problem should be solved to higher accuracy. In subsequent performance experiments we occasionally compute the primal residual for (CA)-BCD and the dual residual for (CA)-BDCD to test for convergence.

\subsection{Performance Experiments}\label{sec:perfexp}
\begin{table}
\begin{center}
\footnotesize
\begin{tabular}{l|c|c|c|c|c}
Algorithm & Name & Features ($d$) & Data Points ($n$) & NNZ$\%$ & residual tolerance ($tol$) \\
\hline
\multirow{3}{*}{BCD} & a9a & $123$ & $32561$ & $11$ & 1e-2\\ \cline{2-6}
& covtype & $54$ & $581012$ & $22$ & 1e-1 \\ \cline{2-6}
& mnist8m & $784$ & $8100000$ & $25$ & 1e-1 \\ \hline
\multirow{3}{*}{BDCD} & news20 & $62061$ &$15935$ & $0.13$ & 1e-2\\ \cline{2-6}
& e2006 & $150360$ & $3308$ & $0.93$ & 1e-2\\ \cline{2-6}
& rcv1 & $47236$ & $3000$ & $0.17$ & 1e-3\\ \hline
\end{tabular}
\end{center}
\caption{LIBSVM datasets used in our performance experiments.} 
\label{tbl:perfdsets}
\end{table}

In Section \ref{sec:numexp} we showed tradeoffs between convergence behavior and algorithm costs for several datasets. In this section, we explore the performance tradeoffs of standard vs. CA variants on datasets obtained from LIBSVM \cite{cc01}. We implemented these algorithms in C/C++ using Intel MKL for (sparse and dense) BLAS routines and MPI \cite{mpi} for parallel processing. While Sections \ref{sec:anal} and \ref{sec:numexp} assumed dense data for the theoretical analysis and numerical experiments, our parallel implementation stores the data in CSR (Compressed Sparse Row) format. We used a Cray XC30 supercomputer (``Edison") at NERSC \cite{edisonhardware} to run our experiments on the datasets shown in Table \ref{tbl:perfdsets}. We used a 1D-column layout for datasets with $n > d$ and a 1D-row layout for $n < d$. We ensured that the parallel file I/O was load-balanced (i.e. each processor read roughly equal bytes) and found that the non-zero entires were reasonably well-balanced\footnote{For datasets with highly irregular sparsity structure, additional load balancing is likely required but we leave this for future work.}. We constrain the running time of (CA-)BCD and (CA-)BDCD by fixing the residual tolerance for each dataset to the values described in Table \ref{tbl:perfdsets}. We ran many of these datasets for smaller tolerances of $1e-8$ and found that our conclusions did not significantly change.

Section \ref{sec:ss} compares the strong scaling behavior of the standard BCD and BDCD algorithms against their CA variants, Section \ref{sec:rt} shows the running time breakdown to illustrate the flops vs. communication tradeoff, and Section \ref{sec:sup} compares the speedups attained as a function of the number of processors, block size and recurrence unrolling parameter, $s$.
\subsubsection{Strong Scaling}\label{sec:ss}

\begin{figure}[t!]
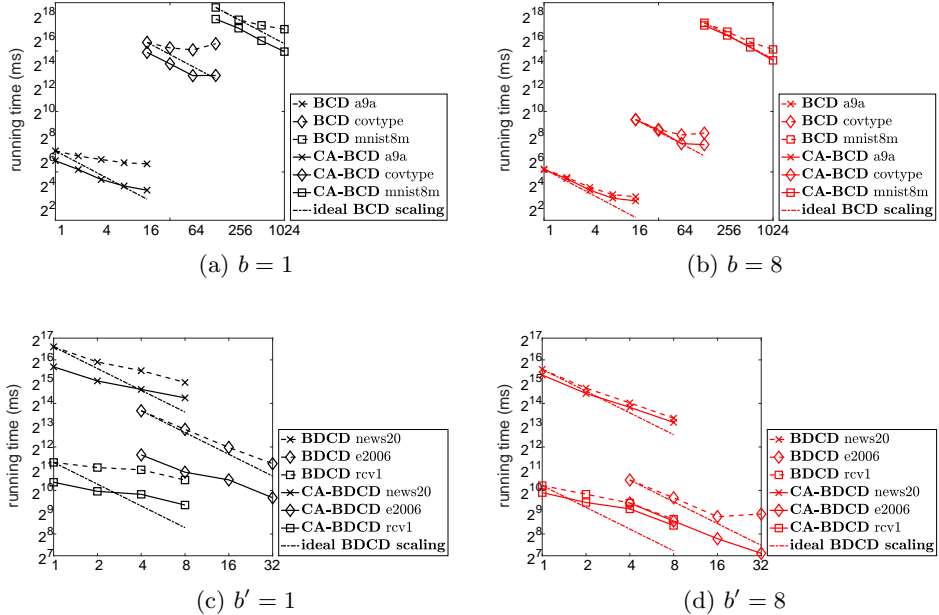

\begin{subfigure}{.49\textwidth}
\centering
\includegraphics[trim = 0in 1.5in 0in 1in, clip,width=\textwidth]{\data/../exp/bcdss_edisonb1}
\caption{$b = 1$}
\label{fig:bcdss_b1}
\end{subfigure}
\begin{subfigure}{.49\textwidth}
\centering
\includegraphics[trim = 0in 1.5in 0in 1in, clip,width=\textwidth]{\data/../exp/bcdss_edisonb8}
\caption{$b = 8$}
\label{fig:bcdss_b4}
\end{subfigure}

\begin{subfigure}{.49\textwidth}
\centering
\includegraphics[trim = 0in 1.5in 0in 1in, clip,width=\textwidth]{\data/../exp/bdcdss_edisonb1}
\caption{$b' = 1$}
\label{fig:bdcdss_b1}
\end{subfigure}
\begin{subfigure}{.49\textwidth}
\centering
\includegraphics[trim = 0in 1.5in 0in 1in, clip,width=\textwidth]{\data/../exp/bdcdss_edisonb8}
\caption{$b' = 8$}
\label{fig:bdcdss_b4}
\end{subfigure}
%
\caption{Strong scaling results for BCD/CA-BCD (top row, Figs. \ref{fig:bcdss_b1}-\ref{fig:bcdss_b4}) and BDCD/CA-BDCD (bottom row, Figs \ref{fig:bdcdss_b1}-\ref{fig:bdcdss_b4}). We report the ideal strong scaling behavior for BCD and BDCD to illustrate the performance improvements gained from the communication-avoiding variants.}
\label{fig:bcdstrong}
\end{figure}

%
%

All strong scaling experiments were conducted with one MPI process per processor (flat-MPI) with one warm-up run and three timed runs. Each data point in Figure \ref{fig:bcdstrong} represents the maximum running time over all processors averaged over the three timed runs. For each dataset in Figure \ref{fig:bcdstrong} we plot the BCD running times, the fastest CA-BCD running times for $s \in \{2, 4, 8, 16, 32\}$, and the ideal scaling behavior. We show the scaling behavior of all datasets for $b \in \{1, 8\}$ to illustrate how the CA-BCD speedups are affected by the choice of block size, $b$. When the BCD algorithm is entirely latency dominated (i.e. Figure \ref{fig:bcdss_b1}), CA-BCD attains speedups between $3.6\times$ to $6.1\times$. When the BCD algorithm is flops and bandwidth dominated (i.e. Figure \ref{fig:bcdss_b4}), CA-BCD attains modest speedups between $1.2\times$ to $1.9\times$. The strong scaling behavior of the BDCD and CA-BDCD algorithms is shown in Figures \ref{fig:bdcdss_b1} and \ref{fig:bdcdss_b4}. CA-BDCD attains speedups between $1.6\times$ to $2.9\times$ when latency dominates and $1.1\times$ to $3.4\times$ when flops and bandwidth dominated.

While we did not experiment with weak scaling, we can observe from our analysis (in Section \ref{sec:anal}) that the BCD and BDCD algorithms achieve perfect weak scaling (in theory). It is likely that the CA-BCD and CA-BDCD algorithms would attain weak-scaling speedups by reducing the latency cost by a factor of $s$, if latency dominates.

\subsubsection{Running Time Breakdown}\label{sec:rt}
\begin{figure}[t!]
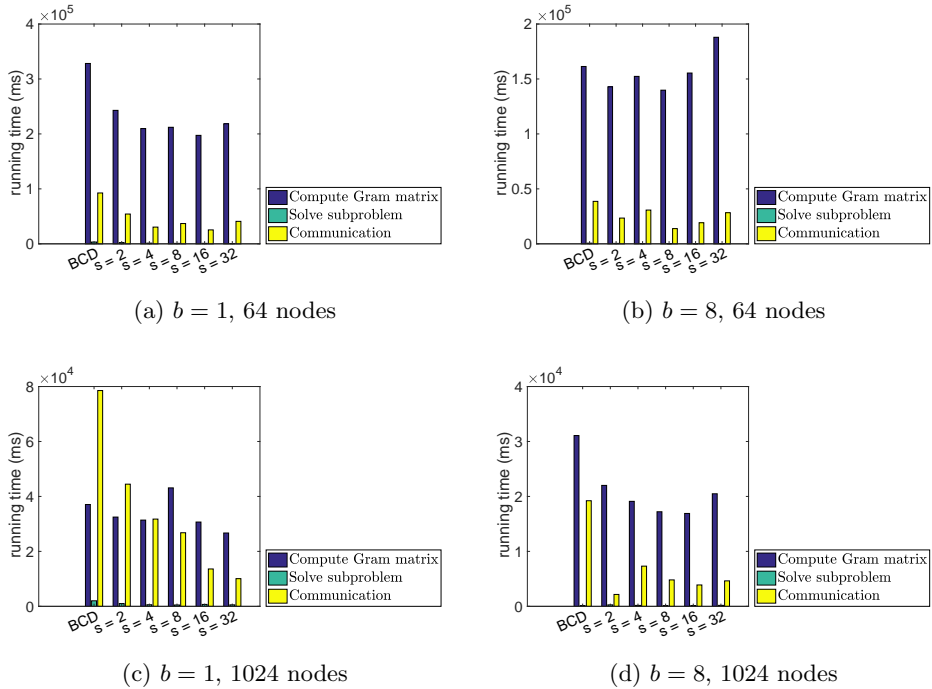

\begin{subfigure}{.49\textwidth}
\centering
\includegraphics[trim = 0.1in 1in 0.1in 1in, clip,width=\textwidth]{\data/../exp/bcdrt_64edisonb1}
\caption{$b = 1$, 64 nodes}
\label{fig:bcdrt_smallb1}
\end{subfigure}
\begin{subfigure}{.49\textwidth}
\centering
\includegraphics[trim = 0.1in 1in 0.1in 1in, clip,width=\textwidth]{\data/../exp/bcdrt_64edisonb8}
\caption{$b = 8$, 64 nodes}
\label{fig:bcdrt_smallb8}
\end{subfigure}

\begin{subfigure}{.49\textwidth}
\centering
\includegraphics[trim = 0.1in 1in 0.1in 1in, clip,width=\textwidth]{\data/../exp/bcdrt_1kedisonb1}
\caption{$b = 1$, 1024 nodes}
\label{fig:bcdrt_largeb1}
\end{subfigure}
\begin{subfigure}{.49\textwidth}
\centering
\includegraphics[trim = 0.1in 1in 0.1in 1in, clip,width=\textwidth]{\data/../exp/bcdrt_1kedisonb8}
\caption{$b = 8$, 1024 nodes}
\label{fig:bcdrt_largeb8}
\end{subfigure}
\caption{Running time breakdown for the mnist8m dataset for $b = 1$ (left column, Figs. \ref{fig:bcdrt_smallb1}-\ref{fig:bcdrt_largeb1}) and $b = 8$ (right column, Figs. \ref{fig:bcdrt_smallb8}-\ref{fig:bcdrt_largeb8}). We report the breakdown for 64 nodes (top row, Figs. \ref{fig:bcdrt_smallb1}-\ref{fig:bcdrt_smallb8}) and for 1024 nodes (bottom row, Figs. \ref{fig:bcdrt_largeb1}-\ref{fig:bcdrt_largeb8}) using the fastest timed run for each algorithm and setting.}

\label{fig:bcdrtbreakdown}
\end{figure}

Figure \ref{fig:bcdrtbreakdown} shows the running time breakdown of BCD and CA-BCD for $s \in \{2, 4, 8, 16, 32\}$ on the mnist8m dataset. We plot the breakdowns for $b \in \{1, 8\}$ at scales of 64 nodes and 1024 nodes to illustrate CA-BCD tradeoffs for different flops vs. communication ratios.
Figures \ref{fig:bcdrt_smallb1} and \ref{fig:bcdrt_smallb8} show the running time breakdown at 64 nodes for $b = 1$ and $b = 8$, respectively. In both cases flops dominate communication and most of the speedup for CA-BCD is due to faster flops. Since BCD with $b = 1$ is memory-bandwidth bound, CA-BCD with $s > 1$ increases the computational complexity and allows each processor to achieve higher flops performance through the use of BLAS-$3$ GEMM operations instead of BLAS-$1$ dot product operations. For $s \geq 8$ CA-BCD begins to saturate memory-bandwidth, therefore, speedup for $s > 8$ is due to reduction in communication time. For $b = 8$, memory-bandwidth is saturated at $s < 8$. The flops running time improves for $s < 8$ since the BLAS-$3$ calls can use larger, more cache-efficient tile sizes. For $s \geq 8$ CA-BCD becomes CPU-bound and does not attain any speedup over BCD. Furthermore, in the $b = 8$ setting, communication is more bandwidth dominated and less communication speedup is expected. On 1024 nodes (Figures \ref{fig:bcdrt_largeb1} and \ref{fig:bcdrt_largeb8}), where communication and latency costs are more dominant, CA-BCD attains larger communication and overall speedups. These experiments suggest that the CA-BCD and CA-BDCD algorithms, for appropriately chosen values of $s$, can attain large speedups when latency is the dominant cost.
%

\subsubsection{Speedup Comparison}\label{sec:sup}
\begin{figure}[t!]
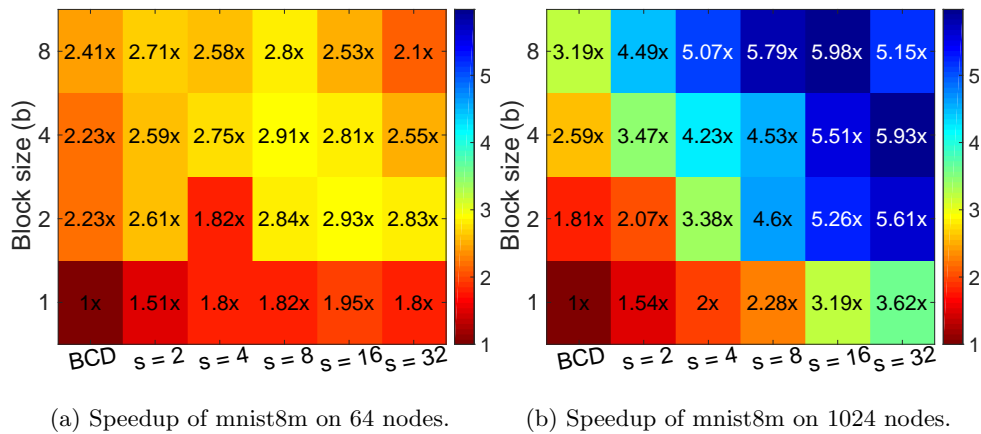

\begin{subfigure}{.49\textwidth}
\centering
\includegraphics[trim = .5in 2.5in 1.in 2.5in,  clip,width=\textwidth]{\data/../exp/bcdheatmap_mnist8m64fixedtol.pdf}
\caption{Speedup of mnist8m on 64 nodes.}
\label{fig:htmnist_small}
\end{subfigure}
\begin{subfigure}{.49\textwidth}
\centering
\includegraphics[trim = .5in 2.5in 1.in 2.5in,  clip,width=\textwidth]{\data/../exp/bcdheatmap_mnist8m1kfixedtol.pdf}
\caption{Speedup of mnist8m on 1024 nodes.}
\label{fig:htmnist_large}
\end{subfigure}
\caption{Heatmaps of the speedups achieved for CA-BCD on the mnist8m dataset for various settings of $b$ and $s$. On the left (Fig. \ref{fig:htmnist_small}) we show speedups for 64 nodes and on the right (Fig. \ref{fig:htmnist_large}) we show speedups for 1024 nodes.}
\label{fig:heatmap}
\end{figure}

Figure \ref{fig:heatmap} summarized the speedups attainable on the mnist8m dataset at 64 nodes and 1024 nodes for several combinations of block sizes ($b$) and recurrence unrolling values ($s$). We normalize the speedups to BCD with $b = 1$. At small scale (Figure \ref{fig:htmnist_small}) we see speedups of $1.95\times$ to $2.91\times$ since flops and bandwidth are the dominant costs. The speedup for larger block sizes is due to faster convergence (i.e. fewer iterations and messages) and due to the use of BLAS-3 matrix-matrix operations. Even at small scale we see that CA-BCD is fastest for all block sizes tested. At large scale, when latency dominates, (Figure \ref{fig:htmnist_large}) we observe greater speedups of $3.62\times$ to $5.98\times$. Once again, we see that CA-BCD is fastest for all block sizes tested. From Figure \ref{fig:bcdss_b4}, we see that BCD and CA-BCD for mnist8m with $b = 8$ would likely scale beyond 1024 nodes. Therefore, we can expect greater speedups for $b = 8$, when latency becomes the dominant cost.

\section{Conclusion and Future Work}\label{sec:conclusion}
%
In this paper, we have shown how to extend the communication-avoiding technique of CA-Krylov subspace methods to block coordinate descent and block dual coordinate descent algorithms in machine learning. We showed that in some settings, BCD and BDCD methods may converge faster than traditional Krylov methods -- especially when the solution does not require high-accuracy. We analyzed the computation, communication and storage costs of the classical and communication-avoiding variants under two partitioning schemes. Our experiments showed that CA-BCD and CA-BDCD are numerically stable algorithms for all values of $s$ tested, experimentally showed the tradeoff between algorithm parameters and convergence. Finally, we showed that the communication-avoiding variants can attain large speedups of up to $6.1\times$ on a Cray XC30 supercomputer using MPI.

While CA-BCD and CA-BDCD appear to be stable, numerical analysis of these methods and proofs of stability would be interesting directions for future work. Extending the communication-avoiding technique to other algorithms (SGD, L-BFGS, Newton's method, etc.), regularization (LASSO, Elastic-net, etc.) and loss functions (SVM, logistic, etc.) would be particularly interesting. 

\section*{Acknowledgements}
AD is supported by a National Science Foundation Graduate Research Fellowship under Grant No. DGE 1106400. This research used resources of the National Energy Research Scientific Computing Center, a DOE Office of Science User Facility supported by the Office of Science of the U.S. Department of Energy under Contract Nos. DE-AC02-05CH11231,  DE-SC0010200, and DE-SC0008700. This work is supported by Cray, Inc. under Grant No. 47277 and the Defense Advanced Research Projects Agency XDATA program. Research partially funded by ASPIRE Lab industrial sponsors and affiliates Intel, Google, Hewlett-Packard, Huawei, LGE, NVIDIA, Oracle, and Samsung.
\nocite{*}
\bibliographystyle{siamplain}
\bibliography{refs}
\end{document}